\documentclass[aip,rsi,preprint,numerical,endfloats*]{revtex4-1} 
\usepackage{hyperref}
\usepackage{graphicx}
\usepackage{amsmath}

\begin{document}
\title{Nanomechanical and topographical imaging of living cells by Atomic Force Microscopy with colloidal probes}


\author{Luca Puricelli}
\author{Massimiliano Galluzzi}
\author{Carsten Schulte}
\author{Alessandro Podest\`a}
\email[]{alessandro.podesta@mi.infn.it}
\author{Paolo Milani}
\affiliation{CIMaINa and Department of Physics, Universit\`a  degli Studi di Milano, Via Celoria 16, 20133 Milano, Italy}


\date{\today}

\begin{abstract}
Atomic Force Microscopy (AFM) has a great potential as a tool to characterize mechanical and morphological properties of living cells; these  properties have been shown to correlate with cells' fate and patho-physiological state in view of the development of novel early-diagnostic strategies. Although several reports have described experimental and technical approaches for the characterization of cellular elasticity by means of AFM, a robust and commonly accepted methodology is still lacking. Here we show that micrometric spherical probes (also known as colloidal probes) are well suited for performing a combined topographic and mechanical analysis of living cells, with spatial resolution suitable for a complete and accurate mapping of cell morphological and elastic properties, and superior reliability and accuracy in the mechanical measurements with respect to conventional and widely used sharp AFM tips. We address a number of issues concerning the nanomechanical analysis, including the applicability of contact mechanical models and the impact of a constrained contact geometry on the measured Young's modulus (the finite-thickness effect). We have tested our protocol by imaging living PC12 and MDA-MB-231 cells, in order to demonstrate the importance of the correction of the finite-thickness effect and the change in Young's modulus induced by the action of a cytoskeleton-targeting drug. 
\end{abstract}



\maketitle  



%
%

%

\section{Introduction}
In the last two decades scientific interest is grown around the study of cells' mechanics~\cite{Zhu2000,Suresh2007,Fletcher2010,Lim2006}, as long as its close connection to several important cellular functions like adhesion, motility, proliferation, differentiation, internal molecular transport or signal transmission, was being demonstrated. Indeed, cells can feel the outer environment and reorganize their cytoskeletal structure, an intricate network of cross-linked proteic filaments~\cite{Alberts2008,Fletcher2010}, in response to variations of their physico-chemical conditions. In general, the mechanical phenotype of cells is related to their own vital parameters~\cite{Jaalouk2009,Diz-Munoz2013,Lamour2010,Bao2003,Discher2005,Paszek2005,Suresh2007,Elson1988,Fletcher2010,Stricker2010}. 

The Atomic Force Microscope~\cite{Muller2008,Alessandrini2005} (AFM) has proved to be a valuable tool for the quantitative characterization of static and frequency-dependent mechanical properties of micro- and nanostructures, including biological specimens~\cite{Weisenhorn1993,Butt2005,Radmacher1993,Alcaraz2003,Kasas2008,Mahaffy2004,Nia2013}, thanks to its ability to sense and apply nanoscale forces, and to capture the three-dimensional topography of samples in different environments, including physiological buffers. Several papers report on the measurement by AFM of modulations in cellular elasticity induced by variations in the environmental conditions, like drugs targeting specific cytoskeletal components~\cite{Rotsch2000,Kasas2005,Oberleithner2006}, by changes in the elasticity and surface energy of the substrate~\cite{Domke2000,Tee2011,Engler2004}, as well as by the correlation between cells' elasticity and their physio-pathological state, including cancer diseases~\cite{Lieber2004,Berdyyeva2005,Lekka1999,Lekka2012,Cross2007,Plodinec2012}. In the light of these considerations, it is not unreasonable to think about future applications of AFM in bio-nano-medicine, with particular regard to the pre-diagnosis of cancer diseases, drug testing, or regenerative medicine. However, despite the considerable results obtained so far, the effective application of AFM as an effective biomedical tool is still hampered by several practical and more fundamental issues, making difficult the comparison of independent experimental results obtained in different experimental sessions or laboratories.

On one side, this is due to the extreme complexity of cellular systems. Indeed, cells are highly organized systems, characterized by a deep heterogeneity and diversity of their physico-chemical properties, even whithin the same specimen.

On the other side, the experimental methodology, regarding both the measurements and the subsequent data-analysis, of nanomechanical AFM-based tests is very complex. Artifacts can easily compromise the accuracy of the results. It is not surprising that the publication of (at least partially) methodological papers has always accompanied more biology-oriented publications (an interesting example, resulting from a recent networking effort within the EU, is represented by Ref.~\onlinecite{COST_paper}). The scenario however is still very fragmented, and a commonly accepted procedure is still missing. Our work represents an effort to contribute to the development of a common methodology.

A fundamental element of nanomechanical measurements of soft samples by AFM is the indenting probe. There is still debate about what are the best AFM probes for the nanomechanical investigation of cells, and the related issues about their spatial resolution limits and the applicability of the contact mechanics models. Commercial sharp tips, with their radius of curvature in the range 5-50 nm, are largely employed thanks to their low cost and their high resolution potential, which enables the concurrent topographical and mechanical investigation of nano-scale cellular components, like stress fibers, blebs and other fine structures~\cite{Rotsch2000,Domke2000,Pletikapic2011,Braunsmann2014,Cartagena2014}. Micrometric spherical probes, also known as colloidal probes (CPs), obtained by attaching spherical microparticles to tipless cantilevers, are not very common, not only because of their much higher cost, but also because of their incompatibility with high spatial resolution. The limited spatial resolution in fact overshadows the great advantages of CPs, i.e. the well-defined geometry, the correspondingly well- (or better-) defined contact mechanics, as well as more subtle issues such as the more uniform strain and stress distribution induced in the sample upon contact, and the smoother averaged output generally provided in force measurements, as we will discuss in more details in the following sections.

A strong point of our work is that the commonly assumed incompatibility of lateral resolution with the use of CPs in combined AFM-based topographic and nanomechanical experiments is a false problem, provided one reconsiders the meaning of cellular elasticity and the technical requirements for accurately characterizing it. These aspects are introduced and discussed in details in the manuscript. We show that CPs can provide enough resolution to accurately and simultaneously map topography and elasticity of living cells, with the ability of distinguishing among the biologically relevant cellular regions (body, nucleus, periphery, lamellipodia,...), and a number of advantages over sharp tips related to the interpretation of mechanical data.

Several aspects of the combined topographic and mechanical analysis are considered in this work, among which the impact of the finite thickness of the cellular specimen on the measured Young's modulus; this leads typically to an overestimation of the elastic modulus~\cite{Costa1999,Harris2011,McKee2011}, due to the presence of the rigid substrate (typically the plastic Petri dish bottom or a glass slide), acting as a spatial constraint to the strain and stress fields induced by the AFM probe in the cell body. We show that this effect is not negligible, and introduces artifacts depending on the local height of the cell and to the intrinsic heterogeneity of its structure. We describe a strategy to take into account the finite-thickness effect, which is based on the recent work by Dimitriadis et al.~\cite{Dimitriadis2002}. We also discuss other issues related to the rather complex data-analysis procedures for extracting quantitative information on cellular elasticity from the (typically huge) raw experimental data set, like the contact-point evaluation, the choice of the proper force (or indentation) interval on the single force curve for the Young's modulus estimation, and the estimation of the global error associated to the average Young's modulus of a population of cells. Noticeably, the accurate estimation of the contact point is an important step of the analysis, not only because it impacts directly on the fitting procedure leading to the Young's modulus estimation~\cite{Crick2006,Domke1998}, but also because this parameter is involved in the finite-thickness effect correction.

We therefore present a protocol for the combined topographical and mechanical analysis by AFM of living cells, as well as biological tissues and thin films, based on the use of micrometer-sized CPs. The protocol is meant to be an open framework within which single aspects of the topographic/nanomechanical characterization activity can be implemented, with the aim of converging towards a robust, accurate, shared methodology. The latter is essential for the quantitative comparison of independent results from different laboratories, and consequently for the effective exploitation of the AFM potential in the biomedical field. As a validation of the proposed procedure, we report on the characterization of the finite-thickness effect and of the modulation of cellular elasticity induced by Cytochalasin-D on living cells.

\section{AFM topographic/mechanical imaging of living cells by colloidal probes}
In this section we outline the major points of our protocol, introducing the issues that characterize the different practices related to the topographical and mechanical analysis of cellular systems, as they have been developed in the last two decades by several authors. Here we discuss in details three major points of nanomechanical analysis of soft samples and in particular of living cells and tissues: the choice of the probe; the quantitative estimation of the impact of the constrained geometry (the finite-thickness effect) that characterizes the mechanical imaging of living cells; the details of the complex data analysis procedure that aims at providing two numbers out of many AFM measurements, namely the average value of the Young's modulus representative of a cell population, and its error. Meanwhile, we notice that when force and sample deformations must be measured by AFM, as in nanomechanical tests, force-distance curves are typically acquired, by recording the cantilever deflection as a function of the distance travelled by the z-piezo during approaching/retracting cycles (details can be found in Ref.~\onlinecite{Butt2005}; latest developments can be found in Ref.~\onlinecite{COST_TD1002}). The acquisition of single force curves can be spatially resolved, by defining a grid of points spanning a finite area including the cell or the system under investigation and recording a set of curves, defining the so-called Force Volume~\cite{Werf1994,Radmacher1996,Heinz1999}. The main advantage of Force Volume and Force Volume -like modes based on the vertical approach of the AFM probe is that the local height of the sample can be inferred by the force curves, so that the simultaneous acquisition of the topographic map and maps of other interfacial properties (elasticity among the others) in one-to-one correspondence is possible. The aforementioned vertical-approach modes have also been modified in order to perform variable-frequency experiments (typically in the 0-500 Hz range), to test the rheological response and viscoelasticity of the samples, including cells~\cite{Radmacher1993,Mahaffy2004,Rico2005,Alcaraz2003,Rother2014}. 

\subsection{Choosing the best probe}\label{section:bestprobe}
The choice of the probe for combined topographic and mechanical analysis of soft samples, living cells in particular, must be done considering different aspects, and will always represent a compromise where pro and cons are balanced with respect to the assumptions and aims of the experimental study one is willing to carry on. Depending on the requirements concerning the spatial resolution of topographic and mechanical maps, the contact mechanics model to adopt, as well as the physical properties that must be characterized, or better the meaning one attributes to them, either standard sharp AFM tips or larger spherical beads can be preferred. The strong point we make in this work is that, provided one re-interprets the concept of cellular elasticity as an effective properties, in the sense of averaged and mesoscopic, then large spherical beads represent the best choice for the AFM probe, in terms of reliability and accuracy of the mechanical results, adaptability to theoretical models, easiness of production and characterization, \textit{as well} as spatial resolution. These points will be discussed in details in the following sections. 
    
\subsubsection{Sharp tips}
The radius of curvature of the AFM probe determines the lateral resolution of the acquired maps. Despite the outstanding capability of resolving nanoscale mechanical (and morphological) heterogeneities of living cells demonstrated by conventional sharp AFM tips~\cite{Rotsch2000,Domke2000,Pletikapic2011,Braunsmann2014,Cartagena2014}, the applicability of hard sharp indenters against very soft and fragile samples like cells has been questioned~\cite{Harris2011,Carl2008,Dimitriadis2002,Mahaffy2004}. For instance, a tendency by sharp tip to overestimate cells' rigidity has been reported and discussed~\cite{Harris2011,McKee2011,Costa1999}, and the applicability to sharp indenters of contact mechanics models have been addressed (the most commonly used non-adhesive models are introduced in Appendix~\ref{app:probe-shapes} and Table~\ref{tab:Probe_shapes}). Numerical simulations have also been performed to investigate and clarify these issues~\cite{Costa1999,Lin2008}.
The critical points in using sharp AFM tip in mechanical measurements can be summarized as follows:

\begin{enumerate}
\item The nanometer-sized radius of curvature implies the application of high stress; the induced strain in soft samples like cells can be very large, even with very small applied forces. 
\item Significant deviations from the nominal geometries are frequent; moreover, evaluation of the relevant geometrical parameters can be difficult and its results inaccurate.
\item Cells are characterized by a heterogeneous and diverse structural complexity, and manifest strong dynamical nanoscale activity, so that the mechanical \textit{nanoscale} properties are a rather poorly-defined concept.
\end{enumerate}

Concerning point 1, Dimitriadis and coworkers~\cite{Dimitriadis2002} have shown how probes with radii of curvature $R \leq 100nm$ can induce very high strains even for forces down to a few pN, which represents the lower limit of force detection for AFM optical beam deflection systems (see Figure 4 in Ref.~\onlinecite{Dimitriadis2002}). Under these conditions, damage of the cell's outer membrane or the underlying cytoskeletal structure is likely; moreover, and more fundamental, the small strain assumption, which is the basis of the most commonly used contact mechanics models, is not satisfied. 

Concerning point 2, i.e. the tip geometry, it is well known that commercial tip shapes often exhibit departures from the ideal ones on which contact mechanics models rely, thus providing an ill-defined contact area and force-indentation response. Contact mechanics models apply therefore only approximately to the case of the contact between a hard sharp indenter and a soft sample like a live cell. In particular, the actual geometry of the commercial probe is only partially accounted for by the existing models, therefore, for as good as one can characterize it (which can be rather time consuming and potentially destructive for the tip), there will always be a residual unaccountable uncertainty. Moreover, the difficulties related to the measurement of the relevant geometrical parameters (such as the tip opening angle) of sharp tips can lead to errors of the order of 15-20\%~\cite{COST_paper}.

In addition to technical issues, point 3 suggests that the cell structure, and in particular its peculiar nanoscale inhomogeneity, must also be considered when choosing the best probe for mechanical tests. Larger standard deviations of results are typically observed when using sharp indenters~\cite{Dimitriadis2002}. This feature can be a direct consequence of the high lateral resolution achievable with standard AFM sharp tips, because these probes are very sensitive to fine nanoscale inhomogeneities of the sample~\cite{Carl2008,Dimitriadis2002}. Moreover, cells' dynamical activity on the nanoscale can introduce significant (time-dependent) fluctuations in the measured Young's modulus values. A higher spatial resolution, despite being an advantage from the topographic point of view, can turn into a disadvantage for the mechanical characterization, providing a less robust and less accurate estimation of the Young's modulus of the cell.

\subsubsection{\label{section:CP}Colloidal spherical probes}
Although several reports on the measurement of mechanical properties of living cells by AFM have been published so far, only a fraction of them is based on the use of CPs, among them see Refs~\onlinecite{Ducker1991,Carl2008,Dimitriadis2002,Mahaffy2004,Harris2011} ; as far as we know, the topographic imaging capabilities of CPs are typically not mentioned in such studies  (with a few remarkable exceptions, as in Ref.~\onlinecite{Alesutan2013}). CPs represent in our opinion a better alternative to sharp tips whenever high resolution is not required, although, as we will show and discuss, they can provide satisfactory spatial resolution for the sake of cell topographic and mechanical characterization; CPs are the key feature of the AFM-based topographic/mechanical imaging protocol of living cells we will describe in the following sections. The main advantages of CPs versus sharp tips can be summarized as follows:

\begin{enumerate}
\item The applied force is distributed across a much wider (micrometric) area, thus significantly reducing stress and strain in the cell.
\item CPs have a well-defined geometry, which can be easily characterized (see Appendix~\ref{app:CP_characterization} and Refs~\onlinecite{Indrieri2011,Mahaffy2004,Neto2001}). The Hertz model (first in Table~\ref{tab:Probe_shapes}) describes accurately the indentation of elastic bodies by spherical CPs.
\item CPs smear out nanoscale inhomogeneities, providing mesoscopic robust values of the Young's modulus of samples~\cite{Carl2008,Dimitriadis2002}.
\item CPs can be functionalized (thanks to their wide surface area) in order to study specific ligand/receptor interactions at the cell's surface.
\item CPs can be purchased, as well as produced directly in the laboratories, according to established protocols (see Appendix~\ref{app:CP} and Refs~\onlinecite{Indrieri2011,Bonaccurso2001,Kuznetsov2012,Gan2007,Mak2006}).\end{enumerate}

\begin{figure}
\includegraphics[width=8.5cm,keepaspectratio]{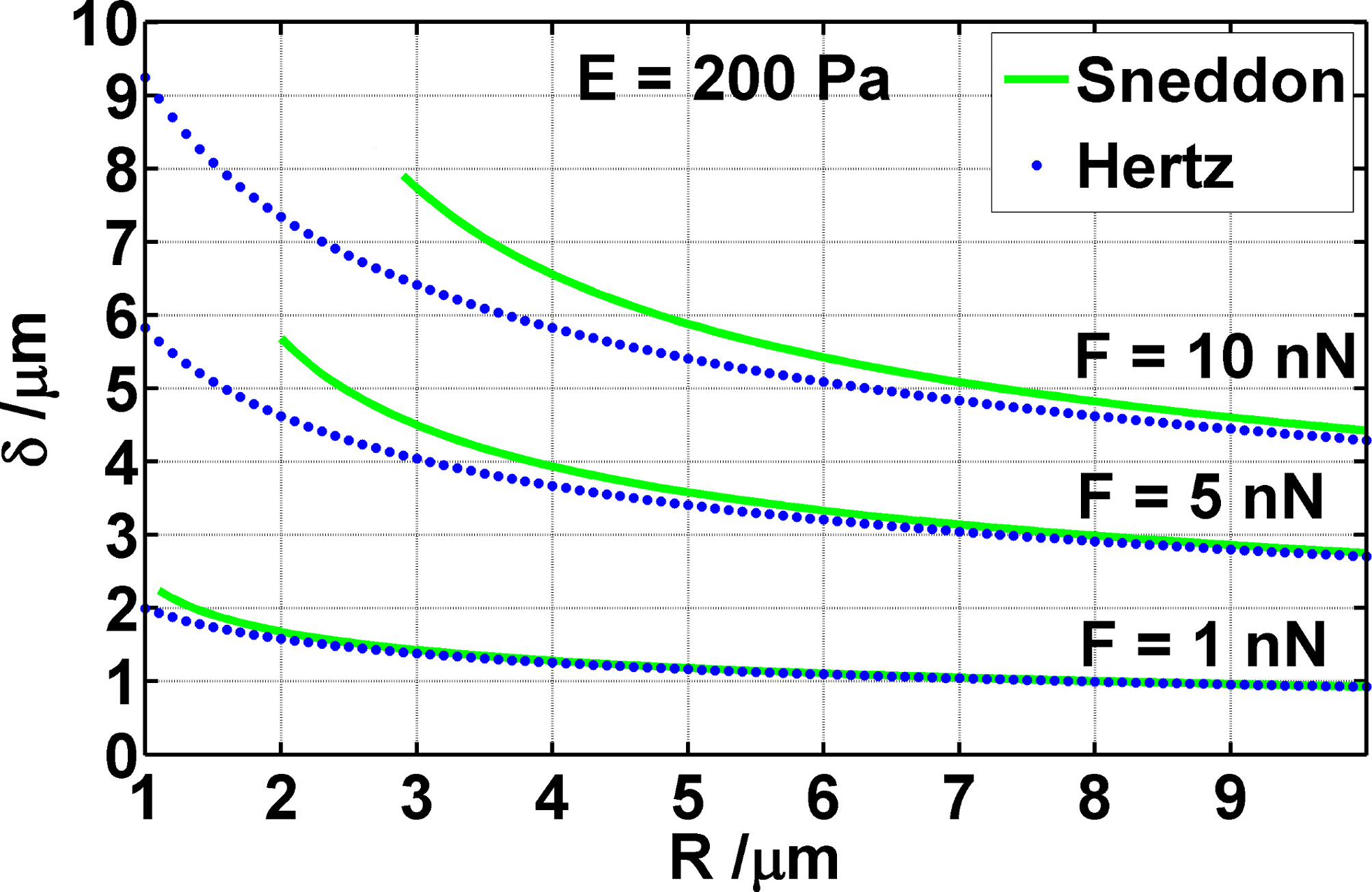}
\caption{\label{fig:Indentation-vs-tipradius}Indentation vs tip radius for the case of a spherical micrometric probe according to the prediction of Hertz and Sneddon models; the Young's modulus (E = 200 Pa) and the selected forces are taken as fixed parameters.}
\end{figure}

\begin{figure}
\includegraphics[width=8.5cm,keepaspectratio]{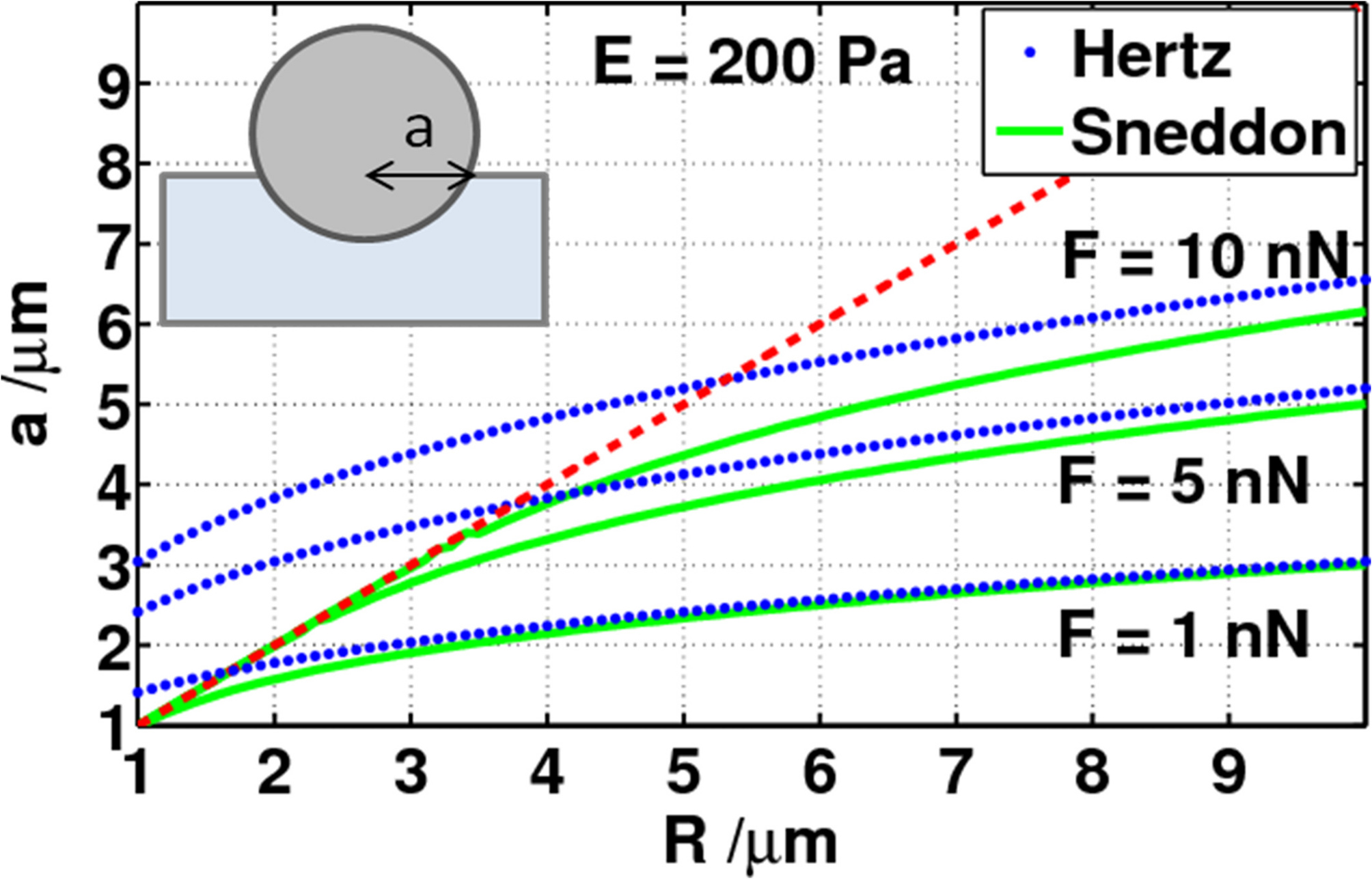}
\caption{\label{fig:contactradius_vs_tipradius}Contact radius (shown in the inset) vs tip radius for the case of a spherical micrometric probe according to the prediction of Hertz and Sneddon models; the Young's modulus (E = 200 Pa) and the selected forces are taken as fixed parameters. The red dotted line marks the region for which the solutions of the Hertz model have physical meaning ($a \le R$).}
\end{figure}

Micrometric CPs satisfy more easily the requirement of small strains at the basis of contact mechanics models (in the linear elastic regime), and at the same time significantly reduce the risk of cell's damage. In order to fully appreciate the advantages of CPs with respect to the applicability of contact mechanics models, we consider the commonly used model for spherical indenter, the Hertz model, first in the list of Table~\ref{tab:Probe_shapes}. This model, firstly proposed by H. Hertz in 1881~\cite{Hertz1881}, describes the local deformations related to the contact of two spheres of radius $R_1$ and $R_2$, characterized by a Young's modulus $E_1$ and $E_2$; the equation shown in Table~\ref{tab:Probe_shapes} and commonly used represents the limiting case of a rigid sphere indenting an elastic half-space ($R_1 \rightarrow +\infty$, $E_1\ll E_2$ or $E_2 \rightarrow +\infty$), which is not exactly the same situation for which the Hertz model had been originally developed (actually the opposite case, a soft indenter on a rigid surface~\cite{Heuberger1996}). Moreover, Hertz calculation is based on the assumption that the contact radius $a$ (see the inset in Figure~\ref{fig:contactradius_vs_tipradius}) is small compared to the radius of the sphere, that is $a\ll R$; this, in turn, implies that the deformation of the sample is much smaller than the radius of the sphere $\delta\ll R$, knowing that contact radius and deformation are related by the equation $\delta=a^{2}/R$ (for more details, see Ref.~\onlinecite{Johnson1985}). Using sharp tips, especially when indenting soft samples like cells, the hypotheses behind the Hertz model can be hardly satisfied~\cite{Heuberger1996}. The model developed by Sneddon, which can be applied to several geometries including the spherical one, does not suffer from the  constraint $\delta\ll R$~\cite{Sneddon1965}, and more appropriately describes the case of a rigid indenter on a deformable surface (details in Appendix~\ref{app:probe-shapes}). Unfortunately, Sneddon's equation for the spherical probe (Eq.~\ref{eq:Sneddon_force}) cannot be cast in an analytic closed form of the kind of Eq.~\ref{eq:Hertz_generalized}, but requires numerical methods to be solved~\cite{Heuberger1996}. Nevertheless, it is possible to prove that when the radius of the spherical indenter is larger than approximately 1-2 $\mu m$, the Hertz model, despite its more severe constraints, represents a very good approximation of the Sneddon model, therefore being appropriate to describe the force-indentation characteristics of soft samples, including living cells. Following Ref.~\onlinecite{Heuberger1996}, we show in Figure~\ref{fig:Indentation-vs-tipradius} the comparison of Hertz and Sneddon models describing the normal indentation of a soft sample as a function of the probe radius, for a Poisson coefficient $\nu = 0.5$ and three different force values (selected in the range typically observed during experiments). We see that for spheres with $R = 5 \mu m$ even in extreme cases ($E=200 Pa$, $F=10 nN$) the relative discrepancy between the two models is well below 10\%, and obviously decreases for milder conditions, so that for $F=1 nN$ the relative discrepancy is only 1.7\%, and further decreases for stiffer samples. Similarly, Figure~\ref{fig:contactradius_vs_tipradius} shows that when $R\gtrsim 5 \mu m$  the unphysical region where $a/R>1$, in the same critical conditions, is never reached in Hertz model (Sneddon model has no such limitations like $a/R\ll1$). In summary, the simple Hertz model represents always a good approximation for the contact mechanics of micrometric CPs, while its application to the case of sharp tips is questionable.

Concerning the other aspects, it has already been shown that micrometric spherical probes with different radii are able to provide comparable Young's modulus' values on both soft gels (Table 1-2 in Ref.~\onlinecite{Dimitriadis2002}) and living cells (Figure 3 in Ref.~\onlinecite{Carl2008}): in the case of gels, they are also in good agreement with macroscopic measurements~\cite{Dimitriadis2002}. The contact geometry of CPs has been shown to be in excellent agreement to theoretical expectations by combined AFM and fluorescence microscopy analysis~\cite{Harris2011}. Moreover, it has been demonstrated that it is possible to detect length and grafting density of the brush layer on living cells~\cite{OCallaghan2011,Sokolov2007}, or even to distinguish between cancerous and healthy cells' brushes~\cite{Iyer2009}, thanks to the sensitivity of micrometric spherical probes provided by their wider surface area.

\subsubsection{Colloidal probes test an effective, yet biologically relevant, elasticity}
It is clear that using micrometric spherical probes we must drop the idea of obtaining nano-scale lateral resolution in both topographic and mechanical maps, definitely loosing information about fine structures like single stress fibers. As far as nanoscale heterogeneities at the membrane and close sub-membrane level are the object of the AFM investigation, sharp commercial tips have to be used; very remarkable results in coupling the nanoscale topographic imaging to the nanomechanical analysis have been obtained using this experimental configuration and published so far as already pointed out. On the other hand, in the light of the previous discussion, we must take into consideration that CPs can represent the best choice in order to obtain mesoscopic, statistically robust values of living cells' Young's modulus. By mesoscopic, it is meant that the modulus should not reflect the mechanical properties of each single nanoscale entity of the complex cellular structure, such as an actin fiber, a microtubule, etc., but instead should represent the average combined collective action of several such entities, in numbers as well as typology. In other words, the effective elasticity of a cell in a given region is the result of the contribution of the membrane, the cytoskeleton and all its components, and possibly of other cellular elements, those that can be found in a reasonably large (mesoscopic) volume of the order of $1 \mu m^3$. This effective mesoscopic elasticity is likely more relevant to clinical, pre-diagnostic applications of AFM nanomechanics than the elasticity of the single nanoscale cellular component, as can be determined by using sharp tips. Once this point of view is accepted, then CPs represent the natural and optimal choice for carrying on a combined topographic and mechanical analysis, as will be definitely confirmed in the next section.

\subsubsection{Lateral resolution of colloidal probes}
\begin{figure*}
\centering
\includegraphics[width=17cm]{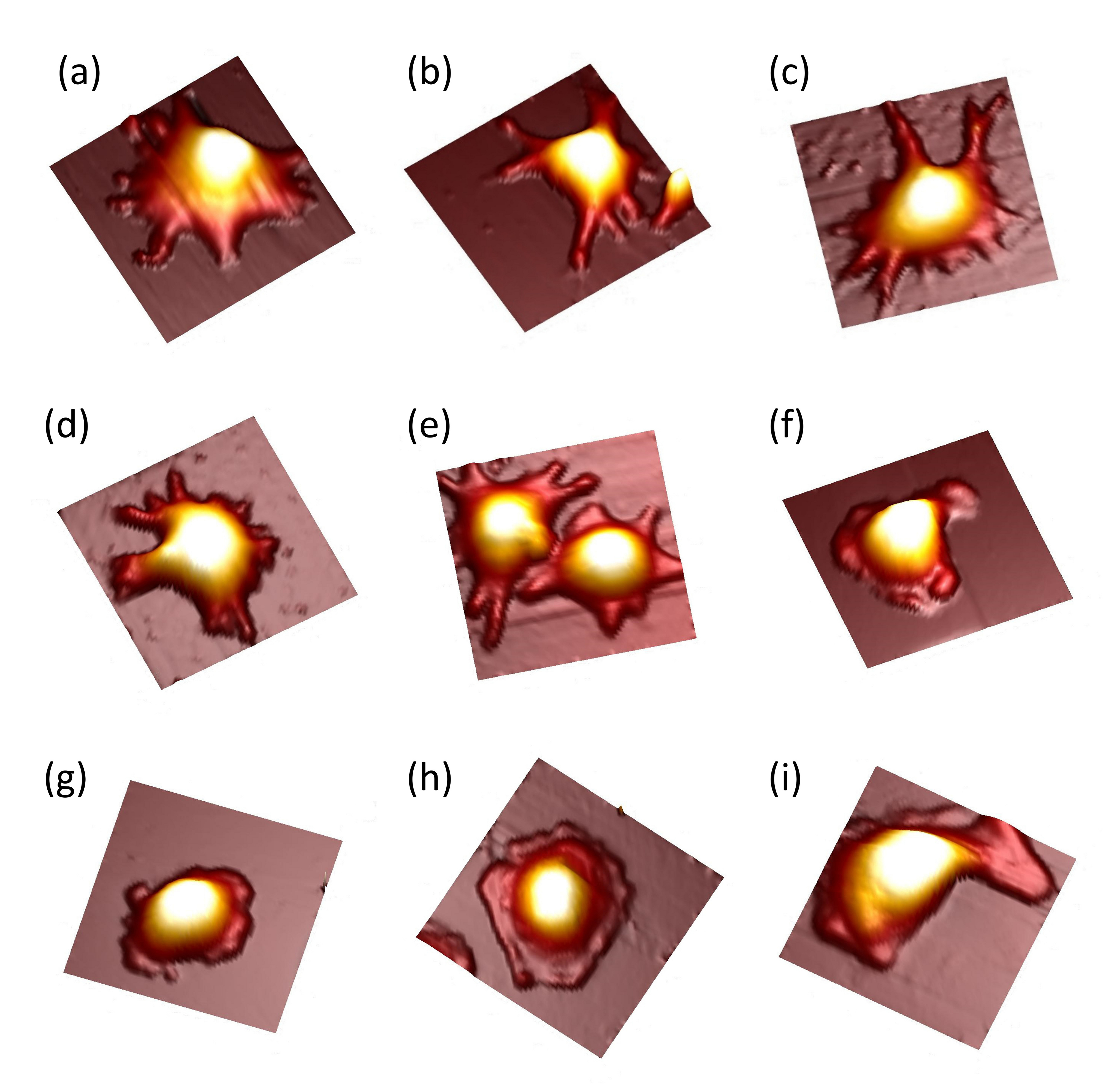}
\caption{Three-dimensional views of AFM topographic maps of living cells recorded in Force Volume mode using CPs with $R\simeq 5\mu m$. Scan size and vertical range are typically $50-90 \mu m$ and $5-6 \mu m$, accordingly. Cells belong to the PC12 (A-E), and MDA-MB-231 (F-I) lines. The main parts of the cell are clearly resolved by the colloidal probe, despite its micrometric dimension.\label{fig:belle_topo_3D_cellule}}
\end{figure*}

As long as one accepts that the effective mesoscopic Young's modulus is the relevant physical observable, the requirements on the lateral resolution of the topographic as well as of the nanomechanical analysis can be relaxed, the minimal resolution becoming the one allowing distinguishing among the major cellular regions, as the cell body, with the underlying nuclear region, and the cell periphery, with lamellipodia, cellular extensions, etc. A lateral resolution of a few microns is therefore more than enough to this purpose. Using CPs with radius up to $5\mu m$, an effective lateral resolution of $1-2 \mu m$ can be obtained, due to the fact that the contact radius $a$ is typically only a fraction of the probe radius (with $R=5 \mu m$ and $\delta=0.5 \mu m$, $a=\sqrt{\delta R}= 1.6 \mu m$; the sampling resolution of the force volume mapping can be comparable or even better). The lateral resolution can be appreciated in Figure~\ref{fig:belle_topo_3D_cellule}, where several three-dimensional views of living cells belonging to the PC12 and MDA-MB-231 lines imaged in Force Volume mode using CPs with radius of appproximately $5 \mu m$ are shown (images have not been corrected for the deformation, as explained in Section~\ref{section:contact-point}). By applying suitable masks to the Force Volume data set, built directly on the topographic map, it is possible to select the force curves corresponding to specific regions of the cell, and perform a site-resolved analysis of cellular elasticity; combining topographic to mechanical analysis is clearly an advantage from this point of view, with respect to performing only the mechanical analysis on pre-selected locations. The other great advantage of combining topographic and mechanical imaging is that the knowledge of the local height of the sample is crucial to implement the correction of the finite-thickness effect, as described in the following section. 

\subsection{Finite-thickness effect}
Beside the requirement of small strains, contact mechanics models are typically based on another important assumption: the sample to be probed should be an elastic half-space, or at least its thickness should be much larger than the maximum indentation value ($h \gg \delta$). This is equivalent to ignore the possible influence of the rigid substrate supporting the sample (i.e. the bottom of a Petri dish, in the case of living cells). For cells adherent to a substrate, this condition is not satisfied, since cells' height typically varies from a few $\mu m$ on the top (the nucleus region), to a few hundreds nm on the peripheral regions, and indentation can easily exceed $1 \mu m$. In literature, this problem has been so far either poorly considered or addressed restricting, as a workaround, the mechanical analysis to the higher cells' areas (typically in the cell body), using small indentations. This solution keeps us from getting mechanical insights on all cellular regions and components, by limiting the mechanical analysis to the shallower regions as the outer membrane with the actin layer; in particular it precludes the investigation of the thinner peripheral regions (as the cytoplasmic protrusions or lamellipodia), being responsible for the cells' motility and rich of focal adhesions points, which are potentially rich in information about cellular dynamics and interactions.

A few reports systematically investigate the influence of sample's thickness on the measured apparent Young's modulus, either experimentally or by means of numerical simulations~\cite{Akhremitchev1999,Mahaffy2004,Oommen2006,Domke1998,Costa1999,Sun1995,Long2011,Santos2012}. Dimitriadis et al.~\cite{Dimitriadis2002} developed an analytic approximate correction for spherical probes in the limit of small indentation, to take into account the aforementioned effect due to the finite thickness of the sample (similar corrections have has been reported for conical indenters~\cite{Santos2012,Gavara2012}; for a study not limited to small indentation, see Ref.~\onlinecite{Long2011}). Dimitriadis et al. analysed the two opposite conditions of a sample free to move (i.e. sliding) on the substrate, and of one that is well-bound to it:

\begin{equation}
F_{free}=\frac{16}{9}ER^{\frac{1}{2}}\delta^{\frac{3}{2}}\left[1+0.884\chi+0.781\chi^{2}
+0.386\chi^{3}+0.0048\chi^{4}\right]\label{eq:ft_corr_free}
\end{equation}

for the case of free-to-move sample and 

\begin{equation}
F_{bound}=\frac{16}{9}ER^{\frac{1}{2}}\delta^{\frac{3}{2}}\left[1+1.133\chi+1.283\chi^{2}
+0.769\chi^{3}+0.0975\chi^{4}\right]\label{eq:ft_corr_bonded}
\end{equation}

for the case of the well-adherent sample. Here, $\chi=\frac{\sqrt{R\delta}}{h}$ is an adimensional variable that combines the critical lengths of the system. Noticeably, being $\delta=a^{2}/R$, we have $\chi = a/h$, i.e. the finite-thickness effect does not depend trivially on the ratio of vertical lengths $\delta$ and $h$, but rather on the ratio of the horizontal dimension of the contact ($a$) to the sample height $h$. The correction therefore takes into account the development of the strain and stress fields into the bulk volume of the sample, not only their vertical extension. As a consequence, we must expect stronger finite-thickness effects for colloidal probe with respect to sharp ones. The validity of Eqs~\ref{eq:ft_corr_free},~\ref{eq:ft_corr_bonded} is confirmed by the comparison with the numerical solutions for the system equations (as shown in Refs~\onlinecite{Dimitriadis2002,Santos2012}), and the good agreement - especially for microspheres with $R= 5 \mu m$ - with the macroscopic Young's modulus value measured on thin test samples of Poly(vinyl alcohol) gels. We can notice that the first term (oustide the square brackets) is nothing but the Hertz model of Table~\ref{tab:Probe_shapes} evalueted for $\nu= 0.5$, whereas the term in the square bracket is an adimensional factor, dependent on the local height and deformation of the sample, quantitatively describing the effect of the rigid substrate on which the cells are adhered, and vanishing as $h$ becomes very large.

A relevant question is whether cells should be considered free-to-move or well-adherent samples. Cells adhere to the underlying substrate through local dynamic structures named focal adhesions (which cells can destroy and rebuild depending on their necessities), therefore they could be classified as inhomogeneously bound samples, i.e. as intermediate cases in between the two limiting conditions. Following the advice of Gavara and Chadwick~\cite{Gavara2012}, we averaged Eqs.~\ref{eq:ft_corr_free} and~\ref{eq:ft_corr_bonded} calculating the arithmetical mean of the coefficients of the powers of the variable $\chi$, obtaining:

\begin{equation}
F=\frac{16}{9}ER^{\frac{1}{2}}\delta^{\frac{3}{2}}\left[1+1.009\chi+1.032\chi^{2}
+0.578\chi^{3}+0.051\chi^{4}\right]\label{eq:ft_corr_average}
\end{equation}

This is the effective equation for the correction of the finite-thickness effect that will be applied in the following sections. Defining the correction factor $\Delta (\chi) \equiv \Delta (\chi (R,\delta,h)) $:

\begin{equation}
\Delta=1+1.009\chi+1.032\chi^{2}+0.578\chi^{3}+0.051\chi^{4}\label{eq:Delta}
\end{equation}

Eq.~\ref{eq:ft_corr_average} can be rewritten as:

\begin{equation}
F=\frac{16}{9}ER^{\frac{1}{2}}\delta^{\frac{3}{2}}\Delta (\chi (R,\delta,h))\label{eq:ft_corr_average_scaled}
\end{equation}

\begin{figure}
\includegraphics[width=8.5cm,keepaspectratio]{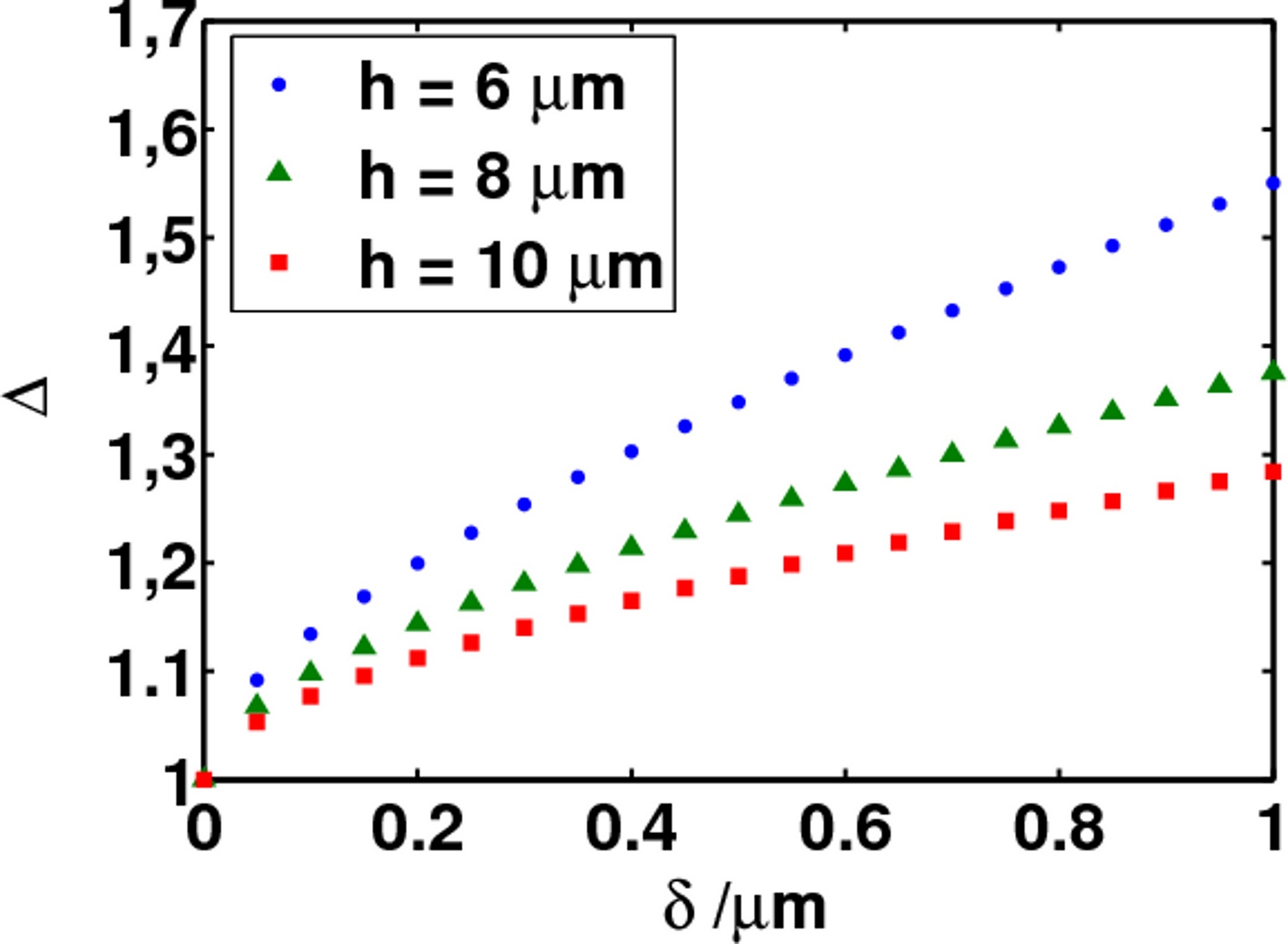}
\caption{\label{fig:delta_vs_indent}Finite-thickness correction adimensional factor $\Delta$ vs indentations $\delta$ for typical cells' height values ($R=5 \mu m$). For typical indentations of several hundreds nm, even in the limit of $h \simeq 10 \mu m$ $\Delta$ is not negligible.}
\end{figure}

Figure~\ref{fig:delta_vs_indent} shows that for a probe of radius $R = 5 \mu m$ and indentations $\delta$ of several hundreds nm, a typical situation in cell's mechanical analysis, the correction factor $\Delta$ is not negligible even for heights $h$ as large as $10 \mu m$.

\subsection{Data analysis}
Here we describe and discuss the data-analysis procedures of the proposed protocol, aiming at defining a robust environment for the extraction of reliable and accurate results from nanomechanical data-sets. All data-analysis procedures and algorithms have been implemented in Matlab environment.

Raw data must be pre-treated and converted so that force vs indentation curves in the proper units are obtained. These preliminary steps are described in Appendix~\ref{app:preliminary_analysis}.

\subsubsection{Linearization of force curves}
Following Carl and Schillers~\cite{Carl2008}, we rewrite Eq.~\ref{eq:ft_corr_average_scaled} in the linearized form:

\begin{equation}
\delta-\delta_{0}=\alpha F^{*}\label{eq:ft-linearized}
\end{equation}

where $\delta_0$ is explicitly introduced in order to account for the contact point, so that Eq.~\ref{eq:ft-linearized} holds for $\delta \geq \delta_0$, ($\delta_0$ is implicitly assumed equal to zero in Table~\ref{tab:Probe_shapes}); $\alpha$ is a constant containing the Young's modulus $E$, given by:

\begin{equation}
\alpha=\left(\frac{9}{16}\frac{1}{ER^{\frac{1}{2}}}\right)^{\frac{2}{3}}
\end{equation}

(with the assumption of Poisson ratio $v=0.5$); $F^*$ represents an effective pseudo-force, i.e. a variable with the dimensions of a force raised to the power of 2/3, given by:

\begin{equation}
F^{*}=\left(\frac{F}{\Delta}\right)^{\frac{2}{3}}\label{eq:pseudo_effective_force}
\end{equation}

where $\Delta$ is the finite-thickness correction factor defined in Eq.~\ref{eq:Delta}. $\delta_0$ and $\alpha$ are left as free parameters while fitting the experimental data.

From a technical point of view, linearization of Eq.~\ref{eq:ft_corr_average_scaled} into Eq.~\ref{eq:ft-linearized} allows extracting the value of the Young's modulus from experimenal data by means of a simple linear fit, which in turn can be done in parallel over a set of thousands force curves simultaneously by applying a matrix linear regression, thus drastically reducing the computational time, with respect to a serial iterative fitting procedure.

\subsubsection{\label{section:lineartrends}Multiple elastic regimes and their detection}
\begin{figure}
\includegraphics[width=8.5cm,keepaspectratio]{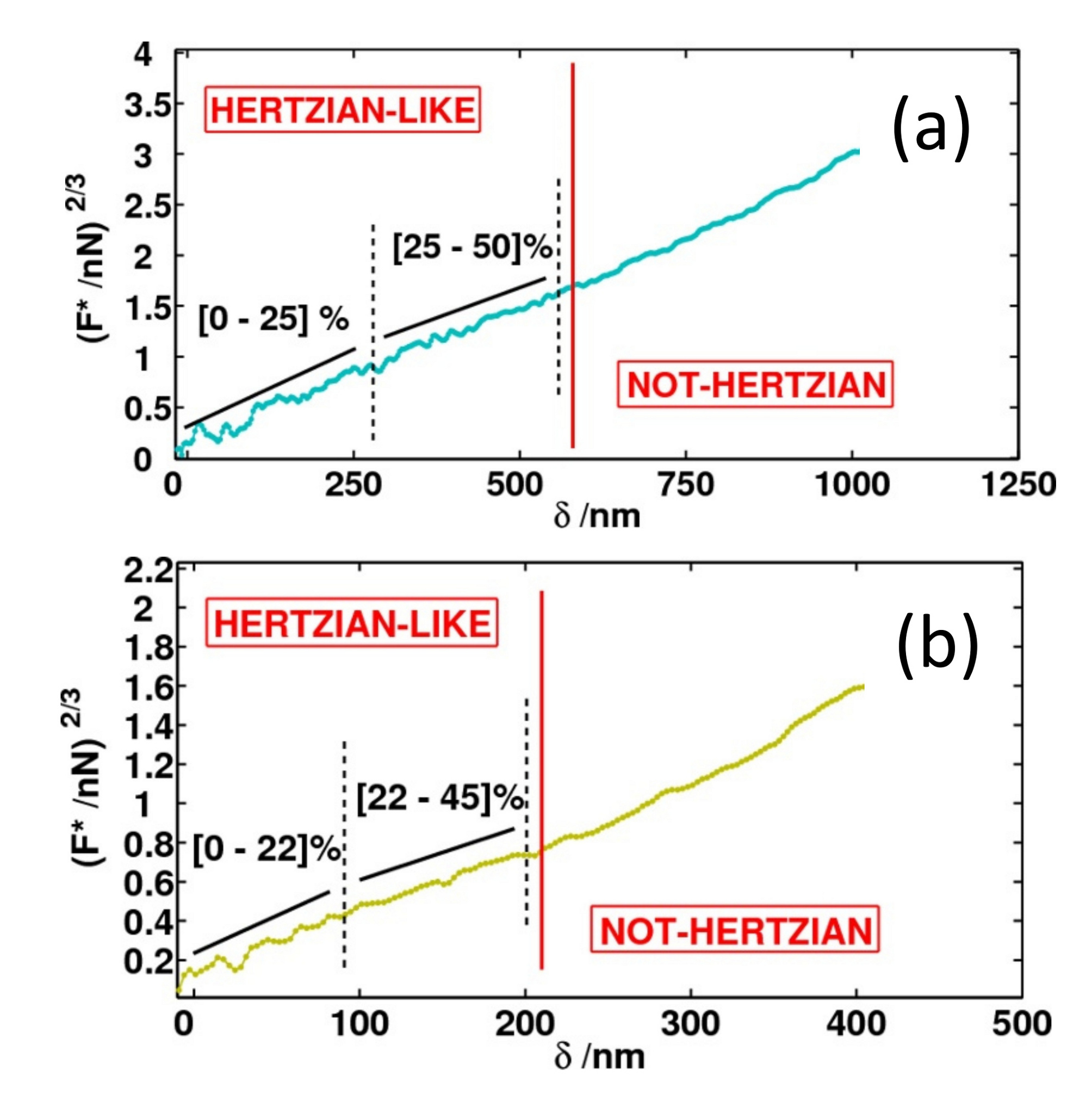}
\caption{\label{fig:curvelin_regimi}Experimental force curves from a PC12 cell linearized according to Eq.~\ref{eq:pseudo_effective_force}: two different elastic regions are clearly visible, together with a non-linear region. The corresponding indentation percentages are reported. a) Force curve taken over the cellular nucleus b) Force curve taken over the cellular peripheral regions or cytoplasmic extensions}
\end{figure}

From the physical point of view, as already pointed out in Ref.~\onlinecite{Carl2008}, linearization of the force-indentation model allows performing a direct inspection on the cell's internal structure: indeed, the presence of multiple linear regimes in the linearized $F
^*(\delta)$ curve can be the signature of different elastic regimes inside the cell, because the slope of the curve depends on the local elastic modulus. In Figure~\ref{fig:curvelin_regimi} the presence of several sloped portion of the curve (from a PC12 cell) is clearly visible, well beyond the noise level. It is also possible to notice a final non-linear region, where the force curve departs from a purely elastic behaviour, and/or where the finite-thickness correction model no longer holds. Trends like the one shown in Figure~\ref{fig:curvelin_regimi} represents an experimental evidence of the fact that cells are heterogeneous and complex systems, where different structural components contribute to the overall mesoscopic elasticity, and/or become active in a particular indentation range, depending also on their spatial distribution inside the cell's body.

When analyzing thousands curves together and automatically, it would be useful to find a quick and automated way for identifying the different elastic ranges, or at least the region where Hertz model (corrected for the finite thickness effect) can provide us with reliable mechanical informations. The implementation of such automatic algorithm is very challenging, for several reasons. First, cells are heterogeneous anisotropic specimens, so that we can observe a large variability of the value and the indentation range of different slopes from curve to curve; moreover, because of local different tihickness, the range of each linear region can change moving from one region of the cell to the other. As a consequence, averaging different curves is not an option, because each curve is not in registry with the others. Second, the presence of noise further complicates the identification of different sloped regions. An interesting and robust solution to this problem was recently suggested by Polyakov et al.~\cite{Polyakov2011}, and it is based on a segmentation algorithm of force curves; the main drawback of the method, as highlighted by the authors themselves, is represented by the computational time required, which could reach several hours for the analysis of a single Force Volume set.

In the light of these considerations, we decided to choose a semi-automatic approach based on the hypothesis that the relative amount of cytoskeletal components is approximately conserved in the cells, irrespective to the region, so that the absolute amount of each components is proportional to the local height of the sample. Under this simplified hypothesis, the relative width of each single elastic range is conserved in the force curves. The relative widths of different elastic regimes can be identified manually on a representative subset of force curves from different cell's regions and from different cells belonging to the same line and measured in the same experimental conditions, then set for all curves and the slopes evaluated by automatic parallel linear regression, as discussed above. The determination of linear ranges is the most time-consuming step of the procedure, but once the percentage ranges are defined it allows analysing automatically all the Force Volume sets from the same  experiment.

\subsubsection{\label{section:contact-point}Contact Point evaluation}
The absence of relevant adhesive interactions between the AFM probe and the cell surface represents an advantage when it allows using the simple Hertz model; the drawback is the lack of a clear jump-in structure in force curves, which helps identifying the contact point $\delta_0$, which represents the origin of the indentation axis. In liquid, noise makes this identification even more difficult. Errors in the determination of the contact point may lead to poor estimation of the Young's modulus~\cite{Crick2006,Dimitriadis2002}. The accurate estimation of the contact point is also important for the correction of the finite-thickness effect, because $\delta_0$ enters the definition of the $\chi$ parameter twice: through the indentation $\delta$, which is calculated with respect to $\delta_0$, and through the sample thickness $h$; indeed, the true height $h$ of the sample is obtained adding the total elastic indentation to the apparent height, measured under compressive load in correspondence of the force setpoint (see Figure 3 of Ref.~\onlinecite{Dimitriadis2002}), and eventually subtracting the average height of the substrate.

It is commonly accepted that the best way to determine the contact point is by means of a fitting procedure~\cite{Dimitriadis2002,Lin2007,Domke1998,Crick2006}.

\begin{figure*}
\includegraphics[width=17cm,keepaspectratio]{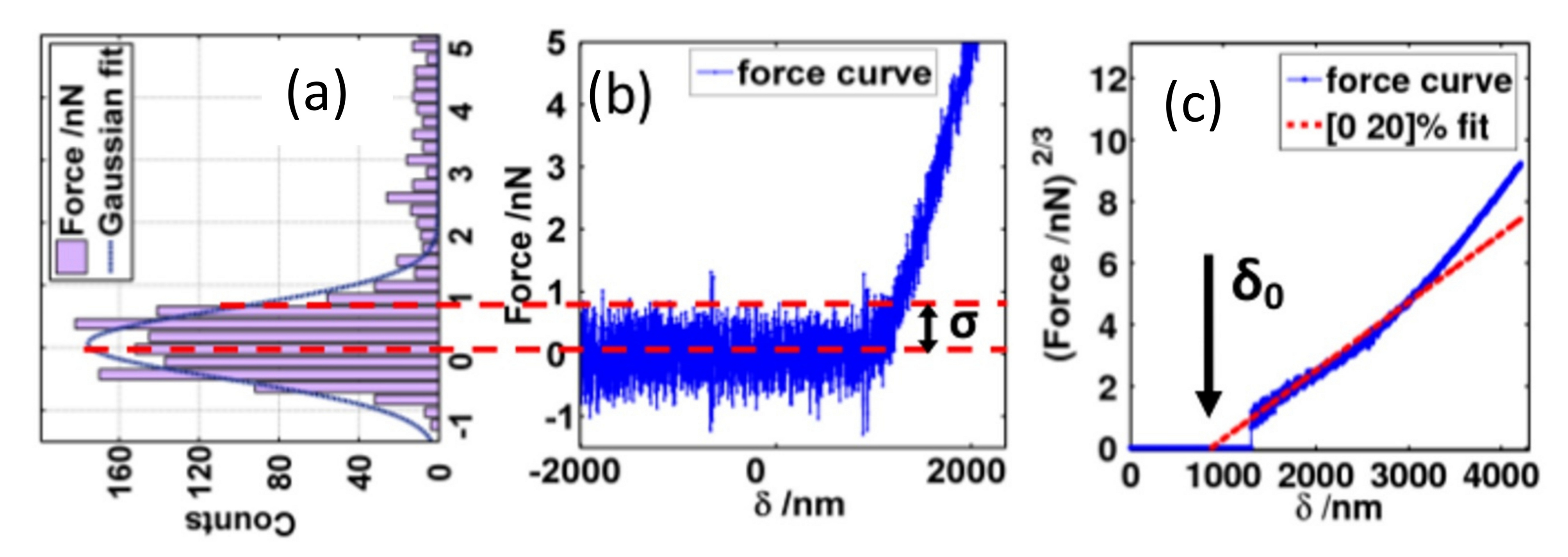}
\caption{\label{fig:cp_finding}Strategy for identifying the region of the force curve to fit with Hertzian model to determine accurately the contact point $\delta_0$. The center and the width $\sigma$ of the main peak in the hystogram of all force values of the force curve are determined by a gaussian fit (a), and used to set a force threshold on the force curve (a magnified view of the force curve close to the contact point is shown in (b)); the first 20\% of the portion of the linearized force curve lying above this threshold is fit by the linearized Hertz model (Eq.~\ref{eq:ft-linearized}), with $\delta_0$ as free parameter (c).}
\end{figure*}

Our strategy for the determination of the contact point $\delta_0$ is the following:

\begin{enumerate}
\item The distribution of all force values of each force curve is evaluated (Figure~\ref{fig:cp_finding}a,b). Force curves are baseline-subtracted (see Appendix~\ref{app:preliminary_analysis}), therefore a well-shaped peak is typically present, representing force values of the non-contact region fluctuating around an average value, typically close to zero. This small residual offset is subtracted to the data. Notice that in the presence of a well-shaped distribution, the gaussian fit can be easily automatized. The width of the distribution, i.e. the standard deviation of the Gaussian curve, is taken as the representative noise level of the force curve.

\item A noise threshold is set N standard deviations (usually N=1 is enough) above the baseline level previously determined (Figure~\ref{fig:cp_finding}b,c). The region of the curve above threshold (the contact part) is considered for fitting.
 
\item The first 20\% of the contact part of the linearized force curve (i.e. the region closest to the contact point) is fitted by the Hertzian model, according to Eq.~\ref{eq:ft-linearized} with $\delta_0$ as free parameter (Figure~\ref{fig:cp_finding}c). The finite-thickness correction can be neglected in this small-indentation region. 
\end{enumerate}

The width of the fitting region is arbitrary. Empirically, analyzing a significant set of curves selected among different cellular regions and cellular samples, we found an optimal percentage equal to 20\% and 30\% for the higher cellular regions (above and near the nucleus) and the thinner ones (cytoplasmic extensions), respectively.
By repeating the fitting procedure varying the percentage width of the fitting interval between 20 and 30\%, and by averaging the standard deviation of the mean of the obtained $\delta_0$ values on different cells, we have estimated the error associated to the contact point as $\epsilon_{\delta_0} \approx 10 nm$; a smaller value would be unreasonable, considered the complexity of the probe/cell contact interface.

\subsubsection{\label{section:final_steps}Calculation of corrected topographic and elastic maps}
Once the contact point $\delta_0$ has been determined with its error, the uncompressed topographic map is calculated, and the finite-thickness effect correction is implemented on force curves. The following actions are performed:

\begin{enumerate}
\item For each force curve, a new indentation axis $\delta'=\delta-\delta_0$ is calculated, so that the contact point is re-located at $\delta'=0$. Force curves are now horizontally aligned with respect to the point of first contact. 
\item The deformation map $\delta_{max}$ is calculated as $\delta_{max}=\delta_{stp}-\delta_0$, where $\delta_{stp}$ is the indentation value on the original axis in correspondence of the force setpoint.
\item The uncompressed topographic map $h$ is calculated by adding the total deformation to the compressed topographic map, then subtracting the average height of the substrate. 
\item The sample thickness $h$ and total indentation $\delta_{max}$ are used to calculate the correction factor $\Delta$ (Eq.~\ref{eq:Delta}) and the effective pseudo-force $F^*$ (Eq.~\ref{eq:pseudo_effective_force}).
\item The values of Young's modulus are extracted via linear regression of linearized curves on the selected indentation range, and the corresponding map is built. 
\end{enumerate}.

When the distance axis of force curves is built using the absolute z-piezo position (z-sensor signal), the uncompressed topographic map $h$ simply corresponds to $\delta_0$ and step 3 can be skipped.

\begin{figure*}
\includegraphics[width=17cm,keepaspectratio]{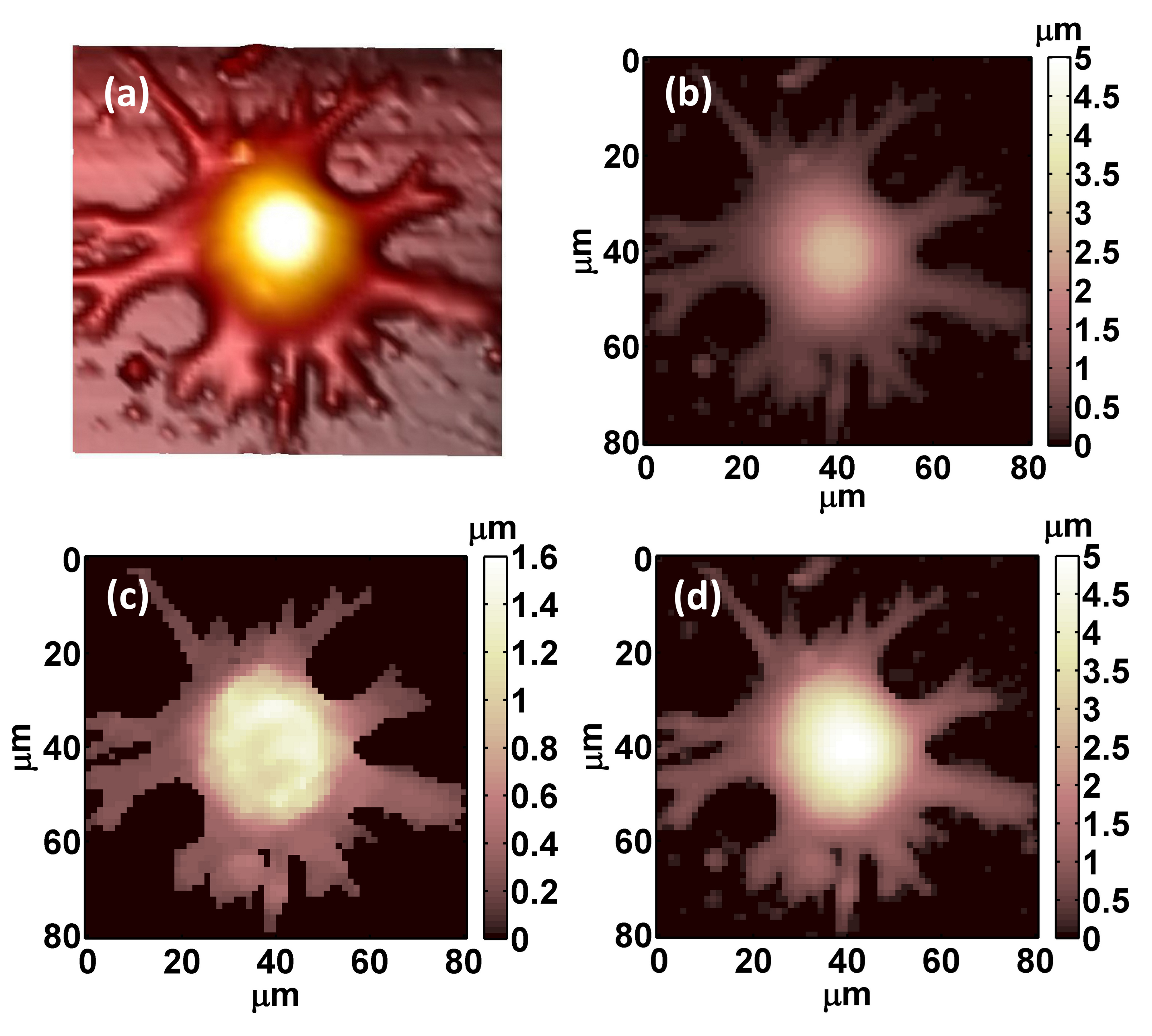}
\caption{Topographic maps for a cell from the line PC12. The maps were acquired using a colloidal probe with radius $R = 5008 \pm 51 nm$, attached to a cantilever with elastic constant $k = 0.32 \pm 0.03 N/m$.  a) 3D view of the compressed height map; b) 2D compressed height map; c) Indentation map: the substrate's height was arbitrarily set to zero; d) Real height map given by the sum of b) and c).\label{fig:PC12_topo}}
\end{figure*}

In Figure~\ref{fig:PC12_topo} we can see an overview of the different steps needed to build the final uncompressed topographic map: starting from the compressed topographic map (Figure~\ref{fig:PC12_topo}a,b), we add the deformation map resulting from the local sample indentation (Figure~\ref{fig:PC12_topo}c) in order to obtain the real height map of the cell (Figure~\ref{fig:PC12_topo}d).

When indenting soft samples as living cells, deformation can be as large as a few microns, as seen in Figure~\ref{fig:PC12_topo}, and the correction of the compressed topographic map is mandatory in order to accomplish the finite-thickness correction of the Young's modulus.

Not considering the identification of the multiple elastic ranges, the method here discussed allows obtaining the combined topographic/mechanical maps of a living cell from a Force Volume set of 64x64=4096 force curves in a few minutes using a 64bit personal computer: the key is represented by the use of Hertz-like linearized expressions like Eq.~\ref{eq:ft-linearized}, which allows analyzing in parallel, by means of matrix linear regression, the whole set of several thousands curves. 

\subsubsection{\label{section:error_analysis}Estimation of the error associated to the Young's modulus}
Several sources of error, as well as the intrinsic variability of the biological specimens, affect the estimation of the Young's modulus  $E_{Cells}$ representative of a population of cells, $N_{Cells}$ of which are actually imaged by AFM (typically $N_{Cells} \simeq 10$). 
The intrinsic cell-to-cell variability is significant even within the same cellular population. Then, at the single-cell level, we must consider several other contributions: the intrinsic variability arising from the heterogeneity of cellular structure (different cellular regions may have different elasticity, the local elasticity itself being the result of several different contributions from different parts of the cytoskeleton); the experimental error related to the mechanical measurement. Experimental errors can be due to uncertainties in the calibration of the AFM probe (cantilever elastic constant, probe radius) and the optical beam detection system (deflection sensitivity), as well as in the evaluation of physical parameters, like the contact point.

In Appendix~\ref{app:error_calculation} we describe in details our strategy to take into account all these uncertainties and estimate the total error, so that the result of the nanomechanical measurement can be expressed as as $E_{Cells} \pm \sigma_{Cells}$.

\section{Application of the protocol to live cell imaging}
We have tested our protocol for topographic/mechanical imaging by CPs on living cells from the lines PC12 (pheochromocytoma of the rat adrenal medulla) and MDA-MB-231 (human breast adenocarcinoma), which have been cultured in vitro and then transferred into the thermostatic AFM fluid cell and kept for a few hours at 37$^\circ C$ (recently, we have also applied our protocol to characterise lamellipodial membrane tension in the newly discovered phenomenon of ligand-independent adhesion signalling by integrins~\cite{Ferraris2014}). We have first investigated the effect of the finite-thickness correction on the determination of the elastic modulus, then we have studied the action of Cytochalsin-D, a cytoskeleton-targeting drug, on cellular elasticity.

\subsection{Evidence of the finite-thickness effect in nanomechanical imaging of living cells}
\begin{figure*}
\centering
\includegraphics[width=17cm,keepaspectratio]{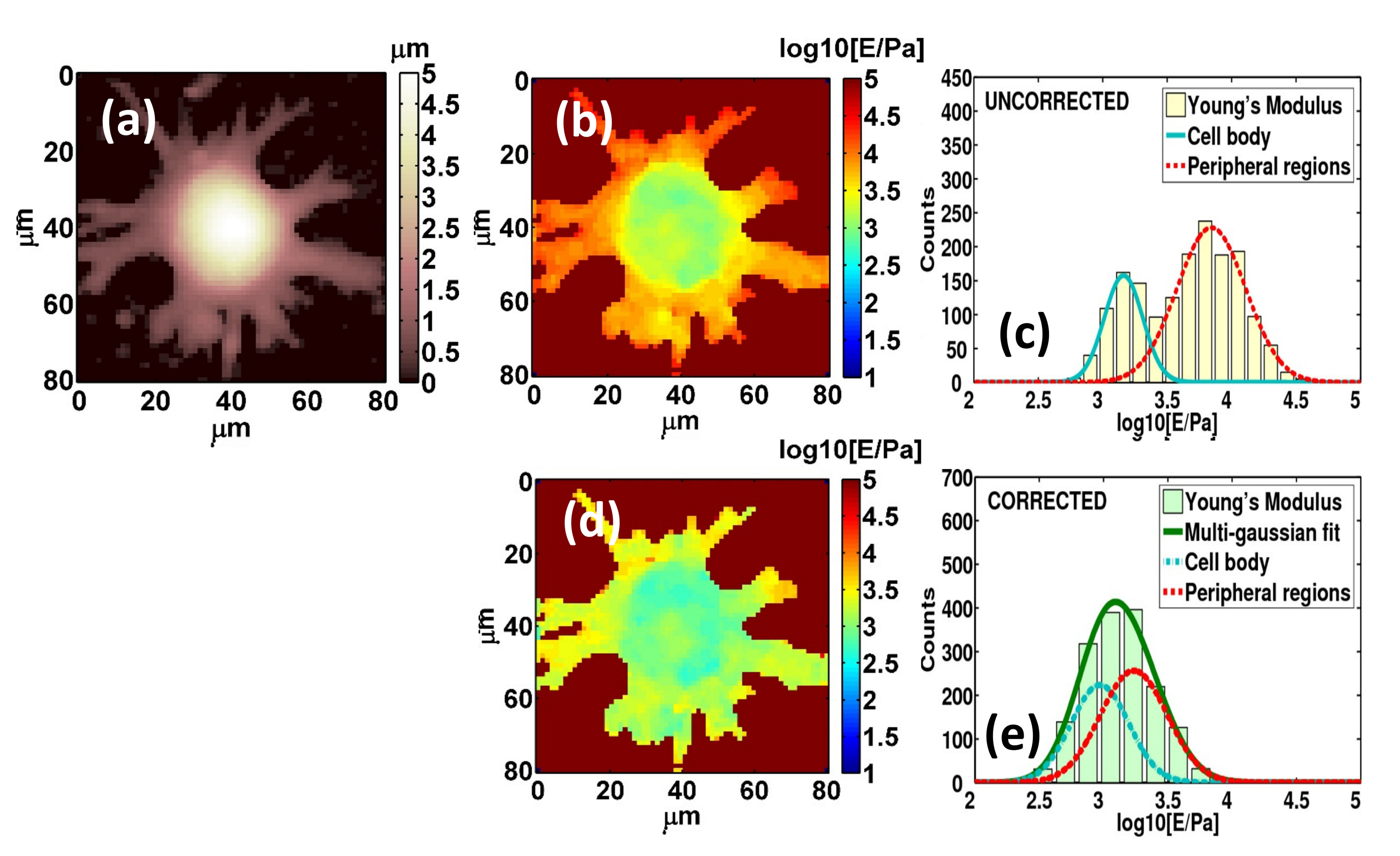}
\caption{The effect of finite thickness of sample in the determination of the Young's modulus. Topographic and elastic maps of a PC12 cell, and corresponding histograms of Young's modulus values (fitted on the $0-40\%$ indentation range), are shown. The value of the Young's modulus of the substrate has been arbitrarily set to 10 GPa. a) Undeformed topographic map, b) Young's modulus map without finite-thickness correction and c) corresponding histogram of the Young's modulus values; d) Young's modulus map with finite-thickness correction and e) corresponding histogram of the Young's modulus values. Young's modulus values are shown in semilog10 scale, with multi-gaussian fit (details in Appendix~\ref{app:mech_exp_details}). Mean values, standard deviations, and the overall error associated to the Force Volume measurement (Section~\ref{section:error_analysis}) are reported in Table~\ref{tab:ft_effect}.\label{fig:PC12_ft}}
\end{figure*}

Figure~\ref{fig:PC12_ft} shows topographic and elastic maps, as well as the corresponding Young's modulus histograms, obtained on PC12 cells before and after the application of the finite-thickness correction. The Hertz model has been fitted to the $0-40\%$ indentation range; indentation was therefore enough deep to sense the presence of the underlying substrate, yet not exceedingly deep, so that Dimitriadis correction could apply with reasonable accuracy.

\begin{table}
\caption{\label{tab:ft_effect}The Young's modulus of a PC12 cell (Figure~\ref{fig:PC12_ft}) calculated with and without the finite-thickness correction. Results are presented in the form: $E \pm \sigma_E (\pm \sigma_{FV}^{Cell})  / Pa$ (see Appendix~\ref{app:error_calculation}).}

\begin{tabular}{p{3cm} p{3cm} p{4cm}}
                     & \textbf{Cell body} & \textbf{Peripheral region}\\
\hline
\textbf{Uncorrected} & {$1706 \pm 671 (\pm 69)$} & {$8165 \pm 3212 (\pm 1062)$}\\
\textbf{Corrected} & {$976 \pm 194 (\pm 30)$} & {$2108 \pm 901 (\pm 169)$}\\
\hline
\end{tabular}
\end{table}

As previously discussed, cells usually appear more rigid because of their finite thickness and the presence of a rigid substrate underneath: for this reason, when the finite thickness correction is applied, we notice, in accordance to Eq.~\ref{eq:ft_corr_average}, an overall decrease of the cell's Young's modulus, and a relative decrease of the rigidity of the thinner peripheral regions with respect to the higher cell body. Young's modulus values (Figure~\ref{fig:PC12_ft}b,d) can be represented in a semilog scale, in order to obtain more compact distributions (Figure~\ref{fig:PC12_ft}c,e). Mean values and standard deviations corresponding to the different modes of the distributions (extracted by a multi-gaussian fit on the semilog10 scale, then converted to the linear scale), and the overall error associated to the Force Volume measurement, are reported in Table~\ref{tab:ft_effect}. When the correction is not applied (Figure~\ref{fig:PC12_ft}c) a bimodal distribution is clearly observed, where the dominant peak represents mainly the contribution of the cell body and the other represents the contribution of the cellular extension; when the correction is applied (Figure~\ref{fig:PC12_ft}e), the two different modes overlap, almost equally weighted, and thus difficult to identify. Nevertheless, the peripheral, thinner, cellular region remains more rigid even after the finite-thickness correction, as can be concluded by crossing the data from Figure~\ref{fig:PC12_ft}c,e.  This is not surprising, as in the thinner cytoplasmic extensions is concentrated the majority of focal adhesion points, which allow the cell to anchor itself to the underlying substrate; it is therefore reasonable that the actin newtork, the principal cytoskeletal component involved in adhesion processes, possesses higher local density and rigidity.

\subsection{The effect of Cytochalsin-D on the Young's modulus of living cells}
\begin{figure*}
\includegraphics[width=17cm,keepaspectratio]{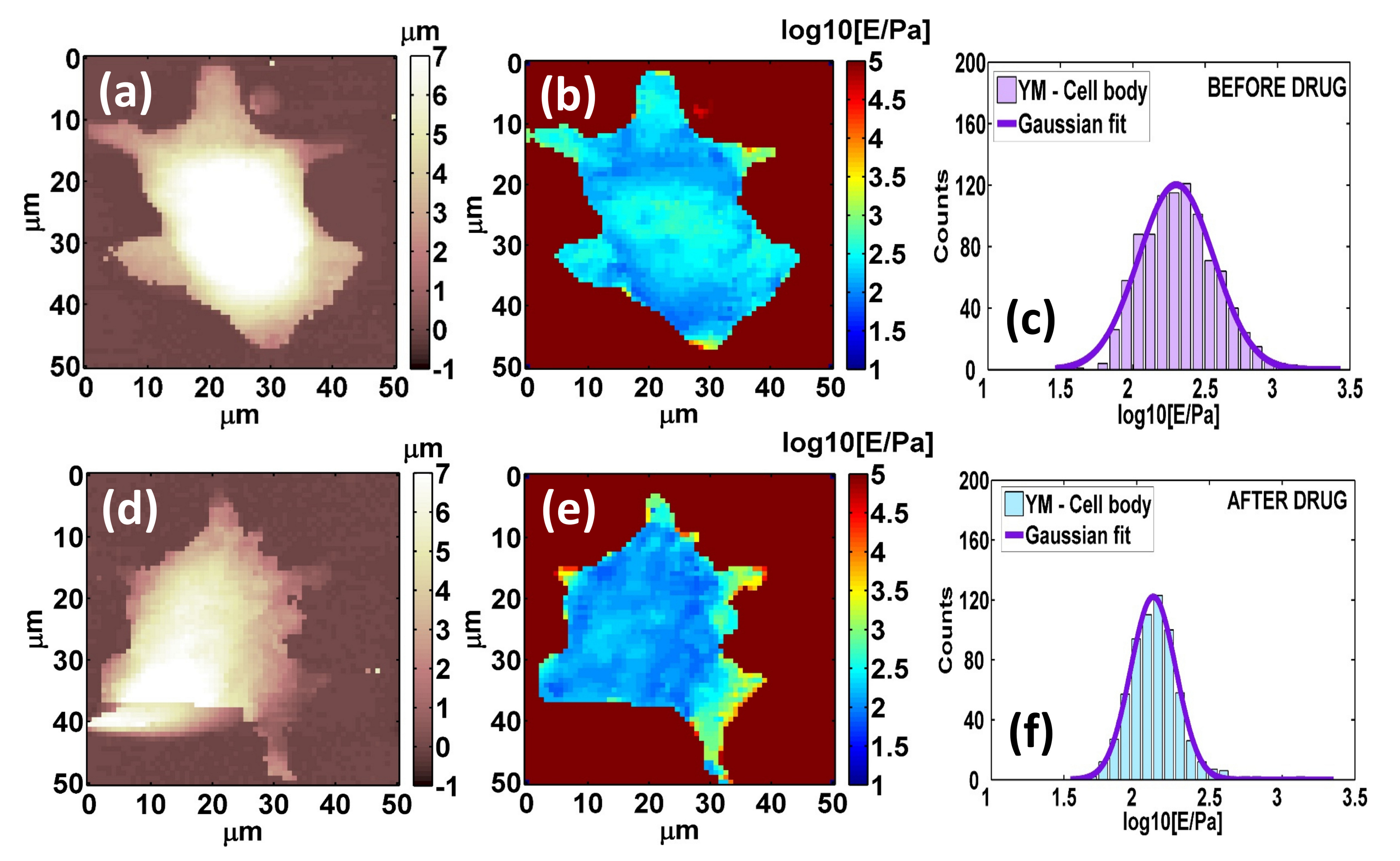}
\caption{\label{fig:MDA_cyto_maps}Combined topographic/mechanical imaging of a cell from the MDA-MB-231 line before and after the introducion of Cytochalasin-D in the cell buffer. The value of the Young's modulus of the substrate has been arbitrarily set to 10 GPa; experimental artifacts have been properly masked when evaluating the Young's modulus histograms, from the cellular body region. a) Topographic map, b) Young's modulus map, c) Histogram of the Young's modulus values before drug's introduction; d) Topographic map, e) Young's modulus map, f) Histogram of the Young's modulus values after drug's introduction. Young's modulus values are reported in histograms in semilog10 scale, with multi-gaussian fit (details in Appendix~\ref{app:mech_exp_details}). Mean values, standard deviations, and the overall error associated to the Force Volume measurement (Section~\ref{section:error_analysis}) are reported in Table~\ref{tab:drug_effect}.}
\end{figure*}

In order to test the validity of the developed protocol to highlight structural changes in the cell's cytoskeleton, we have investigated the effect of a cytoskeleton-targeting drug, Cytochalasin-D, a complex molecule belonging to the class of micotoxins, which is able to penetrate the cell and depolimerize the actin network, binding itself to the active site of the free proteins or breaking the bondings bewteen two consecutive proteins along a filament~\cite{Nair2008}. Therefore, a decrease in cell's rigidity after the introduction of the drug in the culture medium has to be expected; this effect has been demonstrated by several authors thanks to AFM-based nanomechanical studies~\cite{Rotsch2000,Kasas2005,Oberleithner2006}. In Figure~\ref{fig:MDA_cyto_maps} are shown the results of the interaction of Cytochalasin-D with cells from the MDA-MB-231 line; the topographic and elastic maps (corrected for the finite-thickness effect) are shown, as well as the histograms of the Young's modulus values (the corresponding mean values and standard deviations are reported in Table~\ref{tab:drug_effect}) measured on a single cell.The Hertz model has been fitted to the $0-40\%$ indentation range; a relatively shallow indentation range was chosen so to capture the contribution  of the actin network that is coupled to the cell membrane.

\begin{table}
\caption{\label{tab:drug_effect}The Young's modulus of the cellular body of an MDA cell (Figure~\ref{fig:MDA_cyto_maps}) before and after exposure to Cytochalasin-D. Results are presented in the form: $E \pm \sigma_E (\pm \sigma_{FV}^{Cell})  / Pa$ (see Appendix~\ref{app:error_calculation}).}

\begin{tabular}{p{3cm} p{3cm} p{3cm}}
                 &  \textbf{Uncorrected} & \textbf{Corrected}\\
\hline
\textbf{before Cyto D} & {$378 \pm 226 (\pm 19)$} & {$235 \pm 93 (\pm 5)$}\\
\textbf{after Cyto D}  & {$254 \pm 128 (\pm 13)$} & {$142 \pm 42 (\pm 5)$}\\
\hline
\end{tabular}
\end{table}

In this case, the exposure of the cell to the drug has produced not only a significant change in the Young's modulus (compare Figure~\ref{fig:MDA_cyto_maps}b to Figure~\ref{fig:MDA_cyto_maps}e), but  also a morphological change, especially in the thin cytoplasmic protrusions (compare Figure~\ref{fig:MDA_cyto_maps}a to Figure~\ref{fig:MDA_cyto_maps}d); these regions are particularly rich in actin filaments and focal adhesions, which tend to be destroyed, reshaped, and pulled back by the drug. For this reason, the histograms and the mean values reported in Table~\ref{tab:drug_effect} are related only to the cell's body, which undergo minor morphological changes upon drug exposure. The results obtained without applying the finite-thickness correction are qualitatively similar (the cell softens upon interacting with Cytochalasin D), although quantitatively different. When the correction is not applied, the difference between the average values of the Young's modulus before and after the treatment is larger; the standard deviations of data are larger as well, as the apparent modulus depends on the local height of the cell, which is far from being constant. It is worthwhile mentioning that in principle the percentage intervals related to the different slopes in the force curves (see Section~\ref{section:lineartrends}) could change after the exposure to the drug; we actually observe this effect, since the percentage interval corresponding to the linear region changed from $[0-40]\%$ to about $[0-60]\%$ (in order to allow a better comparison, we have used a common indentation range, i.e. $[0-40]\%$).

\section{Conclusions}
We have developed an experimental protocol for the combined topographic and nanomechanical imaging by AFM of soft samples, and in particular of cellular systems, based on the use of micrometric colloidal probes. Once it is accepted that the average mesoscopic Young's modulus of cells is the relevant physical quantity, a reasonable assumption in view of clinical and pre-diagnostic applications, as well as for studying the cell's response to external stimuli, it turns out that CPs represent a better alternative in terms of accuracy and reliability of mechanical measurements than sharp AFM tips. Moreover, we have shown that despite the common belief, CPs provide more-than-enough lateral resolution to carry out a spatially-resolved analysis of the mechanical properties of living cells, where all the important components of the cellular system (the cell body with the nuclear region, the peripheral region, with the cellular extensions, protrusions and/or lapellipodia, etc.) can be clearly identified.

We have implemented a correction of finite-thickness effects in the determination of the Young's modulus, a necessary step in the protocol, due to the softness and limited thickness of the cellular samples. This allows measuring mechanical properties across the whole cellular surface, included the thinnest regions like lamellipodia, cellular extensions or other cytoplasmic extensions, whose importance is justified by their role in cells' motility and richness in focal adhesions. Without the implementation of the finite-thickness effect, nanomechanical analyses should be limited to the highest cellular regions, so to minimize the impact of the constrained contact geometry.

The protocol has been tested by performing combined topographic and mechanical imaging of living cells, from the PC12 and MDA-MB-231 lines. These experiments proved the reliability of the general protocol, and confirmed on one side the importance of the correction of the finite-thickness effect, on the other that CPs not only provide robust estimation of the Young's modulus of soft matter, but also a satisfactory lateral resolution in mechanical mapping of cellular systems.

Summarizing, in view of the standardization process for the extensive application of the AFM in the nano-biomedical field, and considered the many advantages related to the employment of CPs, we think that the ambition of performing a truly nanoscale mapping could be dropped in favour of a decent micrometric resolution, provided by CPs, which is appropriate to map even the smallest cells. The combined use of micrometric spherical probes and vertical approach modes like the Force Volume mode, together with the development and refinement of suitable algorithmic tools for the automation of data analysis and the improvement of AFM probe calibration methods, seems to us the best way towards the optimization of the combined topographic/mechanical imaging on living cells, and the effective exploitation of the potential of AFM in biology and medicine.

\appendix
\section{\label{app:probe-shapes}Force-indentation models}
It is useful to recall the most important contact mechanics models and briefly discuss the physical conditions on which they rely. These models, under the hypothesis of negligible or null adhesion, can be cast (following the outline of Ref.~\onlinecite{Lin2007}) in a simple and generalized equation linking the force $F$ applied by the AFM probe to the deformation (or indentation) $\delta$ of the sample:

\begin{equation}
F=\lambda\delta^{\beta}\label{eq:Hertz_generalized}
\end{equation}

where $\lambda$ is a parameter containing the relevant mechanical informations, i.e. the sample's Young's modulus $E$ and the Poisson ratio $\nu$, as well as the indenter properties, such as R, the radius of curvature, or the tip angle $\alpha$. In Table~\ref{tab:Probe_shapes} the commonly used AFM probe shapes and the corresponding expressions and values for $\lambda$ and $\beta$ are listed.

Sharp tips belong to the last three rows of Table~\ref{tab:Probe_shapes}; in particular, Bilodeau\cite{Bilodeau1992} and Rico~\cite{Rico2005} solutions for pyramidal tips are approximations derived from the Sneddon solution for the conical shape~\cite{Harding1945}. Sneddon solved the Boussinesq axysimmetric problem, calculating exact solutions in the form of Eq.~\ref{eq:Hertz_generalized} for the case of the indentation of an elastic half-space by a solid of revolution whose axis is normal to the sample surface: equations for special shapes like sphere, cone, cylinder and paraboloid of revolution have been derived.
Sneddon calculations are developed in the framework of the classical theory of elasticity, which assumes a linear relationship between stress and strain, which in turn is based on the assumption of small strains; even if the strain upper limit for the validity of linear elasticity is not well defined (cells are sometimes described as elastomers~\cite{X.Zeng2013}, which can bear high strains if compared with other materials) and could change from one cell type to another, it is clear that strains as high as those induced by sharp tips (Figure 4, Ref.~\onlinecite{Dimitriadis2002}) can easily exceed the boundaries of linear elasticity, thus depriving the Young's modulus of physical meaning.

Sneddon model for the indentation $\delta$ of an elastic flat surface by a rigid spherical punch is the following~\cite{Sneddon1965,Heuberger1996}:

\begin{equation}
F=\frac{E}{2\left(1-v^{2}\right)}\left[\left(a^{2}+R^{2}\right)\ln\left(\frac{R+a}{R-a}\right)-2aR\right]\label{eq:Sneddon_force}
\end{equation}

\begin{equation}
\delta=\frac{1}{2}a\ln\left(\frac{R+a}{R-a}\right)\label{eq:Sneddon_indentation}
\end{equation}

where $F$ is the force, $a$ is the contact radius, $R$ is the sphere radius, $E$ and $\nu$ are the Young's modulus and the Poisson coefficient of the surface, accordingly. Eqs~\ref{eq:Sneddon_force}-\ref{eq:Sneddon_indentation} provide the Hertz model (the first in Table~\ref{tab:Probe_shapes}) in the limit of small deformations and large tip radii ($a/R\ll1$)~\cite{Heuberger1996}. Noticeably, while the contact mechanical models in Table~\ref{tab:Probe_shapes} and the Sneddon model have been derived under the assumption that the stress-strain relation is linear (purely elastic behaviour), i.e. under the hypothesis of small strains, the sphere-on-flat model by Sneddon (Eqs~\ref{eq:Sneddon_force}-\ref{eq:Sneddon_indentation}) does not rely on the additional constraint that also the indentation $\delta$ is small compared to the probe radius; the price to pay is the lack of an analytical form for the force vs indentation equation, which makes the Sneddon solution hard to implement in a real experiment. However, it can be shown (Section~\ref{section:CP}) that the Hertz model represents a very good approximation of the Sneddon model for colloidal probe, while this is not the case for sharp tips.  

\begin{table}
\centering
\caption{\label{tab:Probe_shapes}Summary of the most common AFM tips' geometries and the corresponding parameters for Eq.~\ref{eq:Hertz_generalized} (notice that the formula for the conical punch is reported incorrectly in Ref.~\onlinecite{Sneddon1965})}
\begin{tabular}{p{5.5cm} p{2cm} p{1cm}}
\textbf{Probe shape} & $\boldsymbol{\lambda}$ & $\boldsymbol{\beta}$\\
\hline
{Sphere, or paraboloid, of radius R (Hertz~\cite{Hertz1881})} & 
{$\frac{4}{3}\frac{ER^{\frac{1}{2}}}{\left(1-\nu^{2}\right)}$} &
{$\frac{3}{2}$} \\
{Conical punch of tip angle $2\alpha$ (Sneddon~\cite{Harding1945})} & {$\frac{2}{\pi}\frac{E\tan\alpha}{\left(1-\nu^{2}\right)}$} & {$2$}\\
{Four-sided pyramidal punch of tip angle $2\alpha$ (Bilodeau~\cite{Bilodeau1992})} & {$\frac{1.49}{\pi}\frac{E\tan\alpha}{\left(1-\nu^{2}\right)}$} & {$2$} \\
{Four-sided pyramidal punch of tip angle $2\alpha$ (Rico~\cite{Rico2005})} & {$\frac{2.22}{\pi}\frac{E\tan\alpha}{\left(1-\nu^{2}\right)}$} & {$2$}\\
{Flat-ended cylindrical punch of radius R (Sneddon~\cite{Sneddon1965})} & {$2\frac{ER}{\left(1-\nu^{2}\right)}$} & {$1$}
\end{tabular}
\end{table}

\section{\label{app:CP}Production and characterization of colloidal probes}

\subsection{\label{app:CP_production}Production of colloidal probes}
We have developed a protocol for the production and characterization of monolithic borosilicate glass CPs, which is described in details in Ref.~\onlinecite{Indrieri2011}. Probes with similar characteristics can be produced by sintering methods also according to the procedures reported in Refs~\onlinecite{Kuznetsov2012,Bonaccurso2001}. The most important highlights of the protocol and the advantages of the probes that can be produced are reported below:

\begin{enumerate}
\item Borosilicate glass microspheres are attached to tipless silicon cantilevers exploiting only adhesive capillary forces, thus avoiding using epoxy-based adhesives and other chemicals, which may either contaminate the probe and/or the imaging buffer, either keep from cleaning them in aggressive media;
\item Glass spheres are attached to silicon cantilevers by thermal annealing at elevated temperatures ($\simeq 800^\circ C$). The resulting \textit{monolithic} borosilicate glass CPs can be washed/cleaned aggressively after use in order to remove contaminants (probes get readily contaminated upon indenting very soft samples like cells). Considering that the actual geometry and dimensions of the probes can be characterized accurately (more details in Appendix~\ref{app:CP_characterization}), it follows that the same probe can be used and re-used many times, for sake of comparability of obtained data and reproducibility of experiments, on one side, and saving of money, on the other.
\end{enumerate}

Obviously, commercial CPs that are produced by gluing spheres to cantilevers or by using polymeric materials for the micro-spheres do not offer the same advantages of monolithic probes in terms of easiness of cleaning and re-usability.

\subsection{\label{app:CP_characterization}Characterization of colloidal probe radius}
\begin{figure*}
\includegraphics[width=17cm,keepaspectratio]{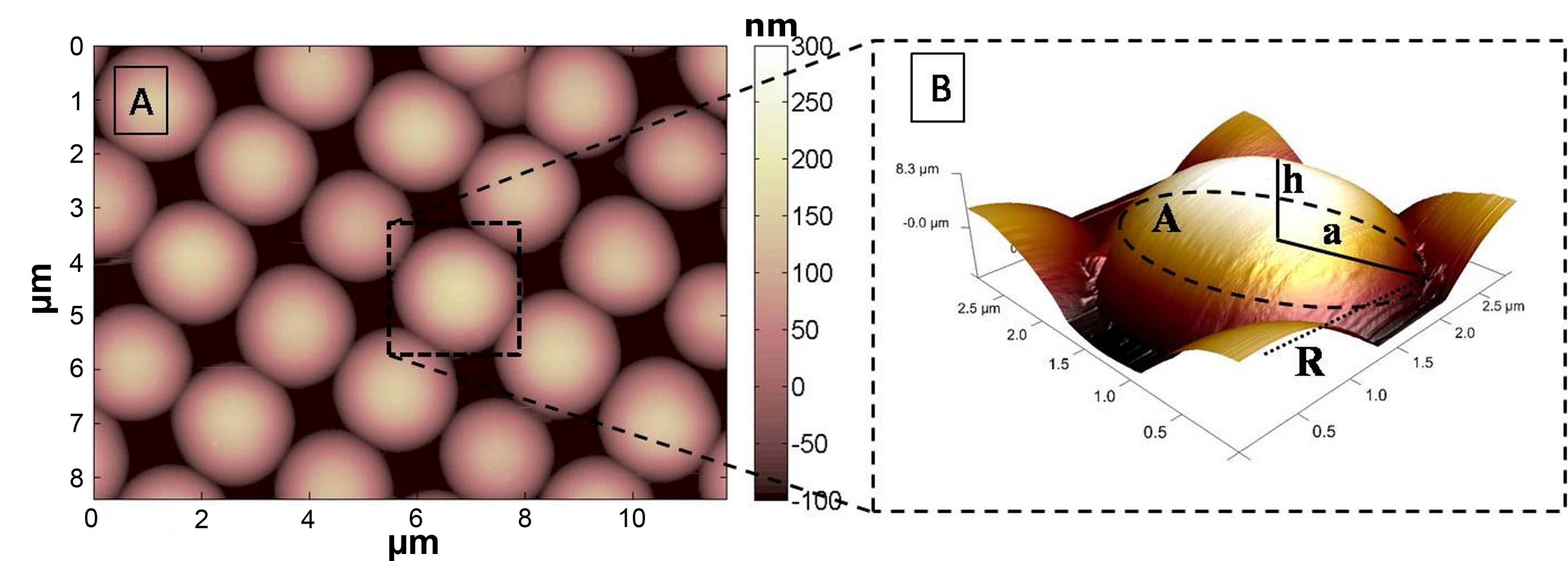}
\caption{\label{fig:Reverse_colloidal}Section of a topographic image acquired by a micrometric spherical probe over an array of spikes. B) Detail of the single spherical cap with the relevant geometrical parameters.}
\end{figure*}

The characterization of the radius of CPs is based on reverse AFM imaging of the probe, which is obtained by imaging a spike grating (like the TGT1 from NT/MDT)(Figure~\ref{fig:Reverse_colloidal}), as described in details in Ref.~\onlinecite{Indrieri2011}.

\begin{figure}
\includegraphics[width=8.5cm,height=6.5cm]{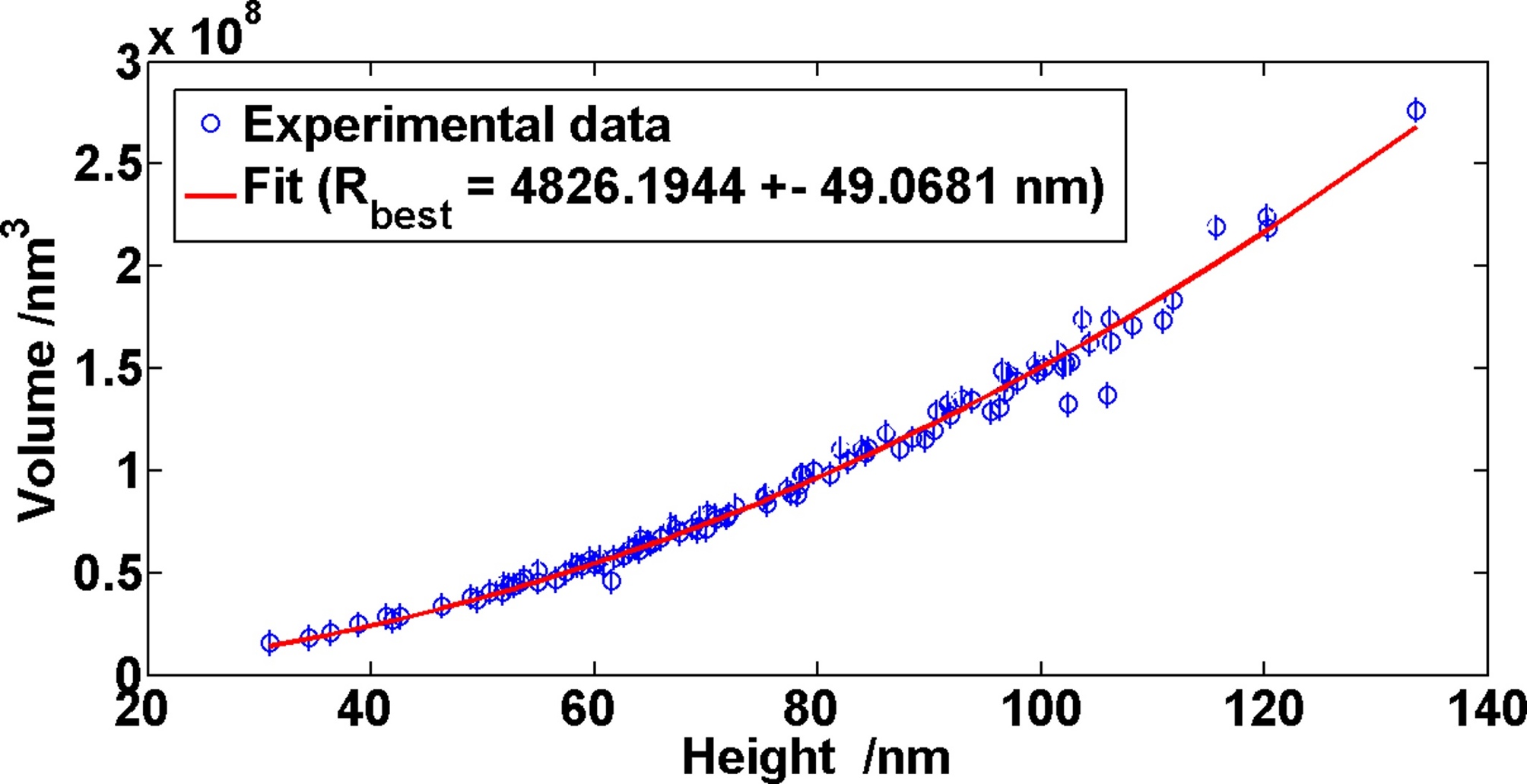}
\caption{\label{fig:hV_fit}Fit of the Volume vs Height relation obtained by a set of spherical caps (see Figure~\ref{fig:Reverse_colloidal}), which provides the radius R of the underlying sphere.}
\end{figure}

A collection of hundreds independent replicas of the probe apical geometry is typically obtained; once the geometrical parameters of each probe replica have been calculated (volume, projected area, height), calibration curves are built (the volume versus height curve is shown in Figure~\ref{fig:hV_fit}) and the probe radius $R$ is extracted by fitting data to the appropriate spherical cap model, in this case $V = (\pi /3) h^2(3R-h)$. The overall accuracy of the probe radius evaluation can be as good as 1\%.

Noticeably, this characterization approach also allows determining the three-dimensional shape of the apical region of the probe, before any spherical cap approximation, which can be useful, for instance, to test the hypothesis of spherical geometry.

\section{Experimental setup and imaging parameters}
\subsection{\label{app:fluid_cell}Thermostatic AFM liquid cell}
Cells need the right environment to live and proliferate, i.e. a proper culture medium, a temperature $T = 37^\circ C$ and an atmosphere with a $5$\% concentration of $CO_2$; in particular, temperature is the most relevant parameter for the stability of cells' conditions, so it should be addressed as a critical aspect for AFM measurements on living cells. In order to be able to run nanomechanical experiments for several hours (to collect a reasonable number of Force Volume maps, for sake of statistics), we have built a thermostatic fluid cell, schematically represented in Figure~\ref{fig:riscaldatore}:

\begin{figure}
\includegraphics[width=8.5cm,height=4.5cm]{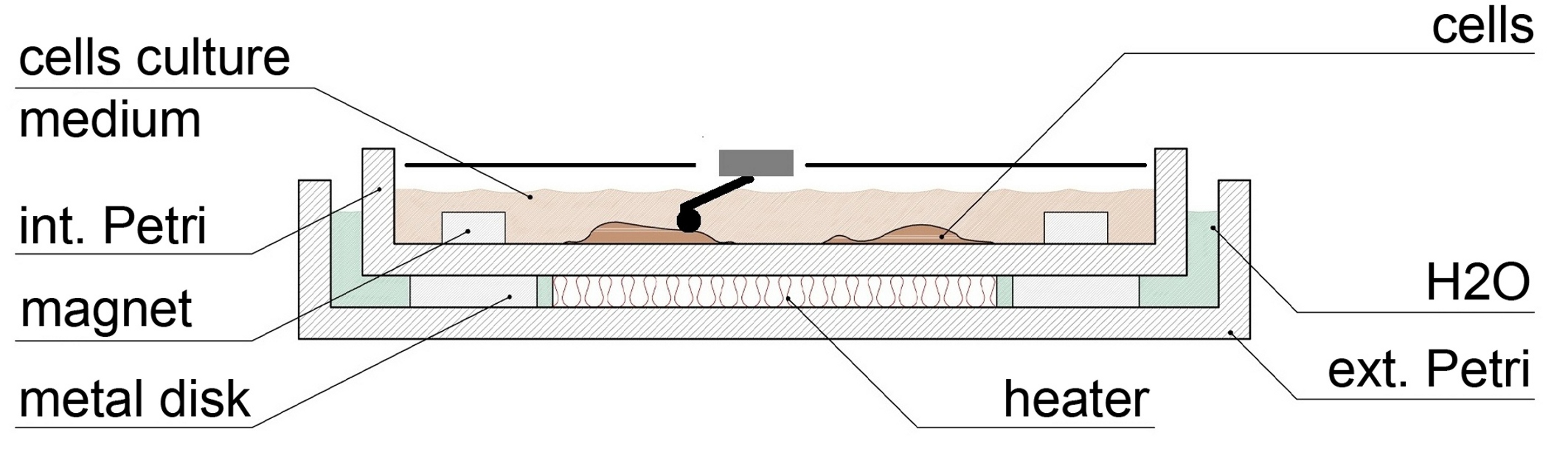}
\caption{\label{fig:riscaldatore}Schematic representation of the handmade thermalized AFM cell.}
\end{figure}

The fluid cell is composed by two nested petri dishes, the outer one working as a thermal bath (distilled water fills the gap between the two), and the innermost one containing the cells and the culture medium. Thermalization is obtained by means of an electrical resistor (inserted into a thin transparent plastic stripe), fixed on the internal surface of the outer petri dish and properly calibrated in order to get the correct relation between the system temperature and the applied voltage. The two petri dishes are fixed one to the other and to the AFM stage using magnets.

A direct consequence of thermalization is the evaporation of the culture medium, which can be harmful for both the AFM optical components and cells, leading to their apoptosis. It is possible to partially solve this problem covering the system with a plastic cap, with a proper hole for the insertion of the AFM head: this expedient suffers from the drawback of limiting the lateral range of the piezoelectric actuators, but on the other side it creates a nearly closed system, characterized by a locally vapour-saturated atmosphere. Comprehensively, this device does not compromise the eyesight of the integrated optical microscope and lets us to measure cells' properties in their physiological conditions for an overall time of about 4-5 hours, before they show clear suffering signals.

\subsection{Cell culture conditions}
PC12 cells were maintained in RPMI-1640 Medium (Sigma-Aldrich) supplemented with $10\%$ horse serum (HS; Sigma-Aldrich), $5\%$ fetal bovine serum (FBS; Sigma-Aldrich), L-glutamine 2mM, 100 units/ml penicillin, 100 μg/mL streptomycin, 1 mM pyruvic acid (sodium salt) and 10 mM Hepes in $5\%$ CO2, $98\%$ air-humidified incubator (Galaxy S, RS Biotech, Irvine, California, USA) at $37^\circ C$. Cells were detached from culture dishes using a solution 1 mM EDTA in HBSS (Sigma-Aldrich), centrifuged at 1000rpm for 5 minutes, and resuspended in culture medium. Subcultures or culture medium exchanges were routinely established every 2nd to 3rd day into Petri dishes ($\Phi = 10 cm$).

The MDA-MB-231 cells were cultured in Dulbecco's modified eagle medium (DMEM; Lonza) supplemented with 10\% fetal bovine serum (FBS; Sigma-Aldrich), L-glutamine 5mM, 100 units/ml penicillin and 100 units/ml streptomycin in 5\% $CO_2$, 98\% air-humidified incubator (Galaxy S, RS Biotech, Irvine, California, USA) at $37^{\circ} C$. Cells were detached from culture dishes using a 0,25\% Trypsin-EDTA in HBSS (Sigma-Aldrich), centrifuged at 1000 rpm for 5 minutes, and resuspended in culture medium. Subcultures or culture medium exchanges were routinely established every 2nd to 3rd day into Petri dishes ($\phi = 10 cm$).

\subsection{\label{app:mech_exp_details}Mechanical measurements and imaging parameters}
All the topographic/mechanical maps were acquired with a Bioscope Catalyst AFM (Bruker) equipped with a Nanoscope V controller, by means of the Force Volume technique, with the following parameters: $64x64$ force curves, each of them characterized by 2048 points, a ramp length $L = 5 \mu m$, a maximum applied force $F \ensuremath{\approx} 10-15 nN$, a global ramp frequency $f = 7.10 Hz$ composed by an approaching velocity $v_{appr} = 43.4 \mu m/s$ and a retracting velocity $v_{retr} = 195 \mu m/s$; this last choice is suggested by the effort to minimize the total time required for the acquisition of a single Force Volume map and the restriction of data analysis to the approaching force curves only. We employed spherical probes with radius $R \ensuremath{\approx}5 \mu m$ and cantilevers with elastic constants $k\ensuremath{\approx} 0.2-0.3 N/m$ (thermal noise calibration), produced and characterized according to the procedures described in Appendix~\ref{app:CP}.
Living cells have been imaged using a home-built thermostatic fluid cell set at $T=37^\circ C$ (see Appendix~\ref{app:fluid_cell}) in their own culture buffer, with the addition of 25 mM HEPES to keep the pH at the physiological value. 

The logarithmic scale is well suited to represent the distribution of Young's modulus values from a single cell, considered the heterogeneity of cell's structure and the scattering of measured values. In semilog10 scale the Young's modulus distributions are approximately gaussian, which suggests that Young's modulus values are distributed lognormally. Peak values and corresponding widths are extracted by applying multi-gaussian fits to Young's modulus histograms in semilog10 scale (the center of the Gaussin corresponds to the $\log_{10}$ of the median value of the Young's modulus). Under the assumption that the distribution is approximately lognormal, the fitting procedure allows getting rid of outliers and noise effectively, while a direct calculation of average values on linear scale would be more strongly affected by their presence. Logarithmic values are then transformed to linear values according to the theory of lognormal distributions; in particular, the mean Young's modulus value $E$ and its standard deviation $\sigma_E$ are calculated as: $E=10^{\mu_{10}+(0.5\ln10)\sigma^2_{10}}$ and $\sigma_E=E\sqrt{10^{\sigma^2_{10}}-1}$, where $\mu_{10}$ and $\sigma_{10}$ are the center and standard deviation of the Gaussian curve in semilog10 scale. Alternatively, the median value and the median absolute deviation (MAD) from the median value can be used to represent average and deviation of Young's modulus values.

\section{\label{app:preliminary_analysis}Data analysis: preliminary operations}
Force vs indentation $F(\delta)$ data must be prepared from raw data in order to extract Young's modulus data by suitable fitting procedures. The standard data pre-processing procedures are listed below (see also Ref.~\onlinecite{Butt2005}):

\begin{itemize}
\item All curves belonging to a Force Volume data set are processed simultaneously, in order to remove linear or sinusoidal baselines, which are often evident in the non-contact region, and are due to laser/cantilever/detector non-ideal alignment~\cite{}, laser interference~\cite{} etc. After baseline subtraction, the non-contact part of the force curve must appear flat. The baseline is fitted to the non-contact part of the curve but is subtracted from the whole curve. 
\item Conversion of the raw cantilever deflection $\Delta V$ measured in Volts by the photodiode into tha true deflection $\Delta d$ in nanometers, through the equations $\Delta d=zsens\cdot\Delta V$ where $zsens$ is the deflection sensitivity in nm/V. $zsens$ is calculated as the inverse slope of the contact region of force curve acquired on a rigid substrate, or better averaged from those force curves in the Force Volume map belonging to the substrate. Determination of $zsens$ parameter is critical, especially with CPs, due to non trivial bending of the cantilever/sphere assembly in the initial part of the contact region~\cite{Weafer2012}, and care must be taken when fitting the linear region (an alternative strategy has been recently proposed~\cite{COST_paper}).
\item Conversion of the horizontal axis from the piezoelectric displacement $z_p$ (the typical abscissa of force curve as recorded by AFM softwares) to the effective tip-sample distance $d$ (which on the negative semi-axis corresponds, with the sign reverted, to the indentation $\delta$, once the abscissa of the contact point has been set to zero): $d=z_p + \Delta d$, where $\Delta d$ is positive or negative depending whether the cantilever is deflected upwards or downwards (see Figure 3 in Ref.~\onlinecite{Dimitriadis2002}, as well as Ref.~\onlinecite{Butt2005}).
\item Calculation of the force $F$ in nN: $F=k\cdot\Delta d$,  where $k$ is the cantilever vertical force constant, in N/m (as calibrated for instance by the thermal noise or Sader methods~\cite{Butt1995,Sader1999}; an alternative strategy has been recently proposed~\cite{COST_paper}). Details on the calibration procedure (as for example on the distinction between intrinsic and effective force constant~\cite{Hutter2005}) can be found here: \url{http://www.physics.uwo.ca/~hutter/calibration/afmcal.html}
\end{itemize}

\section{\label{app:error_calculation}Data analysis: calculation of errors}

The uncertainties related to the cantilever elastic constant and the deflection sensitivity can be reasonably assumed to be of the order of 10\% and 5\%~\footnote{While the statistical error associated to the deflection sensitivity can be rather small, provided several force curves acquired in different location are used, and the corresponding values of the sensitivity averaged, there are subtle and rather poorly accountable systematic errors which can make the total error larger~\cite{Weafer2012}, so that a 5\% estimation is more reasonable (if not optimistic; this issue is discussed in details in a forthcoming publication~\cite{COST_paper})}, respectively. The absolute error on the contact point is about 10 nm, as discussed previously; the error on the probe radius (Appendix~\ref{app:CP_characterization}) is about $\simeq 1\%$ and can be neglected. 

A reliable method for estimating the global error on the Young's modulus deriving by the uncertainties in the calibration of experimental parameters is represented by a MonteCarlo simulation~\cite{Lybanon1985}: starting from a single experimental force curve (bearing its own noise level) in the original form (i.e. neither linearized nor corrected for the finite-thickness effect), a family of noisy curves is generated by adding a suitable noise to the original force curve. To each force value of the original curve a random noise term randomly taken from a gaussian distribution with zero mean and standard deviation $\delta F$ is added.  $\delta F$ is calculated by propagating the experimental uncertainties through the equation $F=k\cdot zsens \cdot \Delta V$:

\begin{equation}
\delta F=\sqrt{\left(\Delta V\cdot zsens\right)^{2}\delta_{k}^{^{2}}+\left(k\cdot\Delta V\right)^{2}\delta_{zsens}^{^{2}}}\label{eq:err_prop_force}
\end{equation}

where $\Delta V$ represents the cantilever deflection detected in Volts by the photodiode and $\delta_k$ and $\delta_{zsens}$ stand for the aforementioned uncertainties on the cantilver elastic constant and the deflection sensitivity. Concerning the error on the contact point, this is accounted for in a similar way, by shifting the indentation axis of the original force curve by a random offset taken from a gaussian distribution with zero mean and standard deviation $\epsilon_{\delta_0} = 10 nm$. For each noisy curve, the correction factor $\Delta$ (Eq.~\ref{eq:Delta}) is calculated, as well as the effective force $F^*$ (Eq.~\ref{eq:pseudo_effective_force}). Typically $N=10^4$ force curves are generated according to this procedure, and fitted in order to obtain N independent values of the Young's modulus, which are expected to be distributed around the best value obtained by fitting the original rescaled curve. From the corresponding statistical distribution we extract the $68$\% confidence interval  $\sigma _{fit}$, simmetrically evaluated around the mean value, which represents the global systematic error of the fitting procedure due to uncertainties in the calibration of the experimental parameters, including the contact point.

Restricting the analysis to those points belonging to a nearly homogeneous regions of the cell (like the area above the nucleus, or the cell's periphery, with the cellular extensions), so to exclude variability due to the intrinsic structural differences between different regions of the cell, we can calculate the global error for the single Force Volume set as:

\begin{equation}
\sigma_{FV}^{Cell}=\sqrt{\sigma_{stat}^{2}+\sigma_{fit}^{2}}\label{eq:sigma_FV_cells}
\end{equation}

where $\sigma_{stat}=\frac{\sigma_{E}}{\sqrt{N}}$ is the standard deviation of the mean of the $N$ Young moduli $E$ (from $N$ force curves) measured on the cellular region, which are distributed with standard deviation $\sigma_{E}$.

As a final step, the average value of the Young's modulus that is representative of the whole cellular population is calculated as the mean value of all cells, $N_{Cells}$ being the number of acquired Force Volume maps (each related to a different cell). The mean values $E_i$ of the Young moduli of single cells is calculated as $E_{Cells}=(\sum{E_i}) /N_{Cells}$, and this value is representative of the whole cellular population. The associated error must take into account both the error associated to the single Force Volume and that arising from the variability within the population. To this purpose, first of all we propagate the error associated to the single Force Volume measurement $\sigma_{FV}^{Cell}$ through the equation for the arithmetic mean, obtaining the total Force Volume error $\sigma_{FV}$:

\begin{equation}
\sigma_{FV}=\frac{1}{N_{Cells}}\sqrt{\sum_{i}\left(\sigma_{FV}^{Cell}\right)^{2}}\label{eq:sigma_FV_tot}
\end{equation}

then we add this error in quadrature to the standard deviation of the mean $\sigma^{mean}_{E_{Cells}}=\frac{\sigma_{E_{Cells}}}{\sqrt{N_{Cells}}}$ of single -cell moduli $E_i$, which are distributed with standard deviation $\sigma_{E_{Cells}}$, in order to take into account the cell-to-cell variability as well. The final total error $\sigma_{Cells}$ associated to the population mean value $E_{Cells}$ is:

\begin{equation}
\sigma_{Cells}=\sqrt{(\sigma^{mean}_{E_{Cells}})^{2}+\sigma_{FV}^2}\label{eq:sigma_TOT}
\end{equation}

The final output of the experimental procedure is therefore given as $E_{Cells} \pm \sigma_{Cells}$ (this typically refers to a specific region of the cell, although in principle an average value representing the cell as a whole can be determined in the same way, and also to a specific indentation range, i.e. to a specific elastic regime of the normalized force curve).

\begin{acknowledgments}
A.P. thanks the COST Action TD1002 for providing a stimulating scientific environment for the discussion of AFM-based nanomechanics of cells and soft matter, and for supporting his networking activities. The authors thank M. Indrieri, D. Piotti, R. Simonetta, F. Fanalista for support in the development and test of the nanomechanical protocol, and C. Lenardi for support in the cell-biology laboratory of CIMaINa.
\end{acknowledgments}

\bibliography{nanomechanics_cells}

\begin{thebibliography}{86}%
\makeatletter
\providecommand \@ifxundefined [1]{%
 \@ifx{#1\undefined}
}%
\providecommand \@ifnum [1]{%
 \ifnum #1\expandafter \@firstoftwo
 \else \expandafter \@secondoftwo
 \fi
}%
\providecommand \@ifx [1]{%
 \ifx #1\expandafter \@firstoftwo
 \else \expandafter \@secondoftwo
 \fi
}%
\providecommand \natexlab [1]{#1}%
\providecommand \enquote  [1]{``#1''}%
\providecommand \bibnamefont  [1]{#1}%
\providecommand \bibfnamefont [1]{#1}%
\providecommand \citenamefont [1]{#1}%
\providecommand \href@noop [0]{\@secondoftwo}%
\providecommand \href [0]{\begingroup \@sanitize@url \@href}%
\providecommand \@href[1]{\@@startlink{#1}\@@href}%
\providecommand \@@href[1]{\endgroup#1\@@endlink}%
\providecommand \@sanitize@url [0]{\catcode `\\12\catcode `\$12\catcode
  `\&12\catcode `\#12\catcode `\^12\catcode `\_12\catcode `\%12\relax}%
\providecommand \@@startlink[1]{}%
\providecommand \@@endlink[0]{}%
\providecommand \url  [0]{\begingroup\@sanitize@url \@url }%
\providecommand \@url [1]{\endgroup\@href {#1}{\urlprefix }}%
\providecommand \urlprefix  [0]{URL }%
\providecommand \Eprint [0]{\href }%
\providecommand \doibase [0]{http://dx.doi.org/}%
\providecommand \selectlanguage [0]{\@gobble}%
\providecommand \bibinfo  [0]{\@secondoftwo}%
\providecommand \bibfield  [0]{\@secondoftwo}%
\providecommand \translation [1]{[#1]}%
\providecommand \BibitemOpen [0]{}%
\providecommand \bibitemStop [0]{}%
\providecommand \bibitemNoStop [0]{.\EOS\space}%
\providecommand \EOS [0]{\spacefactor3000\relax}%
\providecommand \BibitemShut  [1]{\csname bibitem#1\endcsname}%
\let\auto@bib@innerbib\@empty
\bibitem [{\citenamefont {Zhu}, \citenamefont {Bao},\ and\ \citenamefont
  {Wang}(2000)}]{Zhu2000}%
  \BibitemOpen
  \bibfield  {author} {\bibinfo {author} {\bibfnamefont {C.}~\bibnamefont
  {Zhu}}, \bibinfo {author} {\bibfnamefont {G.}~\bibnamefont {Bao}}, \ and\
  \bibinfo {author} {\bibfnamefont {N.}~\bibnamefont {Wang}},\ }\href {\doibase
  10.1146/annurev.bioeng.2.1.189} {\bibfield  {journal} {\bibinfo  {journal}
  {Annual Review of Biomedical Engineering}\ }\textbf {\bibinfo {volume} {2}},\
  \bibinfo {pages} {189} (\bibinfo {year} {2000})}\BibitemShut {NoStop}%
\bibitem [{\citenamefont {SURESH}(2007)}]{Suresh2007}%
  \BibitemOpen
  \bibfield  {author} {\bibinfo {author} {\bibfnamefont {S.}~\bibnamefont
  {SURESH}},\ }\href {\doibase 10.1016/j.actbio.2007.04.002} {\bibfield
  {journal} {\bibinfo  {journal} {Acta Biomaterialia}\ }\textbf {\bibinfo
  {volume} {3}},\ \bibinfo {pages} {413} (\bibinfo {year} {2007})}\BibitemShut
  {NoStop}%
\bibitem [{\citenamefont {Fletcher}\ and\ \citenamefont
  {Mullins}(2010)}]{Fletcher2010}%
  \BibitemOpen
  \bibfield  {author} {\bibinfo {author} {\bibfnamefont {D.~A.}\ \bibnamefont
  {Fletcher}}\ and\ \bibinfo {author} {\bibfnamefont {D.}~\bibnamefont
  {Mullins}},\ }\href {\doibase 10.1038/nature08908} {\bibfield  {journal}
  {\bibinfo  {journal} {NATURE}\ }\textbf {\bibinfo {volume} {463}},\ \bibinfo
  {pages} {485} (\bibinfo {year} {2010})}\BibitemShut {NoStop}%
\bibitem [{\citenamefont {Lim}, \citenamefont {Zhou},\ and\ \citenamefont
  {Quek}(2006)}]{Lim2006}%
  \BibitemOpen
  \bibfield  {author} {\bibinfo {author} {\bibfnamefont {C.}~\bibnamefont
  {Lim}}, \bibinfo {author} {\bibfnamefont {E.}~\bibnamefont {Zhou}}, \ and\
  \bibinfo {author} {\bibfnamefont {S.}~\bibnamefont {Quek}},\ }\href {\doibase
  10.1016/j.jbiomech.2004.12.008} {\bibfield  {journal} {\bibinfo  {journal}
  {Journal of Biomechanics}\ }\textbf {\bibinfo {volume} {39}},\ \bibinfo
  {pages} {195} (\bibinfo {year} {2006})}\BibitemShut {NoStop}%
\bibitem [{\citenamefont {Alberts}\ \emph {et~al.}(2008)\citenamefont
  {Alberts}, \citenamefont {Johnson}, \citenamefont {Lewis}, \citenamefont
  {Raff}, \citenamefont {Roberts},\ and\ \citenamefont {Walter}}]{Alberts2008}%
  \BibitemOpen
  \bibfield  {author} {\bibinfo {author} {\bibfnamefont {B.}~\bibnamefont
  {Alberts}}, \bibinfo {author} {\bibfnamefont {A.}~\bibnamefont {Johnson}},
  \bibinfo {author} {\bibfnamefont {J.}~\bibnamefont {Lewis}}, \bibinfo
  {author} {\bibfnamefont {M.}~\bibnamefont {Raff}}, \bibinfo {author}
  {\bibfnamefont {K.}~\bibnamefont {Roberts}}, \ and\ \bibinfo {author}
  {\bibfnamefont {P.}~\bibnamefont {Walter}},\ }\href@noop {} {\emph {\bibinfo
  {title} {Molecular Biology of the Cell}}}\ (\bibinfo  {publisher} {GARLAND
  SCIENCE, TAYLOR AND FRANCIS GROUP},\ \bibinfo {year} {2008})\BibitemShut
  {NoStop}%
\bibitem [{\citenamefont {Jaalouk}\ and\ \citenamefont
  {Lammerding}(2009)}]{Jaalouk2009}%
  \BibitemOpen
  \bibfield  {author} {\bibinfo {author} {\bibfnamefont {D.~E.}\ \bibnamefont
  {Jaalouk}}\ and\ \bibinfo {author} {\bibfnamefont {J.}~\bibnamefont
  {Lammerding}},\ }\href {\doibase 10.1038/nrm2597} {\bibfield  {journal}
  {\bibinfo  {journal} {Nat Rev Mol Cell Biol}\ }\textbf {\bibinfo {volume}
  {10}},\ \bibinfo {pages} {63} (\bibinfo {year} {2009})}\BibitemShut {NoStop}%
\bibitem [{\citenamefont {Diz-Munoz}, \citenamefont {Fletcher},\ and\
  \citenamefont {Weiner}(2013)}]{Diz-Munoz2013}%
  \BibitemOpen
  \bibfield  {author} {\bibinfo {author} {\bibfnamefont {A.}~\bibnamefont
  {Diz-Munoz}}, \bibinfo {author} {\bibfnamefont {D.~A.}\ \bibnamefont
  {Fletcher}}, \ and\ \bibinfo {author} {\bibfnamefont {O.~D.}\ \bibnamefont
  {Weiner}},\ }\href {\doibase 10.1016/j.tcb.2012.09.006} {\bibfield  {journal}
  {\bibinfo  {journal} {Trends in Cell Biology}\ }\textbf {\bibinfo {volume}
  {23}},\ \bibinfo {pages} {47} (\bibinfo {year} {2013})}\BibitemShut {NoStop}%
\bibitem [{\citenamefont {Lamour}\ \emph {et~al.}(2010)\citenamefont {Lamour},
  \citenamefont {Eftekhari-Bafrooei}, \citenamefont {Borguet}, \citenamefont
  {Soues},\ and\ \citenamefont {Hamraoui}}]{Lamour2010}%
  \BibitemOpen
  \bibfield  {author} {\bibinfo {author} {\bibfnamefont {G.}~\bibnamefont
  {Lamour}}, \bibinfo {author} {\bibfnamefont {A.}~\bibnamefont
  {Eftekhari-Bafrooei}}, \bibinfo {author} {\bibfnamefont {E.}~\bibnamefont
  {Borguet}}, \bibinfo {author} {\bibfnamefont {S.}~\bibnamefont {Soues}}, \
  and\ \bibinfo {author} {\bibfnamefont {A.}~\bibnamefont {Hamraoui}},\ }\href
  {\doibase 10.1016/j.biomaterials.2010.01.099} {\bibfield  {journal} {\bibinfo
   {journal} {BIOMATERIALS}\ }\textbf {\bibinfo {volume} {31}},\ \bibinfo
  {pages} {3762} (\bibinfo {year} {2010})}\BibitemShut {NoStop}%
\bibitem [{\citenamefont {Bao}\ and\ \citenamefont {Suresh}(2003)}]{Bao2003}%
  \BibitemOpen
  \bibfield  {author} {\bibinfo {author} {\bibfnamefont {G.}~\bibnamefont
  {Bao}}\ and\ \bibinfo {author} {\bibfnamefont {S.}~\bibnamefont {Suresh}},\
  }\href {\doibase 10.1038/nmat1001} {\bibfield  {journal} {\bibinfo  {journal}
  {Nature Materials}\ }\textbf {\bibinfo {volume} {2}},\ \bibinfo {pages} {715}
  (\bibinfo {year} {2003})}\BibitemShut {NoStop}%
\bibitem [{\citenamefont {Discher}, \citenamefont {Janmey},\ and\ \citenamefont
  {Wang}(2005)}]{Discher2005}%
  \BibitemOpen
  \bibfield  {author} {\bibinfo {author} {\bibfnamefont {D.~E.}\ \bibnamefont
  {Discher}}, \bibinfo {author} {\bibfnamefont {P.}~\bibnamefont {Janmey}}, \
  and\ \bibinfo {author} {\bibfnamefont {Y.~L.}\ \bibnamefont {Wang}},\ }\href
  {\doibase 10.1126/science.1116995} {\bibfield  {journal} {\bibinfo  {journal}
  {SCIENCE}\ }\textbf {\bibinfo {volume} {310}},\ \bibinfo {pages} {1139}
  (\bibinfo {year} {2005})}\BibitemShut {NoStop}%
\bibitem [{\citenamefont {Paszek}\ \emph {et~al.}(2005)\citenamefont {Paszek},
  \citenamefont {Zahir}, \citenamefont {Johnson}, \citenamefont {Lakins},
  \citenamefont {Rozenberg}, \citenamefont {Gefen}, \citenamefont
  {Reinhart-King}, \citenamefont {Margulies}, \citenamefont {Dembo},
  \citenamefont {Boettiger}, \citenamefont {Hammer},\ and\ \citenamefont
  {Weaver}}]{Paszek2005}%
  \BibitemOpen
  \bibfield  {author} {\bibinfo {author} {\bibfnamefont {M.~J.}\ \bibnamefont
  {Paszek}}, \bibinfo {author} {\bibfnamefont {N.}~\bibnamefont {Zahir}},
  \bibinfo {author} {\bibfnamefont {K.~R.}\ \bibnamefont {Johnson}}, \bibinfo
  {author} {\bibfnamefont {J.~N.}\ \bibnamefont {Lakins}}, \bibinfo {author}
  {\bibfnamefont {G.~I.}\ \bibnamefont {Rozenberg}}, \bibinfo {author}
  {\bibfnamefont {A.}~\bibnamefont {Gefen}}, \bibinfo {author} {\bibfnamefont
  {C.~A.}\ \bibnamefont {Reinhart-King}}, \bibinfo {author} {\bibfnamefont
  {S.~S.}\ \bibnamefont {Margulies}}, \bibinfo {author} {\bibfnamefont
  {M.}~\bibnamefont {Dembo}}, \bibinfo {author} {\bibfnamefont
  {D.}~\bibnamefont {Boettiger}}, \bibinfo {author} {\bibfnamefont {D.~A.}\
  \bibnamefont {Hammer}}, \ and\ \bibinfo {author} {\bibfnamefont {V.~M.}\
  \bibnamefont {Weaver}},\ }\href {\doibase 10.1016/j.ccr.2005.08.010}
  {\bibfield  {journal} {\bibinfo  {journal} {Cancer Cell}\ }\textbf {\bibinfo
  {volume} {8}},\ \bibinfo {pages} {241} (\bibinfo {year} {2005})}\BibitemShut
  {NoStop}%
\bibitem [{\citenamefont {Elson}(1988)}]{Elson1988}%
  \BibitemOpen
  \bibfield  {author} {\bibinfo {author} {\bibfnamefont {E.}~\bibnamefont
  {Elson}},\ }\href@noop {} {\bibfield  {journal} {\bibinfo  {journal} {ANNUAL
  REVIEW OF BIOPHYSICS AND BIOPHYSICAL CHEMISTRY}\ }\textbf {\bibinfo {volume}
  {17}},\ \bibinfo {pages} {397} (\bibinfo {year} {1988})}\BibitemShut
  {NoStop}%
\bibitem [{\citenamefont {Stricker}, \citenamefont {Falzone},\ and\
  \citenamefont {Gardel}(2010)}]{Stricker2010}%
  \BibitemOpen
  \bibfield  {author} {\bibinfo {author} {\bibfnamefont {J.}~\bibnamefont
  {Stricker}}, \bibinfo {author} {\bibfnamefont {T.}~\bibnamefont {Falzone}}, \
  and\ \bibinfo {author} {\bibfnamefont {M.~L.}\ \bibnamefont {Gardel}},\
  }\href {\doibase 10.1016/j.jbiomech.2009.09.003} {\bibfield  {journal}
  {\bibinfo  {journal} {JOURNAL OF BIOMECHANICS}\ }\textbf {\bibinfo {volume}
  {43}},\ \bibinfo {pages} {9} (\bibinfo {year} {2010})}\BibitemShut {NoStop}%
\bibitem [{\citenamefont {Muller}\ and\ \citenamefont
  {Dufrene}(2008)}]{Muller2008}%
  \BibitemOpen
  \bibfield  {author} {\bibinfo {author} {\bibfnamefont {D.~J.}\ \bibnamefont
  {Muller}}\ and\ \bibinfo {author} {\bibfnamefont {Y.~F.}\ \bibnamefont
  {Dufrene}},\ }\href {\doibase 10.1038/nnano.2008.100} {\bibfield  {journal}
  {\bibinfo  {journal} {Nature Nanotechnology}\ }\textbf {\bibinfo {volume}
  {3}},\ \bibinfo {pages} {261} (\bibinfo {year} {2008})}\BibitemShut {NoStop}%
\bibitem [{\citenamefont {Alessandrini}\ and\ \citenamefont
  {Facci}(2005)}]{Alessandrini2005}%
  \BibitemOpen
  \bibfield  {author} {\bibinfo {author} {\bibfnamefont {A.}~\bibnamefont
  {Alessandrini}}\ and\ \bibinfo {author} {\bibfnamefont {P.}~\bibnamefont
  {Facci}},\ }\href {\doibase 10.1088/0957-0233/16/6/r01} {\bibfield  {journal}
  {\bibinfo  {journal} {Measurement Science and Technology}\ }\textbf {\bibinfo
  {volume} {16}},\ \bibinfo {pages} {R65} (\bibinfo {year} {2005})}\BibitemShut
  {NoStop}%
\bibitem [{\citenamefont {{Weisenhorn}}\ \emph {et~al.}(1993)\citenamefont
  {{Weisenhorn}}, \citenamefont {{Khorsandi}}, \citenamefont {{Kasas}},
  \citenamefont {{Gotzos}},\ and\ \citenamefont {{Butt}}}]{Weisenhorn1993}%
  \BibitemOpen
  \bibfield  {author} {\bibinfo {author} {\bibfnamefont {A.~L.}\ \bibnamefont
  {{Weisenhorn}}}, \bibinfo {author} {\bibfnamefont {M.}~\bibnamefont
  {{Khorsandi}}}, \bibinfo {author} {\bibfnamefont {S.}~\bibnamefont
  {{Kasas}}}, \bibinfo {author} {\bibfnamefont {V.}~\bibnamefont {{Gotzos}}}, \
  and\ \bibinfo {author} {\bibfnamefont {H.-J.}\ \bibnamefont {{Butt}}},\
  }\href {\doibase 10.1088/0957-4484/4/2/006} {\bibfield  {journal} {\bibinfo
  {journal} {Nanotechnology}\ }\textbf {\bibinfo {volume} {4}},\ \bibinfo
  {pages} {106} (\bibinfo {year} {1993})}\BibitemShut {NoStop}%
\bibitem [{\citenamefont {Butt}, \citenamefont {Cappella},\ and\ \citenamefont
  {Kappl}(2005)}]{Butt2005}%
  \BibitemOpen
  \bibfield  {author} {\bibinfo {author} {\bibfnamefont {H.-J.}\ \bibnamefont
  {Butt}}, \bibinfo {author} {\bibfnamefont {B.}~\bibnamefont {Cappella}}, \
  and\ \bibinfo {author} {\bibfnamefont {M.}~\bibnamefont {Kappl}},\ }\href
  {\doibase 10.1016/j.surfrep.2005.08.003} {\bibfield  {journal} {\bibinfo
  {journal} {Surface Science Reports}\ }\textbf {\bibinfo {volume} {59}},\
  \bibinfo {pages} {1} (\bibinfo {year} {2005})}\BibitemShut {NoStop}%
\bibitem [{\citenamefont {Radmacher}, \citenamefont {Tillmann},\ and\
  \citenamefont {Gaub}(1993)}]{Radmacher1993}%
  \BibitemOpen
  \bibfield  {author} {\bibinfo {author} {\bibfnamefont {M.}~\bibnamefont
  {Radmacher}}, \bibinfo {author} {\bibfnamefont {R.}~\bibnamefont {Tillmann}},
  \ and\ \bibinfo {author} {\bibfnamefont {H.}~\bibnamefont {Gaub}},\ }\href
  {\doibase 10.1016/s0006-3495(93)81433-4} {\bibfield  {journal} {\bibinfo
  {journal} {Biophysical Journal}\ }\textbf {\bibinfo {volume} {64}},\ \bibinfo
  {pages} {735} (\bibinfo {year} {1993})}\BibitemShut {NoStop}%
\bibitem [{\citenamefont {Alcaraz}\ \emph {et~al.}(2003)\citenamefont
  {Alcaraz}, \citenamefont {Buscemi}, \citenamefont {Grabulosa}, \citenamefont
  {Trepat}, \citenamefont {Fabry}, \citenamefont {Farre},\ and\ \citenamefont
  {Navajas}}]{Alcaraz2003}%
  \BibitemOpen
  \bibfield  {author} {\bibinfo {author} {\bibfnamefont {J.}~\bibnamefont
  {Alcaraz}}, \bibinfo {author} {\bibfnamefont {L.}~\bibnamefont {Buscemi}},
  \bibinfo {author} {\bibfnamefont {M.}~\bibnamefont {Grabulosa}}, \bibinfo
  {author} {\bibfnamefont {X.}~\bibnamefont {Trepat}}, \bibinfo {author}
  {\bibfnamefont {B.}~\bibnamefont {Fabry}}, \bibinfo {author} {\bibfnamefont
  {R.}~\bibnamefont {Farre}}, \ and\ \bibinfo {author} {\bibfnamefont
  {D.}~\bibnamefont {Navajas}},\ }\href {\doibase
  10.1016/s0006-3495(03)75014-0} {\bibfield  {journal} {\bibinfo  {journal}
  {Biophysical Journal}\ }\textbf {\bibinfo {volume} {84}},\ \bibinfo {pages}
  {2071} (\bibinfo {year} {2003})}\BibitemShut {NoStop}%
\bibitem [{\citenamefont {Kasas}\ and\ \citenamefont
  {Dietler}(2008)}]{Kasas2008}%
  \BibitemOpen
  \bibfield  {author} {\bibinfo {author} {\bibfnamefont {S.}~\bibnamefont
  {Kasas}}\ and\ \bibinfo {author} {\bibfnamefont {G.}~\bibnamefont
  {Dietler}},\ }\href {\doibase 10.1007/s00424-008-0448-y} {\bibfield
  {journal} {\bibinfo  {journal} {Pflugers Archiv - European Journal of
  Physiology}\ }\textbf {\bibinfo {volume} {456}},\ \bibinfo {pages} {13}
  (\bibinfo {year} {2008})}\BibitemShut {NoStop}%
\bibitem [{\citenamefont {Mahaffy}\ \emph {et~al.}(2004)\citenamefont
  {Mahaffy}, \citenamefont {Park}, \citenamefont {Gerde}, \citenamefont {Kas},\
  and\ \citenamefont {Shih}}]{Mahaffy2004}%
  \BibitemOpen
  \bibfield  {author} {\bibinfo {author} {\bibfnamefont {R.~E.}\ \bibnamefont
  {Mahaffy}}, \bibinfo {author} {\bibfnamefont {S.}~\bibnamefont {Park}},
  \bibinfo {author} {\bibfnamefont {E.}~\bibnamefont {Gerde}}, \bibinfo
  {author} {\bibfnamefont {J.}~\bibnamefont {Kas}}, \ and\ \bibinfo {author}
  {\bibfnamefont {C.~K.}\ \bibnamefont {Shih}},\ }\href {\doibase
  10.1016/S0006-3495(04)74245-9} {\bibfield  {journal} {\bibinfo  {journal}
  {BIOPHYSICAL JOURNAL}\ }\textbf {\bibinfo {volume} {86}},\ \bibinfo {pages}
  {1777} (\bibinfo {year} {2004})}\BibitemShut {NoStop}%
\bibitem [{\citenamefont {Nia}\ \emph {et~al.}(2013)\citenamefont {Nia},
  \citenamefont {Bozchalooi}, \citenamefont {Li}, \citenamefont {Han},
  \citenamefont {Hung}, \citenamefont {Frank}, \citenamefont {Youcef-Toumi},
  \citenamefont {Ortiz},\ and\ \citenamefont {Grodzinsky}}]{Nia2013}%
  \BibitemOpen
  \bibfield  {author} {\bibinfo {author} {\bibfnamefont {H.~T.}\ \bibnamefont
  {Nia}}, \bibinfo {author} {\bibfnamefont {I.~S.}\ \bibnamefont {Bozchalooi}},
  \bibinfo {author} {\bibfnamefont {Y.}~\bibnamefont {Li}}, \bibinfo {author}
  {\bibfnamefont {L.}~\bibnamefont {Han}}, \bibinfo {author} {\bibfnamefont
  {H.-H.}\ \bibnamefont {Hung}}, \bibinfo {author} {\bibfnamefont
  {E.}~\bibnamefont {Frank}}, \bibinfo {author} {\bibfnamefont
  {K.}~\bibnamefont {Youcef-Toumi}}, \bibinfo {author} {\bibfnamefont
  {C.}~\bibnamefont {Ortiz}}, \ and\ \bibinfo {author} {\bibfnamefont
  {A.}~\bibnamefont {Grodzinsky}},\ }\href {\doibase 10.1016/j.bpj.2013.02.048}
  {\bibfield  {journal} {\bibinfo  {journal} {Biophysical Journal}\ }\textbf
  {\bibinfo {volume} {104}},\ \bibinfo {pages} {1529} (\bibinfo {year}
  {2013})}\BibitemShut {NoStop}%
\bibitem [{\citenamefont {Rotsch}\ and\ \citenamefont
  {Radmacher}(2000)}]{Rotsch2000}%
  \BibitemOpen
  \bibfield  {author} {\bibinfo {author} {\bibfnamefont {C.}~\bibnamefont
  {Rotsch}}\ and\ \bibinfo {author} {\bibfnamefont {M.}~\bibnamefont
  {Radmacher}},\ }\href@noop {} {\bibfield  {journal} {\bibinfo  {journal}
  {BIOPHYSICAL JOURNAL}\ }\textbf {\bibinfo {volume} {78}},\ \bibinfo {pages}
  {520} (\bibinfo {year} {2000})}\BibitemShut {NoStop}%
\bibitem [{\citenamefont {Kasas}\ \emph {et~al.}(2005)\citenamefont {Kasas},
  \citenamefont {Wang}, \citenamefont {Hirling}, \citenamefont {Marsault},
  \citenamefont {Huni}, \citenamefont {Yersin}, \citenamefont {Regazzi},
  \citenamefont {Grenningloh}, \citenamefont {Riederer}, \citenamefont {Forro},
  \citenamefont {Dietler},\ and\ \citenamefont {Catsicas}}]{Kasas2005}%
  \BibitemOpen
  \bibfield  {author} {\bibinfo {author} {\bibfnamefont {S.}~\bibnamefont
  {Kasas}}, \bibinfo {author} {\bibfnamefont {X.}~\bibnamefont {Wang}},
  \bibinfo {author} {\bibfnamefont {H.}~\bibnamefont {Hirling}}, \bibinfo
  {author} {\bibfnamefont {R.}~\bibnamefont {Marsault}}, \bibinfo {author}
  {\bibfnamefont {B.}~\bibnamefont {Huni}}, \bibinfo {author} {\bibfnamefont
  {A.}~\bibnamefont {Yersin}}, \bibinfo {author} {\bibfnamefont
  {R.}~\bibnamefont {Regazzi}}, \bibinfo {author} {\bibfnamefont
  {G.}~\bibnamefont {Grenningloh}}, \bibinfo {author} {\bibfnamefont
  {B.}~\bibnamefont {Riederer}}, \bibinfo {author} {\bibfnamefont
  {L.}~\bibnamefont {Forro}}, \bibinfo {author} {\bibfnamefont
  {G.}~\bibnamefont {Dietler}}, \ and\ \bibinfo {author} {\bibfnamefont
  {S.}~\bibnamefont {Catsicas}},\ }\href {\doibase 10.1002/cm.20086} {\bibfield
   {journal} {\bibinfo  {journal} {CELL MOTILITY AND THE CYTOSKELETON}\
  }\textbf {\bibinfo {volume} {62}},\ \bibinfo {pages} {124} (\bibinfo {year}
  {2005})}\BibitemShut {NoStop}%
\bibitem [{\citenamefont {Oberleithner}\ \emph {et~al.}(2006)\citenamefont
  {Oberleithner}, \citenamefont {Riethmuller}, \citenamefont {Ludwig},
  \citenamefont {Shahin}, \citenamefont {Stock}, \citenamefont {Schwab},
  \citenamefont {Hausberg}, \citenamefont {Kusche},\ and\ \citenamefont
  {Schillers}}]{Oberleithner2006}%
  \BibitemOpen
  \bibfield  {author} {\bibinfo {author} {\bibfnamefont {H.}~\bibnamefont
  {Oberleithner}}, \bibinfo {author} {\bibfnamefont {C.}~\bibnamefont
  {Riethmuller}}, \bibinfo {author} {\bibfnamefont {T.}~\bibnamefont {Ludwig}},
  \bibinfo {author} {\bibfnamefont {V.}~\bibnamefont {Shahin}}, \bibinfo
  {author} {\bibfnamefont {C.}~\bibnamefont {Stock}}, \bibinfo {author}
  {\bibfnamefont {A.}~\bibnamefont {Schwab}}, \bibinfo {author} {\bibfnamefont
  {M.}~\bibnamefont {Hausberg}}, \bibinfo {author} {\bibfnamefont
  {K.}~\bibnamefont {Kusche}}, \ and\ \bibinfo {author} {\bibfnamefont
  {H.}~\bibnamefont {Schillers}},\ }\href {\doibase 10.1242/jcs.02886}
  {\bibfield  {journal} {\bibinfo  {journal} {JOURNAL OF CELL SCIENCE}\
  }\textbf {\bibinfo {volume} {119}},\ \bibinfo {pages} {1926} (\bibinfo {year}
  {2006})}\BibitemShut {NoStop}%
\bibitem [{\citenamefont {Domke}\ \emph {et~al.}(2000)\citenamefont {Domke},
  \citenamefont {Dannohl}, \citenamefont {Parak}, \citenamefont {Muller},
  \citenamefont {Aicher},\ and\ \citenamefont {Radmacher}}]{Domke2000}%
  \BibitemOpen
  \bibfield  {author} {\bibinfo {author} {\bibfnamefont {J.}~\bibnamefont
  {Domke}}, \bibinfo {author} {\bibfnamefont {S.}~\bibnamefont {Dannohl}},
  \bibinfo {author} {\bibfnamefont {W.~J.}\ \bibnamefont {Parak}}, \bibinfo
  {author} {\bibfnamefont {O.}~\bibnamefont {Muller}, \bibfnamefont {D.~J.}},
  \bibinfo {author} {\bibfnamefont {W.~K.}\ \bibnamefont {Aicher}}, \ and\
  \bibinfo {author} {\bibfnamefont {M.}~\bibnamefont {Radmacher}},\ }\href
  {\doibase 10.1016/s0927-7765(00)00145-4} {\bibfield  {journal} {\bibinfo
  {journal} {Colloids and Surfaces B: Biointerfaces}\ }\textbf {\bibinfo
  {volume} {19}},\ \bibinfo {pages} {367} (\bibinfo {year} {2000})}\BibitemShut
  {NoStop}%
\bibitem [{\citenamefont {Tee}\ \emph {et~al.}(2011)\citenamefont {Tee},
  \citenamefont {Fu}, \citenamefont {Chen},\ and\ \citenamefont
  {Janmey}}]{Tee2011}%
  \BibitemOpen
  \bibfield  {author} {\bibinfo {author} {\bibfnamefont {S.~Y.}\ \bibnamefont
  {Tee}}, \bibinfo {author} {\bibfnamefont {J.}~\bibnamefont {Fu}}, \bibinfo
  {author} {\bibfnamefont {C.~S.}\ \bibnamefont {Chen}}, \ and\ \bibinfo
  {author} {\bibfnamefont {P.~A.}\ \bibnamefont {Janmey}},\ }\href {\doibase
  10.1016/j.bpj.2010.12.3744} {\bibfield  {journal} {\bibinfo  {journal}
  {BIOPHYSICAL JOURNAL}\ }\textbf {\bibinfo {volume} {100}},\ \bibinfo {pages}
  {L25} (\bibinfo {year} {2011})}\BibitemShut {NoStop}%
\bibitem [{\citenamefont {Engler}\ \emph {et~al.}(2004)\citenamefont {Engler},
  \citenamefont {Bacakova}, \citenamefont {Newman}, \citenamefont {Hategan},
  \citenamefont {Griffin},\ and\ \citenamefont {Discher}}]{Engler2004}%
  \BibitemOpen
  \bibfield  {author} {\bibinfo {author} {\bibfnamefont {A.}~\bibnamefont
  {Engler}}, \bibinfo {author} {\bibfnamefont {L.}~\bibnamefont {Bacakova}},
  \bibinfo {author} {\bibfnamefont {C.}~\bibnamefont {Newman}}, \bibinfo
  {author} {\bibfnamefont {A.}~\bibnamefont {Hategan}}, \bibinfo {author}
  {\bibfnamefont {M.}~\bibnamefont {Griffin}}, \ and\ \bibinfo {author}
  {\bibfnamefont {D.}~\bibnamefont {Discher}},\ }\href {\doibase
  10.1016/s0006-3495(04)74140-5} {\bibfield  {journal} {\bibinfo  {journal}
  {Biophysical Journal}\ }\textbf {\bibinfo {volume} {86}},\ \bibinfo {pages}
  {617} (\bibinfo {year} {2004})}\BibitemShut {NoStop}%
\bibitem [{\citenamefont {Lieber}\ \emph {et~al.}(2004)\citenamefont {Lieber},
  \citenamefont {Aubry}, \citenamefont {Pain}, \citenamefont {Diaz},
  \citenamefont {Kim},\ and\ \citenamefont {Vatner}}]{Lieber2004}%
  \BibitemOpen
  \bibfield  {author} {\bibinfo {author} {\bibfnamefont {S.~C.}\ \bibnamefont
  {Lieber}}, \bibinfo {author} {\bibfnamefont {N.}~\bibnamefont {Aubry}},
  \bibinfo {author} {\bibfnamefont {J.}~\bibnamefont {Pain}}, \bibinfo {author}
  {\bibfnamefont {G.}~\bibnamefont {Diaz}}, \bibinfo {author} {\bibfnamefont
  {S.-J.}\ \bibnamefont {Kim}}, \ and\ \bibinfo {author} {\bibfnamefont
  {S.~F.}\ \bibnamefont {Vatner}},\ }\href {\doibase
  10.1152/ajpheart.00564.2003} {\bibfield  {journal} {\bibinfo  {journal} {Am J
  Physiol Heart Circ Physiol}\ }\textbf {\bibinfo {volume} {287}},\ \bibinfo
  {pages} {H645} (\bibinfo {year} {2004})}\BibitemShut {NoStop}%
\bibitem [{\citenamefont {Berdyyeva}, \citenamefont {Woodworth},\ and\
  \citenamefont {Sokolov}(2005)}]{Berdyyeva2005}%
  \BibitemOpen
  \bibfield  {author} {\bibinfo {author} {\bibfnamefont {T.~K.}\ \bibnamefont
  {Berdyyeva}}, \bibinfo {author} {\bibfnamefont {C.~D.}\ \bibnamefont
  {Woodworth}}, \ and\ \bibinfo {author} {\bibfnamefont {I.}~\bibnamefont
  {Sokolov}},\ }\href {\doibase 10.1088/0031-9155/50/1/007} {\bibfield
  {journal} {\bibinfo  {journal} {Phys. Med. Biol.}\ }\textbf {\bibinfo
  {volume} {50}},\ \bibinfo {pages} {81} (\bibinfo {year} {2005})}\BibitemShut
  {NoStop}%
\bibitem [{\citenamefont {Lekka}\ \emph {et~al.}(1999)\citenamefont {Lekka},
  \citenamefont {Laidler}, \citenamefont {Gil}, \citenamefont {Lekki},
  \citenamefont {Stachura},\ and\ \citenamefont {Hrynkiewicz}}]{Lekka1999}%
  \BibitemOpen
  \bibfield  {author} {\bibinfo {author} {\bibfnamefont {M.}~\bibnamefont
  {Lekka}}, \bibinfo {author} {\bibfnamefont {P.}~\bibnamefont {Laidler}},
  \bibinfo {author} {\bibfnamefont {D.}~\bibnamefont {Gil}}, \bibinfo {author}
  {\bibfnamefont {J.}~\bibnamefont {Lekki}}, \bibinfo {author} {\bibfnamefont
  {Z.}~\bibnamefont {Stachura}}, \ and\ \bibinfo {author} {\bibfnamefont
  {A.~Z.}\ \bibnamefont {Hrynkiewicz}},\ }\href {\doibase
  10.1007/s002490050213} {\bibfield  {journal} {\bibinfo  {journal} {European
  Biophysics Journal}\ }\textbf {\bibinfo {volume} {28}},\ \bibinfo {pages}
  {312} (\bibinfo {year} {1999})}\BibitemShut {NoStop}%
\bibitem [{\citenamefont {Lekka}\ \emph {et~al.}(2012)\citenamefont {Lekka},
  \citenamefont {Pogoda}, \citenamefont {Gostek}, \citenamefont {Klymenko},
  \citenamefont {Prauzner-Bechcicki}, \citenamefont {Wiltowska-Zuber},
  \citenamefont {Jaczewska}, \citenamefont {Lekki},\ and\ \citenamefont
  {Stachura}}]{Lekka2012}%
  \BibitemOpen
  \bibfield  {author} {\bibinfo {author} {\bibfnamefont {M.}~\bibnamefont
  {Lekka}}, \bibinfo {author} {\bibfnamefont {K.}~\bibnamefont {Pogoda}},
  \bibinfo {author} {\bibfnamefont {J.}~\bibnamefont {Gostek}}, \bibinfo
  {author} {\bibfnamefont {O.}~\bibnamefont {Klymenko}}, \bibinfo {author}
  {\bibfnamefont {S.}~\bibnamefont {Prauzner-Bechcicki}}, \bibinfo {author}
  {\bibfnamefont {J.}~\bibnamefont {Wiltowska-Zuber}}, \bibinfo {author}
  {\bibfnamefont {J.}~\bibnamefont {Jaczewska}}, \bibinfo {author}
  {\bibfnamefont {J.}~\bibnamefont {Lekki}}, \ and\ \bibinfo {author}
  {\bibfnamefont {Z.}~\bibnamefont {Stachura}},\ }\href {\doibase
  10.1016/j.micron.2012.01.019} {\bibfield  {journal} {\bibinfo  {journal}
  {Micron}\ }\textbf {\bibinfo {volume} {43}},\ \bibinfo {pages} {1259}
  (\bibinfo {year} {2012})}\BibitemShut {NoStop}%
\bibitem [{\citenamefont {Cross}\ \emph {et~al.}(2007)\citenamefont {Cross},
  \citenamefont {Jin}, \citenamefont {Rao},\ and\ \citenamefont
  {Gimzewski}}]{Cross2007}%
  \BibitemOpen
  \bibfield  {author} {\bibinfo {author} {\bibfnamefont {S.~E.}\ \bibnamefont
  {Cross}}, \bibinfo {author} {\bibfnamefont {Y.~S.}\ \bibnamefont {Jin}},
  \bibinfo {author} {\bibfnamefont {J.}~\bibnamefont {Rao}}, \ and\ \bibinfo
  {author} {\bibfnamefont {J.~K.}\ \bibnamefont {Gimzewski}},\ }\href {\doibase
  10.1038/nnano.2007.388} {\bibfield  {journal} {\bibinfo  {journal} {NATURE
  NANOTECHNOLOGY}\ }\textbf {\bibinfo {volume} {2}},\ \bibinfo {pages} {780}
  (\bibinfo {year} {2007})}\BibitemShut {NoStop}%
\bibitem [{\citenamefont {Plodinec}\ \emph {et~al.}(2012)\citenamefont
  {Plodinec}, \citenamefont {Loparic}, \citenamefont {Monnier}, \citenamefont
  {Obermann}, \citenamefont {Zanetti-Dallenbach}, \citenamefont {Oertle},
  \citenamefont {Hyotyla}, \citenamefont {Aebi}, \citenamefont {Bentires-Alj},
  \citenamefont {Lim},\ and\ \citenamefont {Schoenenberger}}]{Plodinec2012}%
  \BibitemOpen
  \bibfield  {author} {\bibinfo {author} {\bibfnamefont {M.}~\bibnamefont
  {Plodinec}}, \bibinfo {author} {\bibfnamefont {M.}~\bibnamefont {Loparic}},
  \bibinfo {author} {\bibfnamefont {C.~A.}\ \bibnamefont {Monnier}}, \bibinfo
  {author} {\bibfnamefont {E.~C.}\ \bibnamefont {Obermann}}, \bibinfo {author}
  {\bibfnamefont {R.}~\bibnamefont {Zanetti-Dallenbach}}, \bibinfo {author}
  {\bibfnamefont {P.}~\bibnamefont {Oertle}}, \bibinfo {author} {\bibfnamefont
  {J.~T.}\ \bibnamefont {Hyotyla}}, \bibinfo {author} {\bibfnamefont
  {U.}~\bibnamefont {Aebi}}, \bibinfo {author} {\bibfnamefont {M.}~\bibnamefont
  {Bentires-Alj}}, \bibinfo {author} {\bibfnamefont {R.~Y.~H.}\ \bibnamefont
  {Lim}}, \ and\ \bibinfo {author} {\bibfnamefont {C.~A.}\ \bibnamefont
  {Schoenenberger}},\ }\href {\doibase 10.1038/NNANO.2012.167} {\bibfield
  {journal} {\bibinfo  {journal} {NATURE NANOTECHNOLOGY}\ }\textbf {\bibinfo
  {volume} {7}},\ \bibinfo {pages} {757} (\bibinfo {year} {2012})}\BibitemShut
  {NoStop}%
\bibitem [{\citenamefont {Sch\"ape}\ \emph {et~al.}(2014)\citenamefont
  {Sch\"ape}, \citenamefont {Luque}, \citenamefont {Doschke}, \citenamefont
  {Schillers}, \citenamefont {Walte}, \citenamefont {Uriarte}, \citenamefont
  {Campillo}, \citenamefont {Michanetzis}, \citenamefont {Gostek},
  \citenamefont {Dumitru}, \citenamefont {Herruzo}, \citenamefont {Parot},
  \citenamefont {Galluzzi}, \citenamefont {Podest\`a}, \citenamefont
  {Scheuring}, \citenamefont {Missirlis}, \citenamefont {Garcia}, \citenamefont
  {Odorico}, \citenamefont {Lekka}, \citenamefont {Rico}, \citenamefont
  {Pellequer}, \citenamefont {Oberleithner}, \citenamefont {Navajas},\ and\
  \citenamefont {Radmacher}}]{COST_paper}%
  \BibitemOpen
  \bibfield  {author} {\bibinfo {author} {\bibfnamefont {J.}~\bibnamefont
  {Sch\"ape}}, \bibinfo {author} {\bibfnamefont {T.}~\bibnamefont {Luque}},
  \bibinfo {author} {\bibfnamefont {H.}~\bibnamefont {Doschke}}, \bibinfo
  {author} {\bibfnamefont {H.}~\bibnamefont {Schillers}}, \bibinfo {author}
  {\bibfnamefont {M.}~\bibnamefont {Walte}}, \bibinfo {author} {\bibfnamefont
  {J.~J.}\ \bibnamefont {Uriarte}}, \bibinfo {author} {\bibfnamefont
  {N.}~\bibnamefont {Campillo}}, \bibinfo {author} {\bibfnamefont {G.~P.}\
  \bibnamefont {Michanetzis}}, \bibinfo {author} {\bibfnamefont
  {J.}~\bibnamefont {Gostek}}, \bibinfo {author} {\bibfnamefont
  {A.}~\bibnamefont {Dumitru}}, \bibinfo {author} {\bibfnamefont {E.~T.}\
  \bibnamefont {Herruzo}}, \bibinfo {author} {\bibfnamefont {P.}~\bibnamefont
  {Parot}}, \bibinfo {author} {\bibfnamefont {M.}~\bibnamefont {Galluzzi}},
  \bibinfo {author} {\bibfnamefont {A.}~\bibnamefont {Podest\`a}}, \bibinfo
  {author} {\bibfnamefont {S.}~\bibnamefont {Scheuring}}, \bibinfo {author}
  {\bibfnamefont {Y.}~\bibnamefont {Missirlis}}, \bibinfo {author}
  {\bibfnamefont {R.}~\bibnamefont {Garcia}}, \bibinfo {author} {\bibfnamefont
  {M.}~\bibnamefont {Odorico}}, \bibinfo {author} {\bibfnamefont
  {M.}~\bibnamefont {Lekka}}, \bibinfo {author} {\bibfnamefont
  {F.}~\bibnamefont {Rico}}, \bibinfo {author} {\bibfnamefont {J.-L.}\
  \bibnamefont {Pellequer}}, \bibinfo {author} {\bibfnamefont {H.}~\bibnamefont
  {Oberleithner}}, \bibinfo {author} {\bibfnamefont {D.}~\bibnamefont
  {Navajas}}, \ and\ \bibinfo {author} {\bibfnamefont {M.}~\bibnamefont
  {Radmacher}},\ }\href@noop {} {\enquote {\bibinfo {title} {Standardized
  atomic force microscopy method for measuring mechanical properties of soft
  samples: the dubrovnik procedure},}\ } (\bibinfo {year} {2014}),\ \bibinfo
  {note} {in preparation}\BibitemShut {NoStop}%
\bibitem [{\citenamefont {Pletikapic}\ \emph {et~al.}(2011)\citenamefont
  {Pletikapic}, \citenamefont {Berquand}, \citenamefont {Radic},\ and\
  \citenamefont {Svetlicic}}]{Pletikapic2011}%
  \BibitemOpen
  \bibfield  {author} {\bibinfo {author} {\bibfnamefont {G.}~\bibnamefont
  {Pletikapic}}, \bibinfo {author} {\bibfnamefont {A.}~\bibnamefont
  {Berquand}}, \bibinfo {author} {\bibfnamefont {T.~M.}\ \bibnamefont {Radic}},
  \ and\ \bibinfo {author} {\bibfnamefont {V.}~\bibnamefont {Svetlicic}},\
  }\href {\doibase 10.1111/j.1529-8817.2011.01093.x} {\bibfield  {journal}
  {\bibinfo  {journal} {Journal of Phycology}\ }\textbf {\bibinfo {volume}
  {48}},\ \bibinfo {pages} {174} (\bibinfo {year} {2011})}\BibitemShut
  {NoStop}%
\bibitem [{\citenamefont {Braunsmann}\ \emph {et~al.}(2014)\citenamefont
  {Braunsmann}, \citenamefont {Seifert}, \citenamefont {Rheinlaender},\ and\
  \citenamefont {Sch\"{a}ffer}}]{Braunsmann2014}%
  \BibitemOpen
  \bibfield  {author} {\bibinfo {author} {\bibfnamefont {C.}~\bibnamefont
  {Braunsmann}}, \bibinfo {author} {\bibfnamefont {J.}~\bibnamefont {Seifert}},
  \bibinfo {author} {\bibfnamefont {J.}~\bibnamefont {Rheinlaender}}, \ and\
  \bibinfo {author} {\bibfnamefont {T.~E.}\ \bibnamefont {Sch\"{a}ffer}},\
  }\href {\doibase 10.1063/1.4885464} {\bibfield  {journal} {\bibinfo
  {journal} {Review of Scientific Instruments}\ }\textbf {\bibinfo {volume}
  {85}},\ \bibinfo {pages} {073703} (\bibinfo {year} {2014})}\BibitemShut
  {NoStop}%
\bibitem [{\citenamefont {Cartagena}\ and\ \citenamefont
  {Raman}(2014)}]{Cartagena2014}%
  \BibitemOpen
  \bibfield  {author} {\bibinfo {author} {\bibfnamefont {A.}~\bibnamefont
  {Cartagena}}\ and\ \bibinfo {author} {\bibfnamefont {A.}~\bibnamefont
  {Raman}},\ }\href {\doibase 10.1016/j.bpj.2013.12.037} {\bibfield  {journal}
  {\bibinfo  {journal} {Biophysical Journal}\ }\textbf {\bibinfo {volume}
  {106}},\ \bibinfo {pages} {1033} (\bibinfo {year} {2014})}\BibitemShut
  {NoStop}%
\bibitem [{\citenamefont {Costa}\ and\ \citenamefont {Yin}(1999)}]{Costa1999}%
  \BibitemOpen
  \bibfield  {author} {\bibinfo {author} {\bibfnamefont {K.~D.}\ \bibnamefont
  {Costa}}\ and\ \bibinfo {author} {\bibfnamefont {F.~C.}\ \bibnamefont
  {Yin}},\ }\href@noop {} {\bibfield  {journal} {\bibinfo  {journal} {J Biomech
  Eng}\ }\textbf {\bibinfo {volume} {121}},\ \bibinfo {pages} {462} (\bibinfo
  {year} {1999})}\BibitemShut {NoStop}%
\bibitem [{\citenamefont {{Harris}}\ and\ \citenamefont
  {{Charras}}(2011)}]{Harris2011}%
  \BibitemOpen
  \bibfield  {author} {\bibinfo {author} {\bibfnamefont {A.~R.}\ \bibnamefont
  {{Harris}}}\ and\ \bibinfo {author} {\bibfnamefont {G.~T.}\ \bibnamefont
  {{Charras}}},\ }\href {\doibase 10.1088/0957-4484/22/34/345102} {\bibfield
  {journal} {\bibinfo  {journal} {Nanotechnology}\ }\textbf {\bibinfo {volume}
  {22}},\ \bibinfo {eid} {345102} (\bibinfo {year} {2011})}\BibitemShut
  {NoStop}%
\bibitem [{\citenamefont {McKee}\ \emph {et~al.}(2011)\citenamefont {McKee},
  \citenamefont {Last}, \citenamefont {Russell},\ and\ \citenamefont
  {Murphy}}]{McKee2011}%
  \BibitemOpen
  \bibfield  {author} {\bibinfo {author} {\bibfnamefont {C.~T.}\ \bibnamefont
  {McKee}}, \bibinfo {author} {\bibfnamefont {J.~A.}\ \bibnamefont {Last}},
  \bibinfo {author} {\bibfnamefont {P.}~\bibnamefont {Russell}}, \ and\
  \bibinfo {author} {\bibfnamefont {C.~J.}\ \bibnamefont {Murphy}},\ }\href
  {\doibase 10.1089/ten.teb.2010.0520} {\bibfield  {journal} {\bibinfo
  {journal} {Tissue Engineering Part B: Reviews}\ }\textbf {\bibinfo {volume}
  {17}},\ \bibinfo {pages} {155} (\bibinfo {year} {2011})}\BibitemShut
  {NoStop}%
\bibitem [{\citenamefont {Dimitriadis}\ \emph {et~al.}(2002)\citenamefont
  {Dimitriadis}, \citenamefont {Horkay}, \citenamefont {Maresca}, \citenamefont
  {Kachar},\ and\ \citenamefont {Chadwick}}]{Dimitriadis2002}%
  \BibitemOpen
  \bibfield  {author} {\bibinfo {author} {\bibfnamefont {E.~K.}\ \bibnamefont
  {Dimitriadis}}, \bibinfo {author} {\bibfnamefont {F.}~\bibnamefont {Horkay}},
  \bibinfo {author} {\bibfnamefont {J.}~\bibnamefont {Maresca}}, \bibinfo
  {author} {\bibfnamefont {B.}~\bibnamefont {Kachar}}, \ and\ \bibinfo {author}
  {\bibfnamefont {R.~S.}\ \bibnamefont {Chadwick}},\ }\href {\doibase
  10.1016/S0006-3495(02)75620-8} {\bibfield  {journal} {\bibinfo  {journal}
  {BIOPHYSICAL JOURNAL}\ }\textbf {\bibinfo {volume} {82}},\ \bibinfo {pages}
  {2798} (\bibinfo {year} {2002})}\BibitemShut {NoStop}%
\bibitem [{\citenamefont {Crick}\ and\ \citenamefont {Yin}(2006)}]{Crick2006}%
  \BibitemOpen
  \bibfield  {author} {\bibinfo {author} {\bibfnamefont {S.~L.}\ \bibnamefont
  {Crick}}\ and\ \bibinfo {author} {\bibfnamefont {F.~C.-P.}\ \bibnamefont
  {Yin}},\ }\href {\doibase 10.1007/s10237-006-0046-x} {\bibfield  {journal}
  {\bibinfo  {journal} {Biomech Model Mechanobiol}\ }\textbf {\bibinfo {volume}
  {6}},\ \bibinfo {pages} {199} (\bibinfo {year} {2006})}\BibitemShut {NoStop}%
\bibitem [{\citenamefont {Domke}\ and\ \citenamefont
  {Radmacher}(1998)}]{Domke1998}%
  \BibitemOpen
  \bibfield  {author} {\bibinfo {author} {\bibfnamefont {J.}~\bibnamefont
  {Domke}}\ and\ \bibinfo {author} {\bibfnamefont {M.}~\bibnamefont
  {Radmacher}},\ }\href {\doibase 10.1021/la9713006} {\bibfield  {journal}
  {\bibinfo  {journal} {Langmuir}\ }\textbf {\bibinfo {volume} {14}},\ \bibinfo
  {pages} {3320} (\bibinfo {year} {1998})}\BibitemShut {NoStop}%
\bibitem [{COS(2014)}]{COST_TD1002}%
  \BibitemOpen
  \href {http://www.cost.eu/domains_actions/bmbs/Actions/TD1002} {\enquote
  {\bibinfo {title} {{COST Action TD1002}. {E}uropean network on applications
  of {A}tomic {F}orce {M}icroscopy to {N}ano{M}edicine and {L}ife {S}ciences
  ({AFM4NanoMed\&Bio}).}}\ } (\bibinfo {year} {2014})\BibitemShut {NoStop}%
\bibitem [{\citenamefont {van~der Werf}\ \emph {et~al.}(1994)\citenamefont
  {van~der Werf}, \citenamefont {Putman}, \citenamefont {de~Grooth},\ and\
  \citenamefont {Greve}}]{Werf1994}%
  \BibitemOpen
  \bibfield  {author} {\bibinfo {author} {\bibfnamefont {K.~O.}\ \bibnamefont
  {van~der Werf}}, \bibinfo {author} {\bibfnamefont {C.~A.~J.}\ \bibnamefont
  {Putman}}, \bibinfo {author} {\bibfnamefont {B.~G.}\ \bibnamefont
  {de~Grooth}}, \ and\ \bibinfo {author} {\bibfnamefont {J.}~\bibnamefont
  {Greve}},\ }\href {\doibase 10.1063/1.112106} {\bibfield  {journal} {\bibinfo
   {journal} {Appl. Phys. Lett.}\ }\textbf {\bibinfo {volume} {65}},\ \bibinfo
  {pages} {1195} (\bibinfo {year} {1994})}\BibitemShut {NoStop}%
\bibitem [{\citenamefont {Radmacher}\ \emph {et~al.}(1996)\citenamefont
  {Radmacher}, \citenamefont {Fritz}, \citenamefont {Kacher}, \citenamefont
  {Cleveland},\ and\ \citenamefont {Hansma}}]{Radmacher1996}%
  \BibitemOpen
  \bibfield  {author} {\bibinfo {author} {\bibfnamefont {M.}~\bibnamefont
  {Radmacher}}, \bibinfo {author} {\bibfnamefont {M.}~\bibnamefont {Fritz}},
  \bibinfo {author} {\bibfnamefont {C.~M.}\ \bibnamefont {Kacher}}, \bibinfo
  {author} {\bibfnamefont {J.~P.}\ \bibnamefont {Cleveland}}, \ and\ \bibinfo
  {author} {\bibfnamefont {P.~K.}\ \bibnamefont {Hansma}},\ }\href@noop {}
  {\bibfield  {journal} {\bibinfo  {journal} {Biophys J}\ }\textbf {\bibinfo
  {volume} {70}},\ \bibinfo {pages} {556} (\bibinfo {year} {1996})}\BibitemShut
  {NoStop}%
\bibitem [{\citenamefont {Heinz}\ and\ \citenamefont {Hoh}(1999)}]{Heinz1999}%
  \BibitemOpen
  \bibfield  {author} {\bibinfo {author} {\bibfnamefont {W.~F.}\ \bibnamefont
  {Heinz}}\ and\ \bibinfo {author} {\bibfnamefont {J.~H.}\ \bibnamefont
  {Hoh}},\ }\href {\doibase 10.1016/s0167-7799(99)01304-9} {\bibfield
  {journal} {\bibinfo  {journal} {Trends in Biotechnology}\ }\textbf {\bibinfo
  {volume} {17}},\ \bibinfo {pages} {143} (\bibinfo {year} {1999})}\BibitemShut
  {NoStop}%
\bibitem [{\citenamefont {Rico}\ \emph {et~al.}(2005)\citenamefont {Rico},
  \citenamefont {Roca-Cusachs}, \citenamefont {Gavara}, \citenamefont {Farre},
  \citenamefont {Rotger},\ and\ \citenamefont {Navajas}}]{Rico2005}%
  \BibitemOpen
  \bibfield  {author} {\bibinfo {author} {\bibfnamefont {F.}~\bibnamefont
  {Rico}}, \bibinfo {author} {\bibfnamefont {P.}~\bibnamefont {Roca-Cusachs}},
  \bibinfo {author} {\bibfnamefont {N.}~\bibnamefont {Gavara}}, \bibinfo
  {author} {\bibfnamefont {R.}~\bibnamefont {Farre}}, \bibinfo {author}
  {\bibfnamefont {M.}~\bibnamefont {Rotger}}, \ and\ \bibinfo {author}
  {\bibfnamefont {D.}~\bibnamefont {Navajas}},\ }\href {\doibase
  10.1103/PhysRevE.72.021914} {\bibfield  {journal} {\bibinfo  {journal}
  {PHYSICAL REVIEW E}\ }\textbf {\bibinfo {volume} {72}},\ \bibinfo {pages}
  {021914} (\bibinfo {year} {2005})}\BibitemShut {NoStop}%
\bibitem [{\citenamefont {Rother}\ \emph {et~al.}(2014)\citenamefont {Rother},
  \citenamefont {N{\''{o}}ding}, \citenamefont {Mey},\ and\ \citenamefont
  {Janshoff}}]{Rother2014}%
  \BibitemOpen
  \bibfield  {author} {\bibinfo {author} {\bibfnamefont {J.}~\bibnamefont
  {Rother}}, \bibinfo {author} {\bibfnamefont {H.}~\bibnamefont
  {N{\''{o}}ding}}, \bibinfo {author} {\bibfnamefont {I.}~\bibnamefont {Mey}},
  \ and\ \bibinfo {author} {\bibfnamefont {A.}~\bibnamefont {Janshoff}},\
  }\href {\doibase 10.1098/rsob.140046} {\bibfield  {journal} {\bibinfo
  {journal} {Open Biol}\ }\textbf {\bibinfo {volume} {4}},\ \bibinfo {pages}
  {140046} (\bibinfo {year} {2014})}\BibitemShut {NoStop}%
\bibitem [{\citenamefont {Carl}\ and\ \citenamefont
  {Schillers}(2008)}]{Carl2008}%
  \BibitemOpen
  \bibfield  {author} {\bibinfo {author} {\bibfnamefont {P.}~\bibnamefont
  {Carl}}\ and\ \bibinfo {author} {\bibfnamefont {H.}~\bibnamefont
  {Schillers}},\ }\href {\doibase 10.1007/s00424-008-0524-3} {\bibfield
  {journal} {\bibinfo  {journal} {PFLUGERS ARCHIV-EUROPEAN JOURNAL OF
  PHYSIOLOGY}\ }\textbf {\bibinfo {volume} {457}},\ \bibinfo {pages} {551}
  (\bibinfo {year} {2008})}\BibitemShut {NoStop}%
\bibitem [{\citenamefont {Lin}\ \emph {et~al.}(2008)\citenamefont {Lin},
  \citenamefont {Shreiber}, \citenamefont {Dimitriadis},\ and\ \citenamefont
  {Horkay}}]{Lin2008}%
  \BibitemOpen
  \bibfield  {author} {\bibinfo {author} {\bibfnamefont {D.~C.}\ \bibnamefont
  {Lin}}, \bibinfo {author} {\bibfnamefont {D.~I.}\ \bibnamefont {Shreiber}},
  \bibinfo {author} {\bibfnamefont {E.~K.}\ \bibnamefont {Dimitriadis}}, \ and\
  \bibinfo {author} {\bibfnamefont {F.}~\bibnamefont {Horkay}},\ }\href
  {\doibase 10.1007/s10237-008-0139-9} {\bibfield  {journal} {\bibinfo
  {journal} {Biomech Model Mechanobiol}\ }\textbf {\bibinfo {volume} {8}},\
  \bibinfo {pages} {345} (\bibinfo {year} {2008})}\BibitemShut {NoStop}%
\bibitem [{\citenamefont {Ducker}, \citenamefont {Senden},\ and\ \citenamefont
  {Pashley}(1991)}]{Ducker1991}%
  \BibitemOpen
  \bibfield  {author} {\bibinfo {author} {\bibfnamefont {W.~A.}\ \bibnamefont
  {Ducker}}, \bibinfo {author} {\bibfnamefont {T.~J.}\ \bibnamefont {Senden}},
  \ and\ \bibinfo {author} {\bibfnamefont {R.~M.}\ \bibnamefont {Pashley}},\
  }\href {\doibase 10.1038/353239a0} {\bibfield  {journal} {\bibinfo  {journal}
  {NATURE}\ }\textbf {\bibinfo {volume} {353}},\ \bibinfo {pages} {239}
  (\bibinfo {year} {1991})}\BibitemShut {NoStop}%
\bibitem [{\citenamefont {Alesutan}\ \emph {et~al.}(2013)\citenamefont
  {Alesutan}, \citenamefont {Seifert}, \citenamefont {Pakladok}, \citenamefont
  {Rheinlaender}, \citenamefont {Lebedeva}, \citenamefont {Towhid},
  \citenamefont {Stournaras}, \citenamefont {Voelkl}, \citenamefont
  {Sch\"{a}ffer},\ and\ \citenamefont {Lang}}]{Alesutan2013}%
  \BibitemOpen
  \bibfield  {author} {\bibinfo {author} {\bibfnamefont {I.}~\bibnamefont
  {Alesutan}}, \bibinfo {author} {\bibfnamefont {J.}~\bibnamefont {Seifert}},
  \bibinfo {author} {\bibfnamefont {T.}~\bibnamefont {Pakladok}}, \bibinfo
  {author} {\bibfnamefont {J.}~\bibnamefont {Rheinlaender}}, \bibinfo {author}
  {\bibfnamefont {A.}~\bibnamefont {Lebedeva}}, \bibinfo {author}
  {\bibfnamefont {S.~T.}\ \bibnamefont {Towhid}}, \bibinfo {author}
  {\bibfnamefont {C.}~\bibnamefont {Stournaras}}, \bibinfo {author}
  {\bibfnamefont {J.}~\bibnamefont {Voelkl}}, \bibinfo {author} {\bibfnamefont
  {T.~E.}\ \bibnamefont {Sch\"{a}ffer}}, \ and\ \bibinfo {author}
  {\bibfnamefont {F.}~\bibnamefont {Lang}},\ }\href {\doibase
  10.1159/000354475} {\bibfield  {journal} {\bibinfo  {journal} {Cell Physiol
  Biochem}\ }\textbf {\bibinfo {volume} {32}},\ \bibinfo {pages} {728}
  (\bibinfo {year} {2013})}\BibitemShut {NoStop}%
\bibitem [{\citenamefont {Indrieri}\ \emph {et~al.}(2011)\citenamefont
  {Indrieri}, \citenamefont {Podesta}, \citenamefont {Bongiorno}, \citenamefont
  {Marchesi},\ and\ \citenamefont {Milani}}]{Indrieri2011}%
  \BibitemOpen
  \bibfield  {author} {\bibinfo {author} {\bibfnamefont {M.}~\bibnamefont
  {Indrieri}}, \bibinfo {author} {\bibfnamefont {A.}~\bibnamefont {Podesta}},
  \bibinfo {author} {\bibfnamefont {G.}~\bibnamefont {Bongiorno}}, \bibinfo
  {author} {\bibfnamefont {D.}~\bibnamefont {Marchesi}}, \ and\ \bibinfo
  {author} {\bibfnamefont {P.}~\bibnamefont {Milani}},\ }\href {\doibase
  10.1063/1.3553499} {\bibfield  {journal} {\bibinfo  {journal} {REVIEW OF
  SCIENTIFIC INSTRUMENTS}\ }\textbf {\bibinfo {volume} {82}},\ \bibinfo {pages}
  {023708} (\bibinfo {year} {2011})}\BibitemShut {NoStop}%
\bibitem [{\citenamefont {Neto}\ and\ \citenamefont {Craig}(2001)}]{Neto2001}%
  \BibitemOpen
  \bibfield  {author} {\bibinfo {author} {\bibfnamefont {C.}~\bibnamefont
  {Neto}}\ and\ \bibinfo {author} {\bibfnamefont {V.~S.~J.}\ \bibnamefont
  {Craig}},\ }\href {\doibase 10.1021/la001506y} {\bibfield  {journal}
  {\bibinfo  {journal} {Langmuir}\ }\textbf {\bibinfo {volume} {17}},\ \bibinfo
  {pages} {2097} (\bibinfo {year} {2001})}\BibitemShut {NoStop}%
\bibitem [{\citenamefont {Bonaccurso}(2001)}]{Bonaccurso2001}%
  \BibitemOpen
  \bibfield  {author} {\bibinfo {author} {\bibfnamefont {E.}~\bibnamefont
  {Bonaccurso}},\ }\emph {\bibinfo {title} {Investigation of electrokinetic
  forces on single particles}},\ \href@noop {} {Ph.D. thesis},\ \bibinfo
  {school} {Universitat-Gesamthochschule Siegen, Siegen, Germany.} (\bibinfo
  {year} {2001})\BibitemShut {NoStop}%
\bibitem [{\citenamefont {Kuznetsov}\ and\ \citenamefont
  {Papastavrou}(2012)}]{Kuznetsov2012}%
  \BibitemOpen
  \bibfield  {author} {\bibinfo {author} {\bibfnamefont {V.}~\bibnamefont
  {Kuznetsov}}\ and\ \bibinfo {author} {\bibfnamefont {G.}~\bibnamefont
  {Papastavrou}},\ }\href {\doibase 10.1063/1.4765299} {\bibfield  {journal}
  {\bibinfo  {journal} {Review of Scientific Instruments}\ }\textbf {\bibinfo
  {volume} {83}},\ \bibinfo {pages} {116103} (\bibinfo {year}
  {2012})}\BibitemShut {NoStop}%
\bibitem [{\citenamefont {Gan}(2007)}]{Gan2007}%
  \BibitemOpen
  \bibfield  {author} {\bibinfo {author} {\bibfnamefont {Y.}~\bibnamefont
  {Gan}},\ }\href {\doibase 10.1063/1.2754076} {\bibfield  {journal} {\bibinfo
  {journal} {Review of Scientific Instruments}\ }\textbf {\bibinfo {volume}
  {78}},\ \bibinfo {pages} {081101} (\bibinfo {year} {2007})}\BibitemShut
  {NoStop}%
\bibitem [{\citenamefont {Mak}\ \emph {et~al.}(2006)\citenamefont {Mak},
  \citenamefont {Knoll}, \citenamefont {Weiner}, \citenamefont {Gorschluter},
  \citenamefont {Schirmeisen},\ and\ \citenamefont {Fuchs}}]{Mak2006}%
  \BibitemOpen
  \bibfield  {author} {\bibinfo {author} {\bibfnamefont {L.~H.}\ \bibnamefont
  {Mak}}, \bibinfo {author} {\bibfnamefont {M.}~\bibnamefont {Knoll}}, \bibinfo
  {author} {\bibfnamefont {D.}~\bibnamefont {Weiner}}, \bibinfo {author}
  {\bibfnamefont {A.}~\bibnamefont {Gorschluter}}, \bibinfo {author}
  {\bibfnamefont {A.}~\bibnamefont {Schirmeisen}}, \ and\ \bibinfo {author}
  {\bibfnamefont {H.}~\bibnamefont {Fuchs}},\ }\href {\doibase
  10.1063/1.2190068} {\bibfield  {journal} {\bibinfo  {journal} {Review of
  Scientific Instruments}\ }\textbf {\bibinfo {volume} {77}},\ \bibinfo {pages}
  {046104} (\bibinfo {year} {2006})}\BibitemShut {NoStop}%
\bibitem [{\citenamefont {Hertz}(1881)}]{Hertz1881}%
  \BibitemOpen
  \bibfield  {author} {\bibinfo {author} {\bibfnamefont {H.}~\bibnamefont
  {Hertz}},\ }\href@noop {} {\bibfield  {journal} {\bibinfo  {journal} {J.
  Reine Ang. Math}\ }\textbf {\bibinfo {volume} {92}},\ \bibinfo {pages} {156}
  (\bibinfo {year} {1881})}\BibitemShut {NoStop}%
\bibitem [{\citenamefont {Heuberger}, \citenamefont {Dietler},\ and\
  \citenamefont {Schlapbach}(1996)}]{Heuberger1996}%
  \BibitemOpen
  \bibfield  {author} {\bibinfo {author} {\bibfnamefont {M.}~\bibnamefont
  {Heuberger}}, \bibinfo {author} {\bibfnamefont {G.}~\bibnamefont {Dietler}},
  \ and\ \bibinfo {author} {\bibfnamefont {L.}~\bibnamefont {Schlapbach}},\
  }\href {\doibase 10.1116/1.588525} {\bibfield  {journal} {\bibinfo  {journal}
  {JOURNAL OF VACUUM SCIENCE \& TECHNOLOGY B}\ }\textbf {\bibinfo {volume}
  {14}},\ \bibinfo {pages} {1250} (\bibinfo {year} {1996})},\ \bibinfo {note}
  {8th International Conference on Scanning Tunneling Microscopy and Related
  Methods (STM 95), SNOWMASS, CO, JUL 25-29, 1995}\BibitemShut {NoStop}%
\bibitem [{\citenamefont {Johnson}(1985)}]{Johnson1985}%
  \BibitemOpen
  \bibfield  {author} {\bibinfo {author} {\bibfnamefont {K.~L.}\ \bibnamefont
  {Johnson}},\ }\href@noop {} {\emph {\bibinfo {title} {Contact Mechanics}}}\
  (\bibinfo  {publisher} {CAMBRIDGE UNIVERSITY PRESS},\ \bibinfo {year}
  {1985})\BibitemShut {NoStop}%
\bibitem [{\citenamefont {Sneddon}(1965)}]{Sneddon1965}%
  \BibitemOpen
  \bibfield  {author} {\bibinfo {author} {\bibfnamefont {I.~N.}\ \bibnamefont
  {Sneddon}},\ }\href {\doibase 10.1016/0020-7225(65)90019-4} {\bibfield
  {journal} {\bibinfo  {journal} {INTERNATIONAL JOURNAL OF ENGINEERING
  SCIENCE}\ }\textbf {\bibinfo {volume} {3}},\ \bibinfo {pages} {47} (\bibinfo
  {year} {1965})}\BibitemShut {NoStop}%
\bibitem [{\citenamefont {O'Callaghan}\ \emph {et~al.}(2011)\citenamefont
  {O'Callaghan}, \citenamefont {Job}, \citenamefont {Dull},\ and\ \citenamefont
  {Hlady}}]{OCallaghan2011}%
  \BibitemOpen
  \bibfield  {author} {\bibinfo {author} {\bibfnamefont {R.}~\bibnamefont
  {O'Callaghan}}, \bibinfo {author} {\bibfnamefont {K.~M.}\ \bibnamefont
  {Job}}, \bibinfo {author} {\bibfnamefont {R.~O.}\ \bibnamefont {Dull}}, \
  and\ \bibinfo {author} {\bibfnamefont {V.}~\bibnamefont {Hlady}},\ }\href
  {\doibase 10.1152/ajplung.00342.2010} {\bibfield  {journal} {\bibinfo
  {journal} {Am J Physiol Lung Cell Mol Physiol}\ }\textbf {\bibinfo {volume}
  {301}},\ \bibinfo {pages} {L353} (\bibinfo {year} {2011})}\BibitemShut
  {NoStop}%
\bibitem [{\citenamefont {Sokolov}\ \emph {et~al.}(2007)\citenamefont
  {Sokolov}, \citenamefont {Iyer}, \citenamefont {Subba-Rao}, \citenamefont
  {Gaikwad},\ and\ \citenamefont {Woodworth}}]{Sokolov2007}%
  \BibitemOpen
  \bibfield  {author} {\bibinfo {author} {\bibfnamefont {I.}~\bibnamefont
  {Sokolov}}, \bibinfo {author} {\bibfnamefont {S.}~\bibnamefont {Iyer}},
  \bibinfo {author} {\bibfnamefont {V.}~\bibnamefont {Subba-Rao}}, \bibinfo
  {author} {\bibfnamefont {R.~M.}\ \bibnamefont {Gaikwad}}, \ and\ \bibinfo
  {author} {\bibfnamefont {C.~D.}\ \bibnamefont {Woodworth}},\ }\href {\doibase
  10.1063/1.2757104} {\bibfield  {journal} {\bibinfo  {journal} {APPLIED
  PHYSICS LETTERS}\ }\textbf {\bibinfo {volume} {91}},\ \bibinfo {pages}
  {023902} (\bibinfo {year} {2007})}\BibitemShut {NoStop}%
\bibitem [{\citenamefont {Iyer}\ \emph {et~al.}(2009)\citenamefont {Iyer},
  \citenamefont {Gaikwad}, \citenamefont {Subba-Rao}, \citenamefont
  {Woodworth},\ and\ \citenamefont {Sokolov}}]{Iyer2009}%
  \BibitemOpen
  \bibfield  {author} {\bibinfo {author} {\bibfnamefont {S.}~\bibnamefont
  {Iyer}}, \bibinfo {author} {\bibfnamefont {R.~M.}\ \bibnamefont {Gaikwad}},
  \bibinfo {author} {\bibfnamefont {V.}~\bibnamefont {Subba-Rao}}, \bibinfo
  {author} {\bibfnamefont {C.~D.}\ \bibnamefont {Woodworth}}, \ and\ \bibinfo
  {author} {\bibfnamefont {I.}~\bibnamefont {Sokolov}},\ }\href {\doibase
  10.1038/NNANO.2009.77} {\bibfield  {journal} {\bibinfo  {journal} {NATURE
  NANOTECHNOLOGY}\ }\textbf {\bibinfo {volume} {4}},\ \bibinfo {pages} {389}
  (\bibinfo {year} {2009})}\BibitemShut {NoStop}%
\bibitem [{\citenamefont {Akhremitchev}\ and\ \citenamefont
  {Walker}(1999)}]{Akhremitchev1999}%
  \BibitemOpen
  \bibfield  {author} {\bibinfo {author} {\bibfnamefont {B.~B.}\ \bibnamefont
  {Akhremitchev}}\ and\ \bibinfo {author} {\bibfnamefont {G.~C.}\ \bibnamefont
  {Walker}},\ }\href {\doibase 10.1021/la980585z} {\bibfield  {journal}
  {\bibinfo  {journal} {Langmuir}\ }\textbf {\bibinfo {volume} {15}},\ \bibinfo
  {pages} {5630} (\bibinfo {year} {1999})}\BibitemShut {NoStop}%
\bibitem [{\citenamefont {Oommen}\ and\ \citenamefont
  {Van~Vliet}(2006)}]{Oommen2006}%
  \BibitemOpen
  \bibfield  {author} {\bibinfo {author} {\bibfnamefont {B.}~\bibnamefont
  {Oommen}}\ and\ \bibinfo {author} {\bibfnamefont {K.}~\bibnamefont
  {Van~Vliet}},\ }\href {\doibase 10.1016/j.tsf.2006.01.069} {\bibfield
  {journal} {\bibinfo  {journal} {Thin Solid Films}\ }\textbf {\bibinfo
  {volume} {513}},\ \bibinfo {pages} {235} (\bibinfo {year}
  {2006})}\BibitemShut {NoStop}%
\bibitem [{\citenamefont {Sun}, \citenamefont {Bell},\ and\ \citenamefont
  {Zheng}(1995)}]{Sun1995}%
  \BibitemOpen
  \bibfield  {author} {\bibinfo {author} {\bibfnamefont {Y.}~\bibnamefont
  {Sun}}, \bibinfo {author} {\bibfnamefont {T.}~\bibnamefont {Bell}}, \ and\
  \bibinfo {author} {\bibfnamefont {S.}~\bibnamefont {Zheng}},\ }\href
  {\doibase 10.1016/0040-6090(94)06357-5} {\bibfield  {journal} {\bibinfo
  {journal} {Thin Solid Films}\ }\textbf {\bibinfo {volume} {258}},\ \bibinfo
  {pages} {198} (\bibinfo {year} {1995})}\BibitemShut {NoStop}%
\bibitem [{\citenamefont {Long}\ \emph {et~al.}(2011)\citenamefont {Long},
  \citenamefont {Hall}, \citenamefont {Wu},\ and\ \citenamefont
  {Hui}}]{Long2011}%
  \BibitemOpen
  \bibfield  {author} {\bibinfo {author} {\bibfnamefont {R.}~\bibnamefont
  {Long}}, \bibinfo {author} {\bibfnamefont {M.~S.}\ \bibnamefont {Hall}},
  \bibinfo {author} {\bibfnamefont {M.}~\bibnamefont {Wu}}, \ and\ \bibinfo
  {author} {\bibfnamefont {C.-Y.}\ \bibnamefont {Hui}},\ }\href {\doibase
  10.1016/j.bpj.2011.06.049} {\bibfield  {journal} {\bibinfo  {journal}
  {Biophysical Journal}\ }\textbf {\bibinfo {volume} {101}},\ \bibinfo {pages}
  {643} (\bibinfo {year} {2011})}\BibitemShut {NoStop}%
\bibitem [{\citenamefont {Santos}\ \emph {et~al.}(2012)\citenamefont {Santos},
  \citenamefont {Rebelo}, \citenamefont {Araujo}, \citenamefont {Barros},\ and\
  \citenamefont {de~Sousa}}]{Santos2012}%
  \BibitemOpen
  \bibfield  {author} {\bibinfo {author} {\bibfnamefont {J.~A.~C.}\
  \bibnamefont {Santos}}, \bibinfo {author} {\bibfnamefont {L.~M.}\
  \bibnamefont {Rebelo}}, \bibinfo {author} {\bibfnamefont {A.~C.}\
  \bibnamefont {Araujo}}, \bibinfo {author} {\bibfnamefont {E.~B.}\
  \bibnamefont {Barros}}, \ and\ \bibinfo {author} {\bibfnamefont {J.~S.}\
  \bibnamefont {de~Sousa}},\ }\href {\doibase 10.1039/c2sm07062f} {\bibfield
  {journal} {\bibinfo  {journal} {Soft Matter}\ }\textbf {\bibinfo {volume}
  {8}},\ \bibinfo {pages} {4441} (\bibinfo {year} {2012})}\BibitemShut
  {NoStop}%
\bibitem [{\citenamefont {Gavara}\ and\ \citenamefont
  {Chadwick}(2012)}]{Gavara2012}%
  \BibitemOpen
  \bibfield  {author} {\bibinfo {author} {\bibfnamefont {N.}~\bibnamefont
  {Gavara}}\ and\ \bibinfo {author} {\bibfnamefont {R.~S.}\ \bibnamefont
  {Chadwick}},\ }\href {\doibase 10.1038/NNANO.2012.163} {\bibfield  {journal}
  {\bibinfo  {journal} {NATURE NANOTECHNOLOGY}\ }\textbf {\bibinfo {volume}
  {7}},\ \bibinfo {pages} {733} (\bibinfo {year} {2012})}\BibitemShut {NoStop}%
\bibitem [{\citenamefont {Polyakov}\ \emph {et~al.}(2011)\citenamefont
  {Polyakov}, \citenamefont {Soussen}, \citenamefont {Duan}, \citenamefont
  {Duval}, \citenamefont {Brie},\ and\ \citenamefont
  {Francius}}]{Polyakov2011}%
  \BibitemOpen
  \bibfield  {author} {\bibinfo {author} {\bibfnamefont {P.}~\bibnamefont
  {Polyakov}}, \bibinfo {author} {\bibfnamefont {C.}~\bibnamefont {Soussen}},
  \bibinfo {author} {\bibfnamefont {J.}~\bibnamefont {Duan}}, \bibinfo {author}
  {\bibfnamefont {J.~F.~L.}\ \bibnamefont {Duval}}, \bibinfo {author}
  {\bibfnamefont {D.}~\bibnamefont {Brie}}, \ and\ \bibinfo {author}
  {\bibfnamefont {G.}~\bibnamefont {Francius}},\ }\href {\doibase
  10.1371/journal.pone.0018887} {\bibfield  {journal} {\bibinfo  {journal}
  {PLOS ONE}\ }\textbf {\bibinfo {volume} {6}},\ \bibinfo {pages} {e18887}
  (\bibinfo {year} {2011})}\BibitemShut {NoStop}%
\bibitem [{\citenamefont {Lin}, \citenamefont {Dimitriadis},\ and\
  \citenamefont {Horkay}(2007)}]{Lin2007}%
  \BibitemOpen
  \bibfield  {author} {\bibinfo {author} {\bibfnamefont {D.~C.}\ \bibnamefont
  {Lin}}, \bibinfo {author} {\bibfnamefont {E.~K.}\ \bibnamefont
  {Dimitriadis}}, \ and\ \bibinfo {author} {\bibfnamefont {F.}~\bibnamefont
  {Horkay}},\ }\href {\doibase 10.1115/1.2720924} {\bibfield  {journal}
  {\bibinfo  {journal} {JOURNAL OF BIOMECHANICAL ENGINEERING-TRANSACTIONS OF
  THE ASME}\ }\textbf {\bibinfo {volume} {129}},\ \bibinfo {pages} {430}
  (\bibinfo {year} {2007})}\BibitemShut {NoStop}%
\bibitem [{\citenamefont {Ferraris}\ \emph {et~al.}(2014)\citenamefont
  {Ferraris}, \citenamefont {Schulte}, \citenamefont {Buttiglione},
  \citenamefont {De~Lorenzi}, \citenamefont {Piontini}, \citenamefont
  {Galluzzi}, \citenamefont {Podesta}, \citenamefont {Madsen},\ and\
  \citenamefont {Sidenius}}]{Ferraris2014}%
  \BibitemOpen
  \bibfield  {author} {\bibinfo {author} {\bibfnamefont {G.~M.~S.}\
  \bibnamefont {Ferraris}}, \bibinfo {author} {\bibfnamefont {C.}~\bibnamefont
  {Schulte}}, \bibinfo {author} {\bibfnamefont {V.}~\bibnamefont
  {Buttiglione}}, \bibinfo {author} {\bibfnamefont {V.}~\bibnamefont
  {De~Lorenzi}}, \bibinfo {author} {\bibfnamefont {A.}~\bibnamefont
  {Piontini}}, \bibinfo {author} {\bibfnamefont {M.}~\bibnamefont {Galluzzi}},
  \bibinfo {author} {\bibfnamefont {A.}~\bibnamefont {Podesta}}, \bibinfo
  {author} {\bibfnamefont {C.~D.}\ \bibnamefont {Madsen}}, \ and\ \bibinfo
  {author} {\bibfnamefont {N.}~\bibnamefont {Sidenius}},\ }\href {\doibase
  10.15252/embj.201387611} {\bibfield  {journal} {\bibinfo  {journal} {The EMBO
  Journal}\ } (\bibinfo {year} {2014}),\ 10.15252/embj.201387611}\BibitemShut
  {NoStop}%
\bibitem [{\citenamefont {Nair}\ \emph {et~al.}(2008)\citenamefont {Nair},
  \citenamefont {Joel}, \citenamefont {Wan}, \citenamefont {Lowey},
  \citenamefont {Rould},\ and\ \citenamefont {Trybus}}]{Nair2008}%
  \BibitemOpen
  \bibfield  {author} {\bibinfo {author} {\bibfnamefont {U.~B.}\ \bibnamefont
  {Nair}}, \bibinfo {author} {\bibfnamefont {P.~B.}\ \bibnamefont {Joel}},
  \bibinfo {author} {\bibfnamefont {Q.}~\bibnamefont {Wan}}, \bibinfo {author}
  {\bibfnamefont {S.}~\bibnamefont {Lowey}}, \bibinfo {author} {\bibfnamefont
  {M.~A.}\ \bibnamefont {Rould}}, \ and\ \bibinfo {author} {\bibfnamefont
  {K.~M.}\ \bibnamefont {Trybus}},\ }\href {\doibase 10.1016/j.jmb.2008.09.082}
  {\bibfield  {journal} {\bibinfo  {journal} {JOURNAL OF MOLECULAR BIOLOGY}\
  }\textbf {\bibinfo {volume} {384}},\ \bibinfo {pages} {848} (\bibinfo {year}
  {2008})}\BibitemShut {NoStop}%
\bibitem [{\citenamefont {Bilodeau}(1992)}]{Bilodeau1992}%
  \BibitemOpen
  \bibfield  {author} {\bibinfo {author} {\bibfnamefont {G.~G.}\ \bibnamefont
  {Bilodeau}},\ }\href {\doibase 10.1115/1.2893754} {\bibfield  {journal}
  {\bibinfo  {journal} {JOURNAL OF APPLIED MECHANICS-TRANSACTIONS OF THE ASME}\
  }\textbf {\bibinfo {volume} {59}},\ \bibinfo {pages} {519} (\bibinfo {year}
  {1992})}\BibitemShut {NoStop}%
\bibitem [{\citenamefont {Harding}\ and\ \citenamefont
  {Sneddon}(1945)}]{Harding1945}%
  \BibitemOpen
  \bibfield  {author} {\bibinfo {author} {\bibfnamefont {J.~W.}\ \bibnamefont
  {Harding}}\ and\ \bibinfo {author} {\bibfnamefont {I.~N.}\ \bibnamefont
  {Sneddon}},\ }\href {\doibase 10.1017/s0305004100022325} {\bibfield
  {journal} {\bibinfo  {journal} {Math. Proc. Camb. Phil. Soc.}\ }\textbf
  {\bibinfo {volume} {41}},\ \bibinfo {pages} {16} (\bibinfo {year}
  {1945})}\BibitemShut {NoStop}%
\bibitem [{\citenamefont {Zeng}\ and\ \citenamefont {Li}(2013)}]{X.Zeng2013}%
  \BibitemOpen
  \bibfield  {author} {\bibinfo {author} {\bibfnamefont {X.}~\bibnamefont
  {Zeng}}\ and\ \bibinfo {author} {\bibfnamefont {S.}~\bibnamefont {Li}},\
  }\href {\doibase 10.1061/(ASCE)EM.1943-7889.0000735} {\bibfield  {journal}
  {\bibinfo  {journal} {JOURNAL OF ENGINEERING MECHANICS}\ }\textbf {\bibinfo
  {volume} {140}},\ \bibinfo {pages} {04013003} (\bibinfo {year}
  {2013})}\BibitemShut {NoStop}%
\bibitem [{\citenamefont {Weafer}\ \emph {et~al.}(2012)\citenamefont {Weafer},
  \citenamefont {McGarry}, \citenamefont {{van Es}}, \citenamefont
  {Kilpatrick}, \citenamefont {Ronan}, \citenamefont {Nolan},\ and\
  \citenamefont {Jarvis}}]{Weafer2012}%
  \BibitemOpen
  \bibfield  {author} {\bibinfo {author} {\bibfnamefont {P.~P.}\ \bibnamefont
  {Weafer}}, \bibinfo {author} {\bibfnamefont {J.~P.}\ \bibnamefont {McGarry}},
  \bibinfo {author} {\bibfnamefont {M.~H.}\ \bibnamefont {{van Es}}}, \bibinfo
  {author} {\bibfnamefont {J.~I.}\ \bibnamefont {Kilpatrick}}, \bibinfo
  {author} {\bibfnamefont {W.}~\bibnamefont {Ronan}}, \bibinfo {author}
  {\bibfnamefont {D.~R.}\ \bibnamefont {Nolan}}, \ and\ \bibinfo {author}
  {\bibfnamefont {S.~P.}\ \bibnamefont {Jarvis}},\ }\href {\doibase
  10.1063/1.4752023} {\bibfield  {journal} {\bibinfo  {journal} {Rev Sci
  Instrum}\ }\textbf {\bibinfo {volume} {83}},\ \bibinfo {pages} {093709}
  (\bibinfo {year} {2012})}\BibitemShut {NoStop}%
\bibitem [{\citenamefont {Butt}\ and\ \citenamefont
  {Jaschke}(1995)}]{Butt1995}%
  \BibitemOpen
  \bibfield  {author} {\bibinfo {author} {\bibfnamefont {H.~J.}\ \bibnamefont
  {Butt}}\ and\ \bibinfo {author} {\bibfnamefont {M.}~\bibnamefont {Jaschke}},\
  }\href {\doibase 10.1088/0957-4484/6/1/001} {\bibfield  {journal} {\bibinfo
  {journal} {Nanotechnology}\ }\textbf {\bibinfo {volume} {6}},\ \bibinfo
  {pages} {1} (\bibinfo {year} {1995})}\BibitemShut {NoStop}%
\bibitem [{\citenamefont {Sader}, \citenamefont {Chon},\ and\ \citenamefont
  {Mulvaney}(1999)}]{Sader1999}%
  \BibitemOpen
  \bibfield  {author} {\bibinfo {author} {\bibfnamefont {J.~E.}\ \bibnamefont
  {Sader}}, \bibinfo {author} {\bibfnamefont {J.~W.~M.}\ \bibnamefont {Chon}},
  \ and\ \bibinfo {author} {\bibfnamefont {P.}~\bibnamefont {Mulvaney}},\
  }\href {\doibase 10.1063/1.1150021} {\bibfield  {journal} {\bibinfo
  {journal} {Review of Scientific Instruments}\ }\textbf {\bibinfo {volume}
  {70}},\ \bibinfo {pages} {3967} (\bibinfo {year} {1999})}\BibitemShut
  {NoStop}%
\bibitem [{\citenamefont {Hutter}(2005)}]{Hutter2005}%
  \BibitemOpen
  \bibfield  {author} {\bibinfo {author} {\bibfnamefont {J.~L.}\ \bibnamefont
  {Hutter}},\ }\href {\doibase 10.1021/la047670t} {\bibfield  {journal}
  {\bibinfo  {journal} {Langmuir}\ }\textbf {\bibinfo {volume} {21}},\ \bibinfo
  {pages} {2630} (\bibinfo {year} {2005})}\BibitemShut {NoStop}%
\bibitem [{Note1()}]{Note1}%
  \BibitemOpen
  \bibinfo {note} {While the statistical error associated to the deflection
  sensitivity can be rather small, provided several force curves acquired in
  different location are used, and the corresponding values of the sensitivity
  averaged, there are subtle and rather poorly accountable systematic errors
  which can make the total error larger~\cite {Weafer2012}, so that a 5\%
  estimation is more reasonable (if not optimistic; this issue is discussed in
  details in a forthcoming publication~\cite {COST_paper})}\BibitemShut
  {NoStop}%
\bibitem [{\citenamefont {Lybanon}(1985)}]{Lybanon1985}%
  \BibitemOpen
  \bibfield  {author} {\bibinfo {author} {\bibfnamefont {M.}~\bibnamefont
  {Lybanon}},\ }\href {\doibase 10.1016/0098-3004(85)90032-9} {\bibfield
  {journal} {\bibinfo  {journal} {Computers \& Geosciences}\ }\textbf {\bibinfo
  {volume} {11}},\ \bibinfo {pages} {501} (\bibinfo {year} {1985})}\BibitemShut
  {NoStop}%
\end{thebibliography}%

\end{document}